\documentclass[%
 reprint,
 amsmath,amssymb,
 aps,
]{revtex4-1}

\usepackage{graphicx}
\usepackage{dcolumn}
\usepackage{bm}
\usepackage{color,soul}
\usepackage{amsmath}
\usepackage{xcolor}

\begin{document}


\title{Tunable nanophotonics enabled by chalcogenide phase-change materials}

\author{Sajjad Abdollahramezani$^1$}
\author{Omid Hemmatyar$^1$}
\author{Hossein Taghinejad$^1$}
\author{Alex Krasnok$^2$}
\author{Yashar Kiarashinejad$^1$}
\author{Mohammadreza Zandehshahvar$^1$}
\author{Andrea Al{\`u}$^{2, 3}$}
\author{Ali Adibi$^{1,}$}

\email{ali.adibi@ece.gatech.edu}
\affiliation{$^1$ School of Electrical and Computer Engineering, Georgia Institute of Technology, 778 Atlantic Drive NW, Atlanta, GA 30332-0250, United States}
\affiliation{$^2$ Photonics Initiative, Advanced Science Research Center, City University of New York, New York, NY 10031, USA}
\affiliation{$^3$ Physics Program, Graduate Center, City University of New York, New York, NY 10016, USA}

\date{\today}

\begin{abstract}
Nanophotonics has garnered intensive attention due to its unique capabilities in molding the flow of light in the subwavelength regime. Metasurfaces (MSs) and photonic integrated circuits (PICs) enable the realization of mass-producible, cost-effective, and highly efficient flat optical components for imaging, sensing, and communications. In order to enable nanophotonics with multi-purpose functionalities, chalcogenide phase-change materials (PCMs) have been introduced as a promising platform for tunable and reconfigurable nanophotonic frameworks. Integration of non-volatile chalcogenide PCMs with unique properties such as drastic optical contrasts, fast switching speeds, and long-term stability grants substantial reconfiguration to the more conventional static nanophotonic platforms. In this review, we discuss state-of-the-art developments as well as emerging trends in tunable MSs and PICs using chalcogenide PCMs. We outline the unique material properties, structural transformation, electro-optic, and thermo-optic effects of well-established classes of chalcogenide PCMs. The emerging deep learning-based approaches for the optimization of reconfigurable MSs and the analysis of light-matter interactions are also discussed. The review is concluded by discussing existing challenges in the realization of adjustable nanophotonics and a perspective on the possible developments in this promising area.

\end{abstract}


\maketitle


\section{Introduction}

In recent years, an ever-increasing competition has arisen between two exciting technologies, i.e., electronics and photonics, to realize ubiquitous functionalities from imaging and communication to sensing and medicine \cite{sun2015single, atabaki2018integrating, zeng2014fiber}. Despite the striking progress of electronics in data processing and storage as well as logic operations, photonic platforms are preferred for information transportation, mostly because of the higher speed and larger bandwidth \cite{rickman2014commercialization, thomson2016roadmap, genevet2017recent}. However, to beat the fingerprints of electronics, i.e., integrability and reprogrammability, photonics needs to evolve into the next generation of miniaturized systems incorporating reconfigurable platforms with adjustable properties to fully manipulate the key features of photons, the information carrier in photonic platforms. 

Metasurfaces (MSs) and photonics integrated circuits (PICs) are currently the two mature nanophotonics platforms the manipulation of light for on-chip and free-space applications, respectively. MSs, the two-dimensional analog of metamaterials, have attracted significant attention due to their unprecedented ability to control incident electromagnetic fields in the subwavelength regime \cite{yu2011light,staude2013tailoring,arbabi2015dielectric}. Owing to their judiciously engineered optical scatterers, or the so-called meta-atoms, arranged in a periodic or aperiodic texture, the amplitude, phase, polarization, and frequency of the impinging light can be spatially and spectrally manipulated, making a big step towards the realization of the next-generation flat optics \cite{khorasaninejad2016metalenses,ding2018bifunctional,yesilkoy2019ultrasensitive}. A myriad of novel phenomena and optical functionalities have thus been demonstrated including beam shaping and steering \cite{arbabi2016multiwavelength}, polarization generation \cite{arbabi2018full}, large-angle holography \cite{lee2018complete}, directional lasing \cite{ha2018directional}, analog computing \cite{abdollahramezani2017dielectric}, quantum emission \cite{tran2017deterministic}, nonlinear generation \cite{krasnok2018nonlinear}, structural coloration \cite{hemmatyar2019full}, and biosensing \cite{krasnok2018spectroscopy}.
Similarly, complementary metal-oxide semiconductor (CMOS)-compatible silicon (Si) and silicon nitride (SiN) PICs have experienced phenomenal transformations over the past decade. Several key infrastructures such as modulators, (de)multiplexers, filters, and detectors featuring low optical attenuation, high optical mode confinement, wide operational bandwidth from visible (vis) to beyond infrared (IR), and immunity to electromagnetic interference have enabled a wave of ubiquitous applications including data communication, spectroscopy, and sensing. 

However, the main challenge with the developed static MS configurations is that their assigned functionality remains fixed once fabricated, hindering many practical free-space applications that need real-time tuning \cite{jahani2016all,kuznetsov2016optically,baranov2017all,staude2017metamaterial,ding2017gradient,su2018advances,ding2018review,kamali2018review,neshev2018optical,tseng2018metalenses,luo2018subwavelength,he2018high,wen2018geometric,chen2019metasurface,zang2019chromatic,staude2019all,sung2019progresses,vaskin2019light,chen2019empowered,koshelev2019nonradiating,fan2019constructing}. On the other hand, most on-chip photonic demonstrations are based on the application-specific designs of PICs. To orient the potential of PICs for emerging applications of microwave photonics, neuromorphic computing, and quantum computation, high reconfigurable and general-purpose platforms similar to the field-programmable gate array (FPGA) in electronics are indispensable \cite{zhuang2015programmable,capmany2016microwave,harris2017quantum,perez2017multipurpose,shen2017deep,tait2017neuromorphic,wang2018multidimensional,kim2019hybrid}.

There exist several approaches in which external stimuli such as electrical current, electrostatic fields, heat, chemical reactions, mechanical forces,  magnetic fields, and optical pumping are applied to adapt the functionality of MSs dynamically. Such approaches rely on the incorporation of two-dimensional (2D) materials (including graphene and transition-metal dichalcogenides) \cite{sherrott2017experimental}, highly doped semiconductors [such as Si and germanium (Ge)] \cite{lewi2015widely}, transparent conductive oxides (such as indium tin oxide) \cite{kafaie2018dual}, phase-transition materials (such as vanadium dioxide) \cite{kim2019phase}, metal hydrides (such as hydrogenated palladium and magnesium) \cite{palm2018dynamic}, liquid crystals \cite{komar2017electrically}, micro-electromechanical devices \cite{arbabi2018mems}, elastic platforms \cite{kamali2016highly}, and plasmonic‐metal/semiconductor heterostructures \cite{zuev2016fabrication,taghinejad2018ultrafast,lepeshov2019hybrid} with nanophotonics frameworks. On the other hand, PICs primarily rely on intrinsic thermo-optics and electro-optics effects of the Si and SiN to enable reconfigurable building blocks. However, the modest range of these properties has spurred the development of hybrid integrated platforms employing secondary materials including electro-optics polymers (such as organic polymers) \cite{koos2009all}, 2D materials (such as graphene) \cite{phare2015graphene}, phase-transition materials \cite{markov2015optically}, noble metals (such as Au) \cite{haffner2015all}, and semiconductors (such as germanium) \cite{feng2012high}.

Amongst existing approaches, the integration of phase-change materials (PCMs), specially chalcogenide alloys, with nanophotonic platforms offers the most promising path to reconfigurable optical functionalities. This stems from a unique property of chalcogenide PCMs that allows for reliable and repeatable switching of its optical/electrical attributes over billions of switching cycles.
The most critical concern for the realization of fully reconfigurable PCM-based MSs and PICs is the development of a reliable approach for the phase conversion in the PCM element. So far, most reports have relied on the direct thermal conversion of plain or patterned arrays of PCM inclusions by external stimuli, such as bulky thermal heaters or focused light from a continuous wave laser with a wide beamwidth covering the whole surface of the structure. It is worth mentioning that these approaches are one-way, meaning that only full or partial crystallization of initially amorphous PCMs can be achieved. To exploit the potential of PCMs to the fullest extent, reprogrammable pixelated PCM-based MSs and PICs capable of locally controlling the amplitude and phase profiles of the scattered or guiding light are of great importance. To enable tuning at the pixel level, employing short electrical currents or laser pulses as external stimuli for selective switching of the PCM state within the MS unit-cells is essential. This is also applicable to the reconfigurable PICs where optical functionalities are adjusted through the local modification of the state of incorporated PCM nanostructures. The existing laser switching methods and optical set-ups for characterization of dynamic PCM-based nanophotonic devices were outlined in Ref.~\cite{behera2017laser}. At present, optical switching for addressing individual subwavelength PCM patterns appears more accessible; however, an extensive effort is required to adopt an electrical alternative.

Due to the fast-developing research in dynamic nanophotonics, several review articles have been published that extensively investigated the pros and cons of the available tuning approaches \cite{raeis2017metasurfaces,ferrera2017dynamic,wuttig2017phase,miller2018optical,bang2018recent,nemati2018tunable,krasnok2019active,hail2019optical,paniagua2019active,cao2019fundamentals,cui2019tunable,zou2019resonant,he2019tunable,taghinejad2019all,ding2019dynamic,shaltout2019spatiotemporal,kang2019recent,miller2018optical,perez2018programmable,harris2018linear, wright2019integrated}. In this article, we specifically outline the current achievements and ongoing developments in hybrid PCM-based nanophotonics technology, including both on-demand MS and PIC platforms. We mainly focus on the application of PCM-based MSs in the mid-infrared (mid-IR) and near-infrared (near-IR) frequency regimes, and sparsely discuss applications in the visible (vis) spectral range. 

The paper is organized as follows. We first introduce the fundamental properties of chalcogenide PCMs. Next, we review reconfigurable multifunctional MSs (both plasmonic and dielectric) for global and local tailoring of amplitude and phase of light. Then, we discuss the emergence of chalcogenide PCMs in the PIC technology, which is envisioned for the next generation of high-speed CMOS-compatible computation and communications systems. Next, we outline the emerging field of deep learning in analysis, design, and optimization of dynamic PCM-based nanophotonics. Finally, we conclude the article with an outlook and a perspective on the rapidly growing field of reconfigurable nanophotonics using nonvolatile chalcogenide PCMs.

\section{Chalcogenide-Based PCMs: Material Properties}

Reversible switching of a material’s phase between amorphous and crystalline states and its application in data storage date back to the 1960s in chemical and metallurgical studies of ovonic threshold switching in disordered structures \cite{ovshinsky1968reversible}. Since then, search for new PCMs has led to a diverse portfolio of potential candidates that includes elemental materials such as Si \cite{hajto1994metal} and compounds such as transition-metal oxides \cite{argall1968switching} and chalcogenide glasses \cite{owen1973electronic,wuttig2005phase}. For most applications, an ideal PCM should offer high-speed and low-power phase-switching, a large number of switching cycles, long-term thermal stability of the amorphous phase, and a large optical/electrical contrast between two phases. Considering these criteria, chalcogenide glasses based on germanium (Ge)-antimony (Sb)-telluride (Te) (or shortly GST) alloys, and particularly Ge$_2$Sb$_2$Te$_5$, stand out in the pool of PCM materials. Phase transition in GST alloys is accompanied by a large refractive index contrast (e.g., $\Delta$n $\approx$ 1.5 at 405 nm wavelength) and resistivity change ($\sim$3 orders of magnitude), features that have found immediate applications in commercialized rewritable optical disks \cite{meinders2006optical,yamada1987high} and electronic memories \cite{owen1973electronic,neale1973application}. Most importantly, the property change in GST is non-volatile, meaning that the static power-consumption for data storage is virtually zero. 

\begin{figure}
	\centering
	\includegraphics[trim={0cm 0cm 0cm 0cm},width=0.5\textwidth, clip]{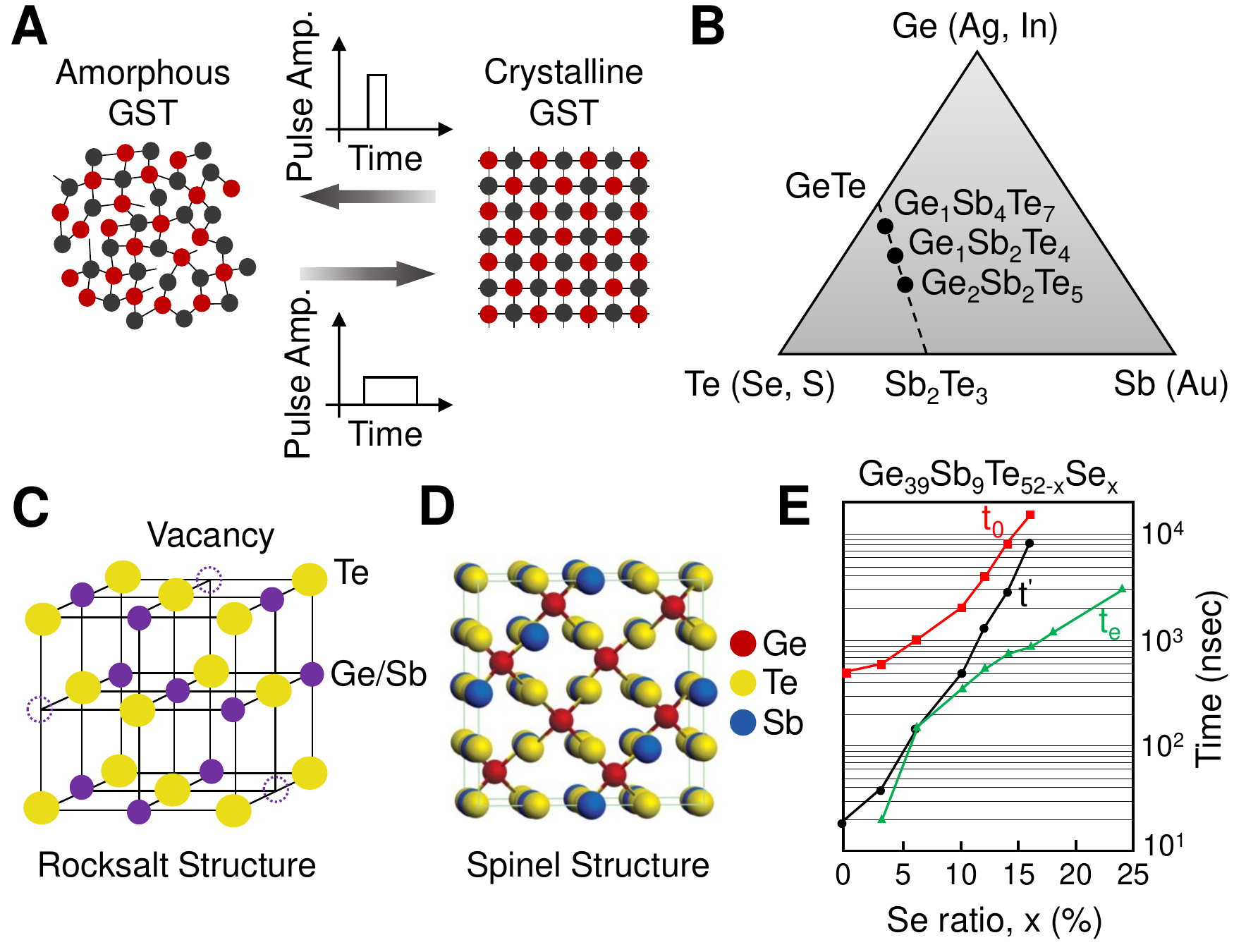}
	\caption{Principles of phase transformation in GST-based PCMs. (A) Schematic illustration of the reversible switching of the GST phase between amorphous and crystalline states. A long pulse with a moderate amplitude enables amorphous-to-crystalline conversion. A short pulse with a large amplitude enables crystalline-to-amorphous conversion via the melt-quenching process. (B) The phase diagram of the Ge-Sb-Te system. The dashed line shows the pseudo-binary line, which connects GeTe and Sb$_2$Te$_3$ compositions. Most popular GST-based PCMs can be identified on this line. (C, D) Rocksalt (cubic) and spinel structures of GST with octahedral and tetrahedral Ge coordination, respectively. As shown in C, most GST-based PCMs contain up to $\sim$25\% vacancy-type defects at Ge/Sb sits.(E) Role of Se doping in modifying the switching mechanism (hence, speed) of a representative GST alloy. t$_{0}$: nucleation time of as-deposited film, t’: nucleation time of the pre-conditioned film, and t$_c$: time needed for the full crystallization of the film. Panels (C-E) are adopted from \cite{yamada2000structure}, \cite{welnic2006unravelling}, and \cite{coombs1995laser}, respectively.}
	\label{figH1}
\end{figure}

As Figure~\ref{figH1}A schematically shows, applying a long- and high-energy pulse to amorphous GST (a-GST) heats the material up to its glass-transition temperature ($T_g$), which drives the phase transition to its crystalline state (c-GST). Conversely, applying a short- and higher-energy pulse heats the c-GST above its melting temperature ($T_m$) and then suddenly cools it down, yielding a-GST by the melt quenching of the c-GST. A typical cooling rate of 1 $^{\circ}$C/nsec is needed for the melt quenching of most GST-based PCMs \cite{wuttig2017phase}. We note that many materials can be amorphized via the melt-quenching process, but they mostly display a negligible optical constant upon the phase transition (e.g., GaAs). Therefore, a systematic approach needs to be taken for the identification of PCM materials that can fulfill the requirements of practical applications. Given the complexities of phase-transition mechanisms and difficulties of theoretically predicting the material properties in the amorphous state, we limit our discussions to only major GST-based alloys. However, at the end of this section we set a semi-empirical guideline for searching for new PCMs. 

The phase diagram shown in Figure~\ref{figH1}B helps classify different PCMs based on Ge-Sb-Te alloys. GeTe was the first chalcogenide-based PCM material to show a relatively fast crystallization with a large optical contrast \cite{chen1986compound}. Subsequently, several GST alloys such as Ge$_1$Sb$_4$Te$_7$, Ge$_2$Sb$_2$Te$_5$, and Ge$_1$Sb$_2$Te$_4$ were identified \cite{yamada1987high,yamada1991rapid} all along a pseudo-binary line that connects the GeTe and Sb$_2$Te$_3$ compounds. Moving from GeTe towards Sb$_2$Te$_3$ on the pseudo-binary line, crystallization speed increases, $T_g$ and $T_m$ decrease, and data retention (i.e., the retention of the amorphous state) decreases. In other words, Sb$_2$Te$_3$ offers the fastest crystallization speed, but its amorphous state is unstable. In contrast, GeTe offers a very stable amorphous phase, but its crystallization dynamic is relatively slow. Therefore, a compromise between the crystallization speed and the amorphous stability can be made by selecting a ternary composition close to the center of the pseudo-binary line. For example, Ge$_2$Sb$_2$Te$_5$ offers a fast crystallization speed ($<$20 ns) with a moderate $T_g$ (100-150 $^{\circ}$C) that ensures the long-term data retention ($\sim$10 years). It is also important to note that deviation from the canonical stoichiometry (i.e., (GeTe)$_\textrm{n}$(Sb$_2$Te$_3$)$_\textrm{m}$; n, m: integers) reduces the switching speed, primarily because the crystallization proceeds through a slow phase-segregation step \cite{chen1986compound,coombs1995laser}. This concept was behind the addition of extra Sb to the Ge$_1$Sb$_2$Te$_4$ alloy to make the phase switching speed compatible with slow CD-writers in the early days of optical data recording \cite{yamada1991rapid}. Similarly, slow crystallization speeds ($\mu$sec regimes) were observed in nonstoichiometric Ge$_{15}$Te$_{85}$ doped with Sb and S \cite{ovshinsky1968reversible}.

In the crystalline phase, GST alloys form a cubic lattice similar to that of a rocksalt structure (Figure~\ref{figH1}C) in which Te atoms occupy one lattice site and the Ge/Sb atoms randomly occupy the remaining lattice sites. However, fast crystallization cycles locally distort the ideal cubic structure shown in Figure~\ref{figH1}C.  Moreover, most studied GST-based PCMs contain up to $\sim$25\% vacancy-type defects in the Ge/Sb sublattice \cite{nonaka2000crystal,yamada2000structure}, an important characteristic that seems to be directly linked to the switching properties of GST alloys. We note, however, that the cubic structure is a metastable phase (around $\sim$150 $^{\circ}$C) that can further transition to a stable hexagonal structure once the Ge$_2$Sb$_2$Te$_5$ alloy is heated to $\sim$310 $^{\circ}$C, a transition that has been argued as a disadvantage for the 225 stoichiometry of GST alloys (i.e., Ge$_2$Sb$_2$Te$_5$) \cite{friedrich2000structural}.

Unlike the crystalline phase, the arrangement of atoms in the amorphous phase is the subject of debate. Most studies have assumed that a-GST is locally similar to the crystalline phase, but it lacks the long-range atomic order observed in the crystalline phase. However, this notion has been challenged in several experimental X-ray spectroscopies that echo a pronounced difference in the local arrangement of atoms in amorphous and crystalline phases, that is: “the Ge atoms that occupy octahedral positions in the crystalline phase switch to tetrahedral coordination in the amorphous phase.” \cite{welnic2006unravelling,kolobov2004understanding}. To explain the discrepancy between these two narratives, Welnic \textit{et al.} conducted a density functional theory (DFT) study to identify possible atomic arrangements in Ge$_1$Sb$_2$Te$_4$ as a representative stoichiometric GST alloy \cite{welnic2006unravelling}. This study revealed that the Ge$_1$Sb$_2$Te$_4$ alloy may establish two different crystal structures: (1) the distorted cubic structure with “octahedral” Ge bonds (as shown in Figure~\ref{figH1}C) and (2) a spinel structure in which Ge atoms form “tetrahedral” bonds with Te atoms, while Sb/Te atoms occupy atomic sites identical to those in the rocksalt structure (see Figure~\ref{figH1}D). Interestingly, the ground-sate energy of the cubic structure is only $\sim$30 meV per atom less than that of the spinel structure. Thus, the difference in reported local orders of a-GST may stem from the competition between two energetically similar structures with significantly different local atomic arrangements. In addition, the assumption of the spinel structure can successfully explain the increase in the larger bandgap of the amorphous phase as compared to the crystalline state with the cubic structure.

Switching the phase from a-GST to c-GST generally involves two events: nucleation of small crystallites and the subsequent growth of these crystalline domains \cite{coombs1995laser}. To the first order, the nucleation process is governed by the thermodynamics of the phase change (i.e., a temperature-controlled process), while the speed of the crystal growth is primarily governed by the kinetics of the phase change (i.e., the atomic motion of elements). At high temperatures, above $T_m$ ($\sim$600 $^{\circ}$C for Ge$_2$Sb$_2$Te$_5$), GST is in a liquid (molten) phase with highly mobile atoms. Thus, crystallization of GST is thermodynamically allowed once the temperature drops below the melting point, and the probability of crystalline nucleation increases as the temperature further drops, reaching the maximum probability at the glass-transition temperature, $T_g$. However, at low temperatures close to $T_g$, the crystalline nuclei cannot efficiently grow because the atomic mobility is extremely small. Thus, fast crystallization of a-GST occurs at intermediate temperatures between $T_g$ and $T_m$, which is 150-250 $^{\circ}$C for the cubic structure and 300-350 $^{\circ}$C for the hexagonal structure. It is uniquely interesting that a-GST alloys remain amorphous for over 10 years at room temperature, but the very same material can be crystallized in only $\sim$20 ns upon heating up to only a few hundred degrees of Celsius, a feature that can be hardly found in any other PCM outside chalcogenide glasses.

Here, we would like to emphasize that the crystallization of melt-quenched GST films is strikingly different than the crystallization of as-deposited films. In a detailed study, Coombs \textit{et al.} showed that crystallization of as-deposited films proceeds through a nucleation-dominated process, while in the melt-quenched films crystallization is growth dominated \cite{coombs1995laser}. Apparently, the melt-quenching process leaves some subcritical crystalline domains in the amorphous matrix, so that the slow nucleation step can be bypassed in the subsequent re-crystallization cycles. Therefore, crystallization of a melt-quenched GST alloy is significantly faster than that of an as-deposited amorphous film. However, this study suggests that the application of a short pulse can pre-condition an as-deposited film so that it mimics the fast switching speed of a melt-quenched film. For instance, using a pre-conditioning laser pulse of 100 nsec width, the crystallization time of the GeTe film can be reduced from 1 $\mu$sec to only 30 nsec, which reflects the switching of the crystallization mechanism from the nucleation-dominated mode to the growth-dominated mode. We note that the presence of residual crystalline nuclei in melt-quenched films could be explained based on the low interfacial energy between amorphous matrix and crystalline domains, as demonstrated in undercooling experiments performed on GST alloys \cite{friedrich2001morphology,kalb2005kinetics}.

The switching behavior and the optical/electronic property contrast, induced by the phase switching, can be customized by changing the composition of a chalcogenide PCM, primarily through the substitutional doping of isoelectronic elements (i.e., parenthesis in Figure~\ref{figH1}B) in ternary GST alloys. For example, the doping of GST alloys with selenium (Se) atoms forms a quaternary Ge-Sb-Se-Te PCM that is referred to as GSST \cite{zhang2019broadband}. A recent study shows that an optimized addition of Se to the Ge$_2$Sb$_2$Te$_5$ alloy can significantly reduce the optical loss in the near- to mid-IR spectral range. Indeed, the optimized GSST alloy (i.e., Ge$_2$Sb$_2$Se$_4$Te$_1$) shows a broadband transparency in the 1-18.5 $\mu$m wavelength range, while offering a large refractive-index change ($\Delta$n $\approx$ 2) without a loss penalty (i.e., $\Delta$k $\approx$ 0). DFT calculations show that increasing the Se content widens the bandgap of the GSST alloy, leading to a smaller optical loss in the IR regime. In addition, compared to the GST alloys, GSST alloys show a smaller density of states close to band edges, which reduces the free-carrier absorption loss in the IR regime. Interestingly, the substitution of the Te by Se does not change the crystal structure of GSST as it still forms cubic (metastable) and hexagonal (stable) structures similar to the GST case.

Increasing the Se content in GSST monotonically increases the crystallization temperature, which translates into a better stability of the amorphous phase. However, this stability comes at a major drawback that is explained in further detail in Ref.~\cite{coombs1995laser}, that is: the crystallization becomes significantly slower as the Se replaced the Te in a GST alloy. As shown in Figure~\ref{figH1}E, the time needed for the onset of nucleation in as-deposited GSST (i.e., t$_0$) and pre-conditioned GSST (i.e., t') and the total time needed for the completion of the GSST crystallization (i.e., t$_c$) monotonically increase with the Se content in the GSST. In fact, the t$_c$ increases from $\sim$20 ns in GST to $\sim$3 $\mu$sec in a GSST alloy with only $\sim$25\% Se content, showing more than a 100-fold reduction in the crystallization speed. An even slower crystallization speed can be expected in results reported in Ref.~\cite{zhang2019broadband}, where Se content is even larger and reaches ~ 80\% in the optimized composition (i.e., Ge$_2$Sb$_2$Se$_4$Te$_1$). It is also interesting that the influence of pre-conditioning on the crystallization speed becomes weaker as the Se content is increasing (compare t$_0$ and t' in Figure~\ref{figH1}E). The slower dynamics in GSST is attributed to the nucleation-dominated crystallization which is intrinsically sluggish \cite{coombs1995laser}. 

The large optical contrast in GST-based PCMs stems from the pronounced change of local atomic arrangements, in which the Ge coordination switches between octahedral (in crystalline) and tetrahedral (in amorphous) bonds. Such an atomic change is reflected in the $\sim$7\% volumetric expansion of the GST following the crystalline-to-amorphous conversion, as experimentally evidenced in X-ray diffraction measurements \cite{njoroge2002density}. In addition, quantum mechanical calculations, using the Fermi’s golden rule, show that such a local structural alteration leads to a large change in the elements of the matrix that governs optical transitions between two initial and final states in GST-based PCMs \cite{welnic2007origin}. Considering the central role of Ge atoms in the coordination change, one can expect that a larger Ge content in GST alloys provides a larger optical contrast, as can be seen in a comparison between, for instance, Ge$_8$Sb$_2$Te$_11$ with $\Delta$n+i$\Delta$k $\approx$ -1.48+i1.35 and Ge$_2$Sb$_2$Te$_5$ with $\Delta$n+i$\Delta$k $\approx$ -1.2+i1.05 at 405 nm wavelength \cite{yamada2002phase}. However, the exact mechanism behind this trend is not well understood. Welnic \textit{et al.} suggest that a lower density of vacancies in GST alloys enhances the optical contrast, and the addition of the Ge may help in achieving this goal \cite{welnic2006unravelling}. However, Kolobov \textit{et al.} argue that the presence of the vacancies at Ge/Sb sites is an intrinsic part of the GST structure and the addition of the Ge will not reduce the vacancy density \cite{kolobov2004understanding}. Experimental studies support this argument, as the addition of extra Ge/Sb atoms did not reduce vacancy sites and instead led to the phase segregation and the accumulation of extra Ge/Sb atoms at the grain boundaries \cite{yamada2000structure,privitera2003crystallization}. It is very important to notice that the origin of the optical contrast in PCMs is strictly different from that of the covalent semiconductors (e.g., Si and GaAs) in which local atomic arrangements remain intact, and the optical contrast stems primarily from the smearing of electronic states and the formation of tail states in the bandgap of the amorphous phase.

\begin{figure}
	\centering
	\includegraphics[trim={0cm 0cm 0cm 0cm},width=0.5\textwidth, clip]{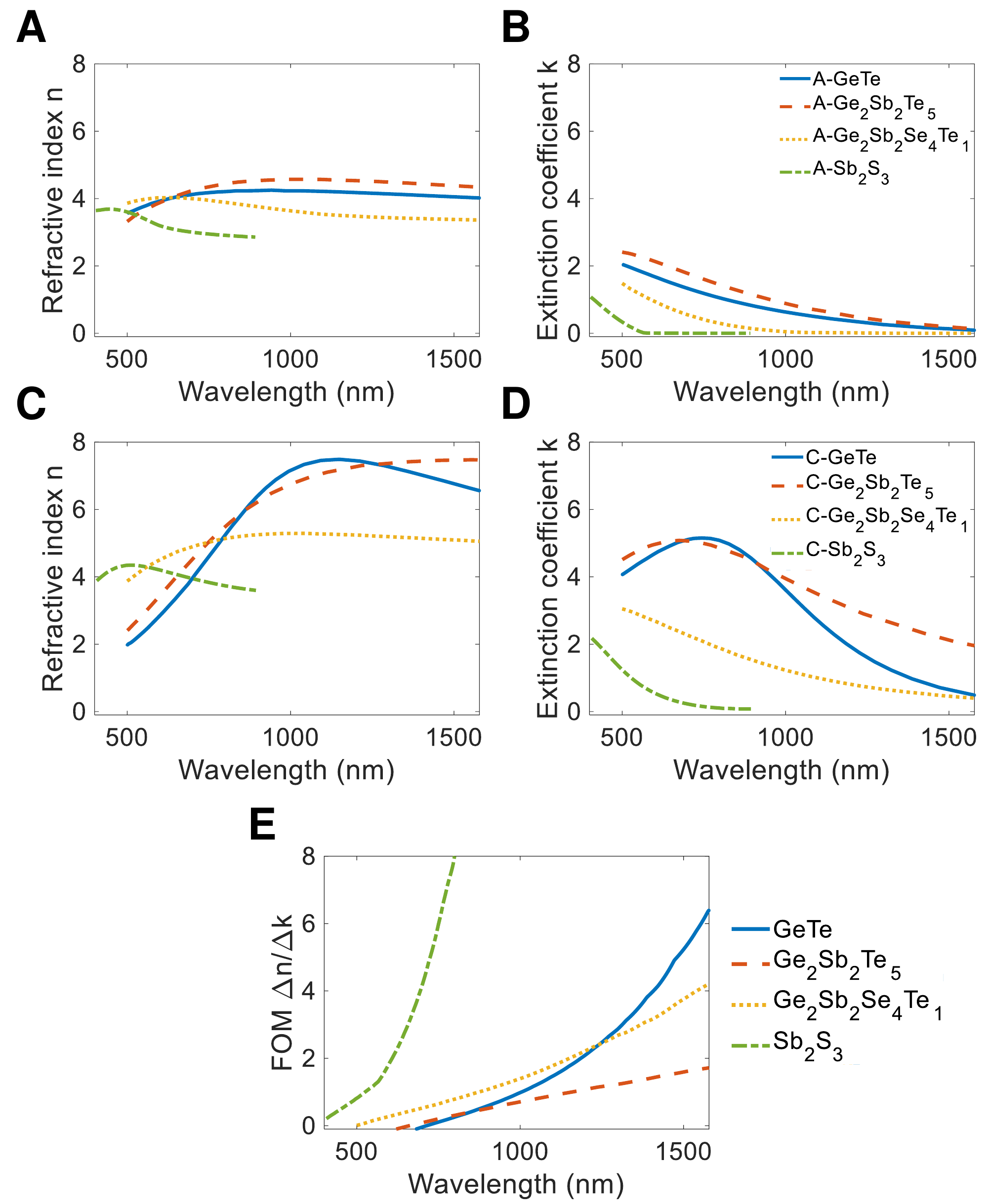}
	\caption{Optical constants of popular PCMs for optical applications. (A, C) Refractive index (n), and (B, D) extinction coefficient (k) of several PCMs in amorphous and crystalline states, respectively. (E) The FOM of PCMs listed in panels A-D. $\Delta$n and $\Delta$k are defined as the relative change of n and k between amorphous and crystalline states. References:  GeTe and Ge$_2$Sb$_2$Te$_5$ \cite{wuttig2017phase}, Ge$_2$Sb$_2$Se$_4$Te$_1$ \cite{zhang2018broadband}, and Sb$_2$S$_3$ \cite{dong2019wide}.}
	\label{figH2}
\end{figure}

Reevaluating the portfolio of successful PCMs shows common characteristics that can be further employed for the identification and, yet better, prediction of alternative PCMs. First, all identified PCMs show distorted octahedral structures. Second, a relatively large density of intrinsic vacancies seems to be mandatory for the structural stability. Third, unlike most semiconductors with sp$^3$-hybridized bonds (i.e., tetrahedral coordination), atomic bonds in PCMs display p-type characteristics. The p-type bonding (i) guarantees the 6-fold coordination in the crystalline state, (ii) is prone to distortion (because it is weaker than sp$^3$ bonds) which is necessary for the crystal stabilization, and (iii) can be easily broken for the fast switching of bonds from octahedral to tetrahedra on amorphization. Interestingly, p-type bonding has been shown to occur in alloys with more than 4 valance electrons \cite{luo2004dependence}. Thus, considering these similarities, the search for new PCMs can be narrowed down to the alloys of group 15 and 16 with octahedra structures, more than 4 valance electrons, and proper T$_g$ and T$_m$ values. Also, special attention should also be paid to the specific stoichiometry of PCM alloys to eliminate the phase segregation during phase-transition cycles, if the fast switching speeds are desired. Though, for an ultimate search, the development of a robust model for the amorphous phase seems mandatory for the prediction of the novel PCM materials using the first principle calculations.

The holy grail of the optical modulation/switching is the large change in optical constants, and more specifically the refractive index of active regions in photonic devices. In Figure~\ref{figH2}, we have presented real and imaginary parts of the refractive index for some promising chalcogenide-based PCMs, including two promising binary compounds, GeTe and Sb$_2$S$_3$. Ideally, a PCM with a large $\Delta$n and a small $\Delta$k is desired. As described by the well-known Kramers-Kronig relation, however, the real and imaginary parts of the refractive index are not independently controllable. Thus, the relative ratio of the index change to the loss change (i.e., $\Delta$n/$\Delta$k) serves as a figure of merit (FOM) for comparing various PCMs. As shown in Figure~\ref{figH2}E, for operation in visible, NIR, and telecommunication wavelength (i.e., 1.55 $\mu$m), Sb$_2$S$_3$, GeTe, and Ge$_2$Sb$_2$Se$_4$Te$_1$ provide the best FOMs, respectively. We note that such a conclusion is solely made based on the FOM. However, based on the switching speed, for instance, Ge$_2$Sb$_2$Te$_5$ offers a faster response than both Ge$_2$Sb$_2$Se$_4$Te$_1$ and GeTe. Therefore, in the selection of PCM materials, the ultimate application should be considered.

We would like to mention that the inclusion of PCM materials in optical device platforms (discussed below) necessitates several fabrication considerations beyond the intrinsic material properties. First, most GST-based PCMs are prone to rapid oxidation upon long ambient exposures. Thus, PCMs are usually capped by a protecting layer (e.g., SiO$_2$, Ta$_2$O$_5$, ZnS, ITO, etc.) that can withstand high-temperature phase switching events as well as fulfill the requirements of conversion stimuli (i.e., electrical vs optical). Second, in some cases, an irreversible diffusion of metal atoms into PCMs  may be expected, which degrades the phase switching behavior in hybrid metal-PCM devices (e.g., plasmonic platforms). Addition of a diffusion-barrier layer or the use of alternative plasmonic metals (e.g., TiN) \cite{lu2019inter} can fix this issue. We note that the addition of capping/diffusion-barrier layers may affect the details of the phase conversion in PCMs, probably because of modifying the nucleation/growth of crystalline domains as the interfacial energies at amorphous-crystalline, amorphous-capping layer, and the crystalline-capping layer will be modified. Such effects may be behind the reported change of the crystallization temperature following the capping of the Ge$_2$Sb$_2$Te$_5$ alloy with ZnS-SiO$_2$ capping layers \cite{friedrich2001morphology}. Finally, device architectures should support fast thermal time-constants to allow for the rapid temperature change ($\sim$10$^9$ $^{\circ}$C/sec) needed for the amorphization via the melt-quenching, a factor that calls for the use of materials with proper thermal coefficients (e.g., thermal conductance, heat capacitance, etc.).

\section{Active Amplitude control with tunable phase-change metasurfaces}

\begin{figure*}
	\centering
	\includegraphics[trim={0cm 1.8cm 0cm 0cm},width=\textwidth, clip]{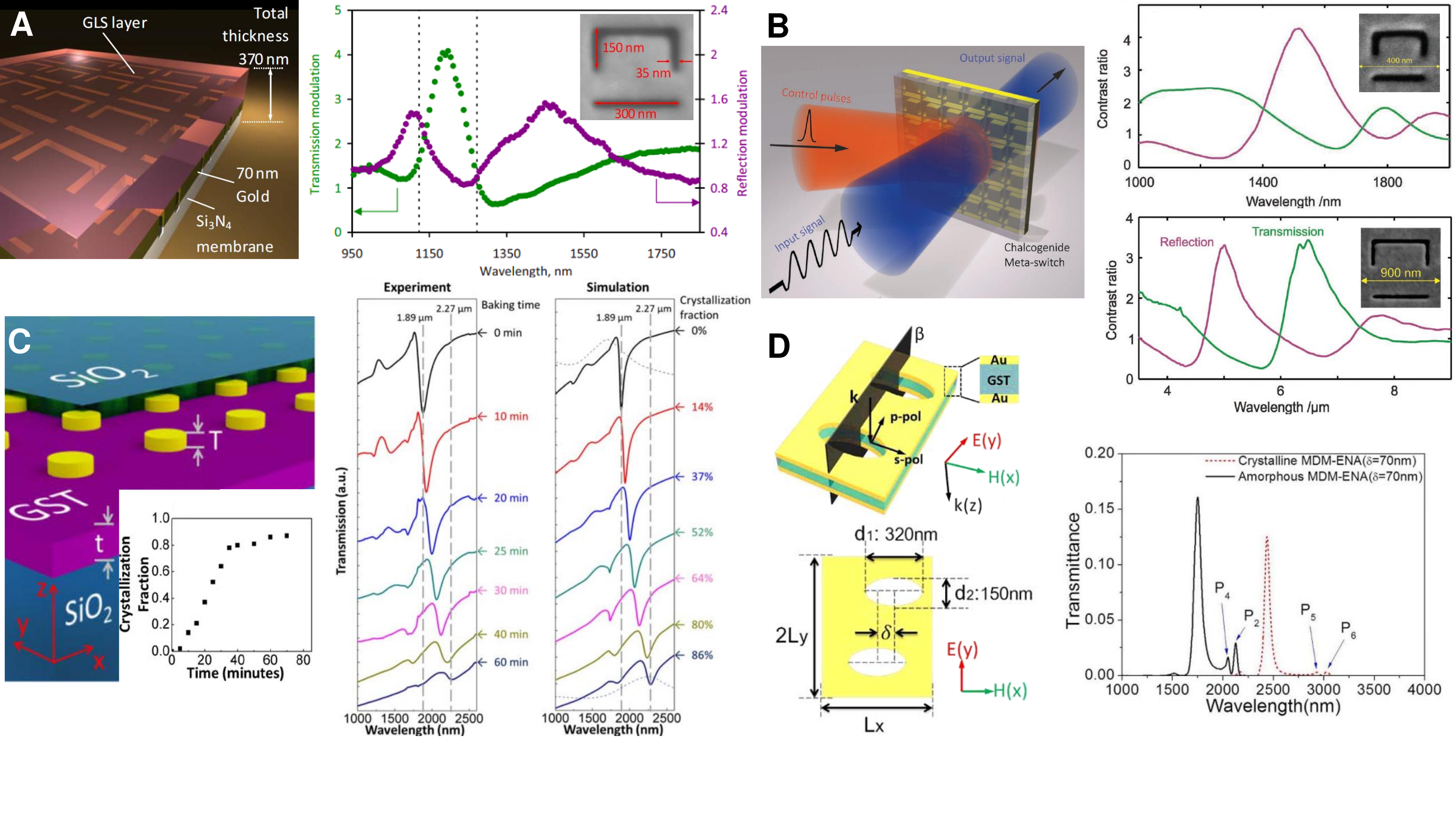}
	\caption{Pioneering reports on dynamic plasmonic/PCM based MSs for the global control of the amplitude response. 
	(A) Electro-optic switching \cite{samson2010metamaterial}. 
	Left: a hybrid structure consists of an array of patterned Au meta-atoms deposited on a SiN membrane covered with GLS. 
	Right: the spectral response of the transmission/reflection modulation contrast ratio associated with the phase transition of GLS (inset: scanning electron microscope (SEM) image of the fabricated meta-atom.)
	(B) Bidirectional and all-optical amplitude modulation \cite {gholipour2013all}.
	Left: artistic impression of a PCM-based MS for the all-optical modulation. A laser pulse controls the phase of overlying PCM film between amorphous and crystalline states for the high-contrast transmission and reflection switching of the information-carrying laser. 
	Right: near-IR (top) and mid-IR (bottom) transmission/reflection modulation contrast ratios associated with the phase switching of the GST film. The SEM image of fabricated meta-atoms are shown in the insets. 
	(C) Active tuning of lattice plasmon resonances \cite{chen2013hybrid}.
	Left: the sketch of the GST-Au hybrid plasmonic MS for controlling the near- to mid-IR resonant wavelengths (inset: fractional crystallization as a function of the heating time).
	Right: experimental measurements and simulation results from the continuously tuned MS, which is thermally treated at different baking time at the fixed temperature of 135 $^{\circ}$C.
	(D) The dynamic tuning of Fano resonances \cite{cao2014fast}.
	Left: illustrations of asymmetric elliptical nanohole meta-atoms for the fast tuning of the supported Fano resonance mode. 
	Right: the transmission spectrum of the whole MS in cases of amorphous and crystalline states.}
	\label{figS1}
\end{figure*}

\subsection{Hybrid plasmonic/PCM metasurfaces for global amplitude control}

To highlight the unique capabilities of PCMs in meta-optic applications, early studies leveraged an easy-to-fabricate layer of PCM as the functional material. Although later rather complex patterned PCM nanostructures were utilized, so far, a vast majority of demonstrations of optical functionalities has been limited to the global control of PCMs across the entire MS. Herein, a comprehensive review of developed hybrid plasmonic and dielectric MSs using PCMs for global amplitude control is presented. Throughout this review, the so-called modulation depth (MD) and extinction ratio (ER) are considered as quantitative representations of the amplitude modulation of the reflected, transmitted, or absorbed light. For the reflected light, MD is defined as as the difference between the maximum reflected power (P$_{\textrm{max}}$) and the minimum reflected power (P$_{\textrm{min}}$) normalized by the incident power value (P$_{\textrm{inc}}$) at a specific wavelength, i.e.,
MD = (P$_{\textrm{max}}$ - P$_{\textrm{min}}$)/P$_{\textrm{inc}}$. Moreover, ER is described in terms of the logarithm of the ratio of the maximum to the minimum reflected power, i.e., 
ER = -10 log$_{10}$(P$_{\textrm{max}}$/P$_{\textrm{min}}$).
It is noteworthy that based on the operation principle of the device under test, P$_{\textrm{max}}$ and P$_{\textrm{min}}$ are associated with either amorphous, semi-crystalline, or fully crystalline state of PCMs.

Hybrid PCM-plasmonic MSs have been considered as a promising set of candidates for the real-time control of the key properties of light due to the high sensitivity of metallic nanoantennas to the ambient changes. Such striking sensitivity originates from the spanned nature of localized surface plasmon resonance modes which are concentrated outside of nanoantennas. Based on the first-order perturbation theory, the resonance frequency change $\Delta \omega$ is governed by 
$\Delta \omega / \omega_{R} = \iiint \Delta \epsilon E^{2} dr^{3}/\iiint \epsilon E^{2} dr^{3}$, in which $E$ is the total electric field and $\Delta \epsilon$ is the induced change of the dielectric permittivity and the integration spans over the unit-cell of a MS \cite{johnson2002perturbation}. Accordingly, the phase transition in the PCM in contact with the meta-atoms significantly influences the effective permittivity of the unit-cell and thus the resonance behavior, which enables the spatio-temporal and/or spectro-temporal control of the incident light.

In a pioneering work by Samson \textit{et al.}, an innovative electro-optic switch by employing a homogeneous plasmonic MS composed of a periodic array of Au asymmetrically split resonators covered with a thin layer of GLS is implemented (see Figure~\ref{figS1}A) \cite{samson2010metamaterial}. By applying electrical pulses, $\sim$10 ms duration and $>$45 V amplitude, to the electrodes connected to the embedded 200-nm-thick GLS film and the patterned metallic layer, up to 150 nm blueshift in the narrowband Fano resonance is observed in the near-IR wavelength range. This large tuning range stems from the  significant refractive-index contrast of GLS after its uniform phase transition from the amorphous to the crystalline state. Such an ultrathin configuration provides a transmission modulation with a contrast ratio of 4:1 upon phase-change of GLS. We note, however, that the reversible electrical switching was not reported in this work. 

In a following work \cite {gholipour2013all}, Gholipour \textit{et al.} enhanced the resonance shift by replacing GLS with a layer of Ge$_2$Sb$_2$Te$_5$ embedded between two ultrathin supporting layers of ZnS-SiO$_{2}$ (see Figure~\ref{figS1}B). They have shown a global reversible phase transition across a large area of the 15 nm-thick GST layer by applying a single laser pulse with varying durations and peak intensities. As a proof of concept, a 50 $\mu$m beamwidth laser pulse of 0.1 mW$\mu$m$^{-2}$ intensity with the duration of 50 ns and 0.25 mW$\mu$m$^{-2}$ with the duration of 100 ns have been exploited for amorphization and crystallization of the GST layer, respectively. Such an all-optical, bidirectional, and non-volatile MS offers a remarkable resonance shift of 200 nm with 2.5-fold MD in the transmitted near-IR light and 1.3 $\mu$m resonance shift with 4-fold contrast ratio in the mid-IR band, respectively.

To demonstrate the potential of PCMs for the design of highly tunable MS optics, Chen \textit{et al.} explored the use of intermediate phase transitions in Ge$_2$Sb$_2$Te$_5$ integrated into a plasmonic crystal \cite{chen2013hybrid}. Their proposed structure consists of a 20-nm-thick PCM layer sandwiched between an array of Au disks and a quartz substrate, as shown in Figure~\ref{figS1}C. To realize the stepwise nonvolatile tuning of the lattice resonance over a 500 nm range (from near- to mid-IR), the fabricated sample was uniformly baked on a hot plate for different time periods (up to 60 minutes) but at a fixed temperature of 135 $^{\circ}$C. To reveal the relation between the crystallization fraction and the annealing time, the experimental results (i.e., optical reflection spectra) were theoretically reconstructed in full-wave simulations, a mapping scheme that gives an intuitive understanding between the degree of the phase change in the PCM layer and the optical response of the MS.

Following these pioneering reports, several other groups proposed alternative types of hybrid PCM-plasmonic architectures for the realizations of dynamic metadevices. Cao \textit{et al.} published several theoretical proposals using Ge$_2$Sb$_2$Te$_5$ for the tunable perfect absorption in the visible and mid-IR regimes \cite{cao2014broadband, cao2013mid}. They numerically demonstrated a tri-layer fishnet MS comprising of metal-PCM-metal elliptical nanoholes in a broken symmetry architecture and realized a narrow band Fano resonance (see Figure~\ref{figS1}D) \cite{cao2014fast}. To show the potential of the proposed configuration for high-speed all-optical switching applications, a comprehensive photothermal model was used to study the temporal variation of the temperature as the PCM film was exposed to ultrafast laser pulses. Their results showed that a low-intensity pump light with the fluence of 9.6 $\mu$W$\mu$m$^{-2}$ and pulse duration of 360 ps can convert the phase of a 160-nm-thick GST layer from the as-deposited to the crystalline state in a one-way fashion. Such a phase transition resulted in a 1 $\mu$m tuning range of the supported Fano resonance mode with sharp transparency peaks in the mid-IR regime (see Figure~\ref{figS1}D). In another work, the same group introduced a tri-layer planar chiral MS in the shape of a gammadion combined with a thin GST layer. Using this structure, they reported a large frequency shift of 58\% (in mid-IR) for the circular dichroism, defined as the difference in the transmittance of right-handed and left-handed circularly-polarized light \cite{cao2013strongly}. Their opto-thermal model predicts $\sim$5 ns crystallization transition time to uniformly heat the 24-nm-thick GST layer to 883K through a low light intensity of 0.016 mW$\mu$m$^{-2}$.

\begin{figure*}[t]
	\centering
	\includegraphics[trim={0cm 5.3cm 0cm 0cm},width=\textwidth, clip]{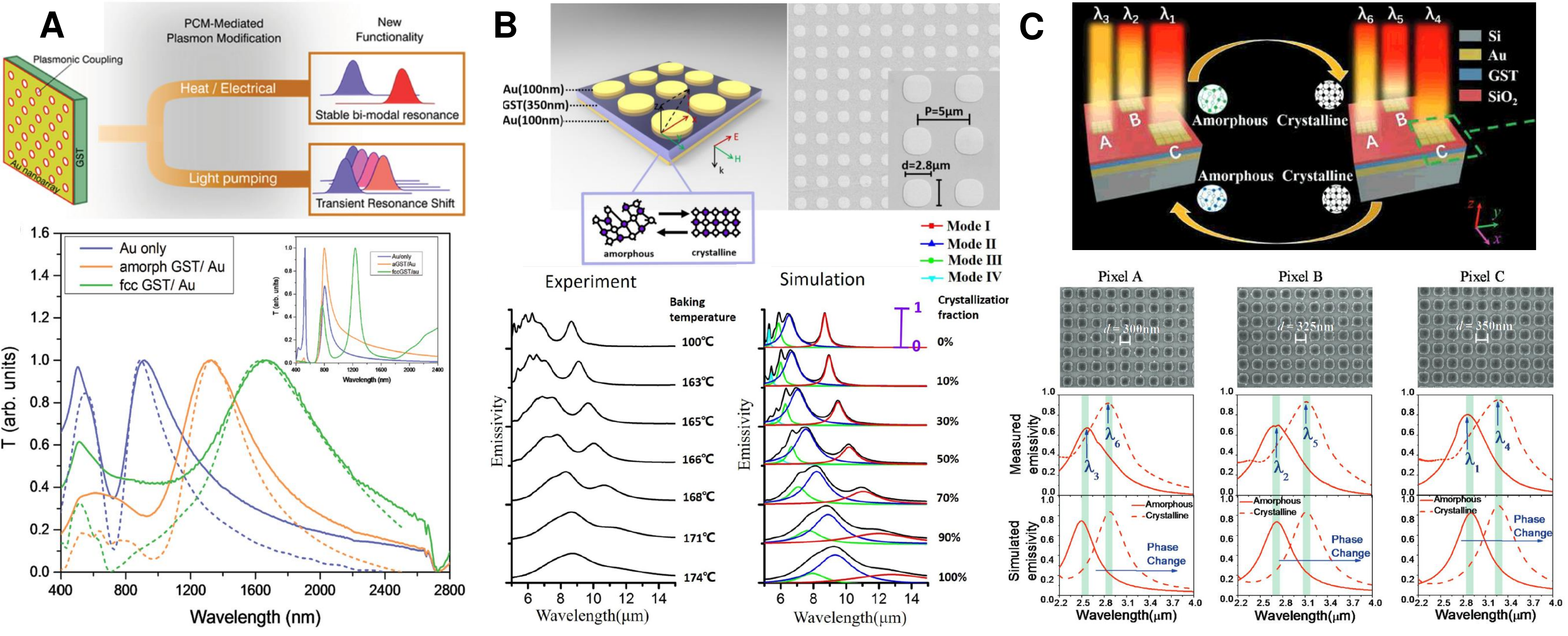}
	\caption{Common hybridized plasmonic MSs using PCMs for the global amplitude manipulation near- and mid-IR regimes. 
	(A) Ultrafast wide-range tuning of optical resonances \cite{rude2016ultrafast}.
	Top: a hybrid PCM-plasmonic MS for controlling the bimodal and transient EOT response using external stimuli. 
	Bottom: calculated transmission spectra obtained from numerical simulations (dashed curves) and experimental measurements (solid curves) for two states of GST compared with the results of a bare perforated Au array as a reference.
	(B) Dynamic thermal emission \cite{qu2017dynamic}. 
	Top: an artistic schematic and a SEM image of the fabricated PCM-plasmonic thermal emitter.
	Bottom: experimental measurements of continuously tuned emissivities of the thermal emitter annealed at different temperatures for a fixed period of 60 seconds (left). Emissivities and modes evolution calculated from numerical simulations at corresponding crystallization fractions (right). 
	(C) Multispectral thermal emission \cite{cao2018tuneable}. 
	Top: a 3D perspective view of the PCM-based thermal emitter MS.
	Bottom: SEM images, experimentally measured emissivity, and numerically calculated emissivity of the hybrid MS with an arrays of Au nanosquares with different widths and identical lattice constants.}
	\label{figS2}
\end{figure*}

New hybrid MS designs were explored by Rud{\'e} \textit{et al.} for the demonstration of ultrafast (i.e., ps regime), broadband (up to 385 nm spectral shift), and large (up to 60\% MD) optical tuning of the optical transmission in the vis-near-IR spectral range \cite{rude2016ultrafast}. As shown in Figure~\ref{figS2}A, the investigated platform was formed by structuring a 2D array of nanoholes in an Au film underneath a thin layer of Ge$_2$Sb$_2$Te$_5$ ($\sim$20 nm). A remarkable modulation tuning of the extraordinary optical transmission (EOT) response using both thermal- and current-induced structural transitions was thoroughly discussed.

To promote the applicability of reconfigurable MSs in the C-band communications, Carrillo \textit{et al.} demonstrated a 1D array of hybrid nanogratings for the amplitude modulation with a MD of $\sim$77\% and an ER of 20 dB \cite{carrillo2016design}. The proposed design benefits from an ITO capping layer for protecting a 60-nm-thick GST layer from the environmental oxidation while still allowing both optical and electrical access. Through a comprehensive sensitivity analysis considering the side effects of fabrication imperfections and critical design parameters on the metadevice performance, a systematic study of the optimization of the structural design was proposed. Furthermore, to explore the in situ switching of the PCM state, electro-thermal simulations using electrical pulses, applied between the top and bottom layers of the MS, were carried out. Regardless of the plasmonic metal type (Au, Al, W, and TiN), numerical results show that a 2.4 V and 50 ns RESET pulse with a 15/5 ns rise/fall time can uniformly amorphize the GST layer. On the other hand, for crystallization, a SET pulse of 1.4 V and 100 ns with a 30 ns rise/fall time was employed. It is worth noting that during the amorphization, the uniformly distributed temperature exceeds the melting point of GST (i.e., 600 $^{\circ}$C), while during the crystallization cycle, the GST layer experiences $\sim$425 $^{\circ}$C temperature, which is necessary for a rapid (a few tens of nanosecond) process. The same group later presented an experimental demonstration of a more practical architecture for the on-demand quality factor control \cite{carrillo2018reconfigurable}.

Dual-functional opto-electric nanoscale devices using PCMs have gained significant attention recently. Raeis Hosseini \textit{et al.} took the advantage of a uniquely tunable metadevice framework with a bifunctional behavior; as a tunable MS for perfect absorption of visible light and as a resistive switching memory device for data storage \cite{raeis2019dual}. The structure consists of an array of Ag nanosquares on a 25-nm-thick Ge$_2$Sb$_2$Te$_5$ film deposited on the top of an Al-coated Pt mirror. Following the phase transformation, a narrowband ($\sim$50 nm) to wideband ($\sim$400 nm) perfect absorption in the visible spectral range and bipolar resistive switching with high ON/OFF ratio ($\sim$10$^{6}$) occurs.

In contrary to the conventional metal-insulator-metal (MIM) tri-layer reflective MSs, Dong \textit{et al.} experimentally demonstrated a transmissive MS by eliminating the bottom metal reflector \cite{dong2018tunable}. The fabricated MS exhibits a transmission dip that can be switched from 3 to 6 $\mu$m upon the phase transformation in a 65-nm-thick GST film on a hotplate. The sparse distribution of the nanosquare array significantly reduces the device capacitance and facilitates the nsec-order electrical switching speed of the GST layer. The proposed structure enables collinear filter design necessary for the spectrally selective microscale mid-IR hyperspectral imaging.

The ability to control the thermal emission from an object with subwavelength thickness has attracted a growing interest to a wide range of applications, including radiative cooling and energy harvesting \cite{julian2019all}. According to the Kirchhoff’s law, the absorptivity of material equals its emissivity at equilibrium conditions, thus MS absorbers can be employed as thermal emitters within the mid-IR spectral band \cite{baranov2019nanophotonic}. Qu \textit{et al.} utilized an MIM architecture (as shown in Figure~\ref{figS2}B) incorporating a 350-nm-thick Ge$_2$Sb$_2$Te$_5$ layer to dynamically manipulate the thermal emission with zero-static power \cite{qu2017dynamic}. By baking the sample on a hot plate for a fixed time period of 60 seconds at varying temperatures, the emissivity, bandwidth, and peak wavelength can be finely tuned. In a follow-up work \cite{qu2018polarization}, they developed a dual-band thermal emitter with an engineered array of Au nanoellipses to excite higher-order magnetic resonances with a perpendicular polarization along short and long axes. The polarization of the thermal emission is rotated by 90$^{\circ}$ at the 9.55 $\mu$m peak wavelength upon the partial conversion of the PCM film ($\sim$40\% crystallization fraction). Inspired by these initial studies, several exciting technologies such as wide-angle and near-perfect dynamic thermal camouflage devices \cite{du2017control} and switchable and wavelength-selective thermal emitters \cite{qu2018thermal,du2018wavelength} were demonstrated. 

More recently, multispectral thermal emission using a PCM-plasmonic MS in the mid-IR wavelength range (2-3 $\mu$m) was studied \cite{cao2018tuneable}. The MIM architecture is formed by the arrangement of three distinct pixels where each pixel consists of uniform Au nanosquares with a different width designed to radiate at a distinct wavelength (see Figure~\ref{figS2}C). The developed multiphysics heat transfer model shows that the reversible switching of perfect emissivity between two distinct wavelengths can be achieved by the phase transition in a 50-nm-thick Ge$_2$Sb$_2$Te$_5$ within only 300 ns.  

\begin{figure*}[t]
	\centering
	\includegraphics[trim={0cm 0cm 0cm 0cm},width=\textwidth, clip]{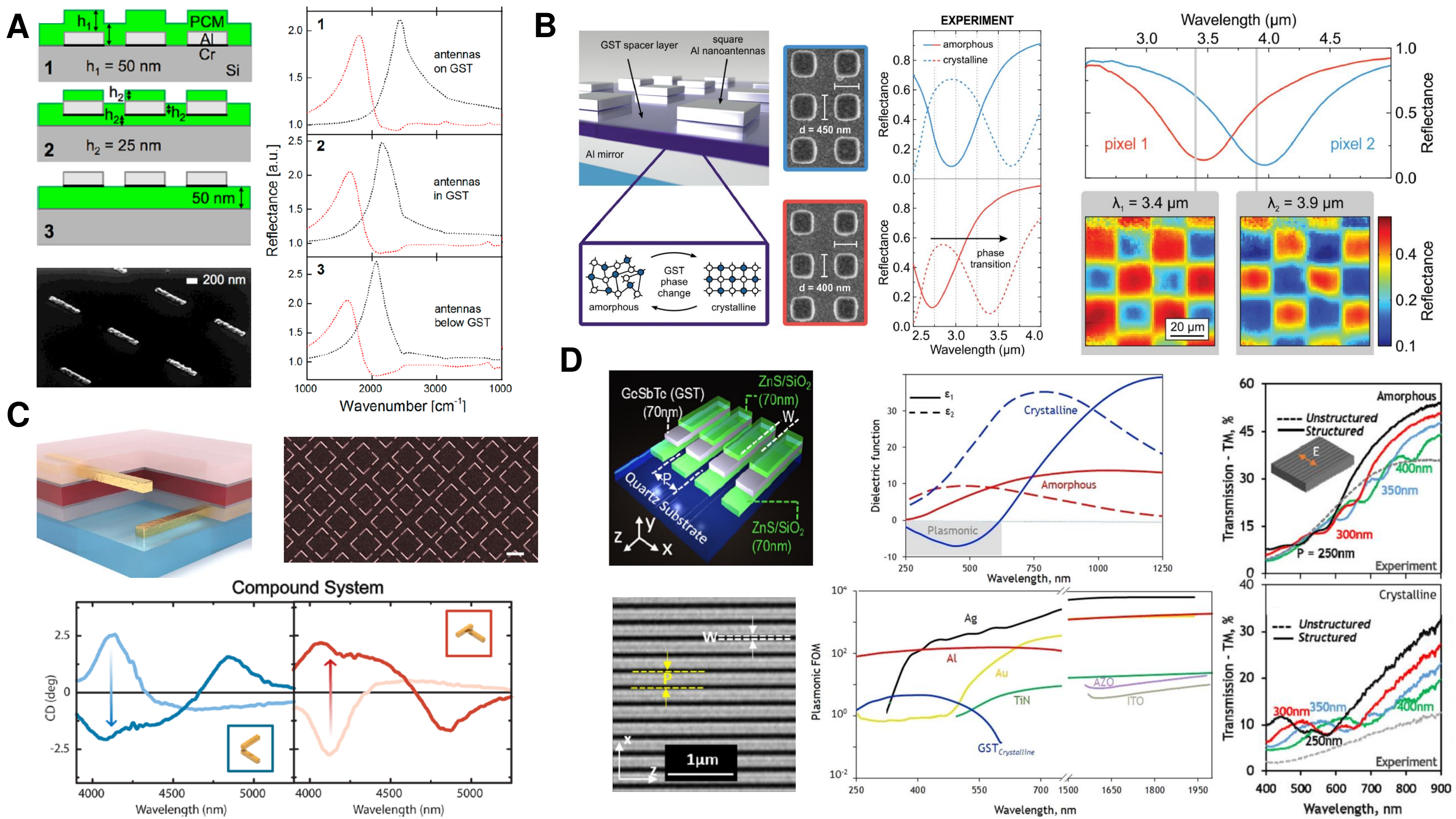}
	\caption{Innovative hybrid MSs for adaptive meta-optic applications.
	(A) Low loss mid-IR resonance tuning \cite{michel2013using}.
	Left: Schematic of the hybrid MS for resonance tuning. Three different layouts with Al nanoantennas below, in the middle, and on top of the PCM layer (top). SEM image of the fabricated sample (bottom). 
	Right: Simulated (dotted lines) and experimental (solid lines) reflectance spectra illustrating the resonance shifting characteristics for different sample layouts with amorphous (black) and crystalline (red) states. The strongest shift can be observed in case 1. 
	(B) Multispectral thermal imaging \cite{tittl2015switchable}.
	Left: Schematic of the switchable perfect absorber metadevice and SEM micrographs of fabricated perfect absorber metadevices with nanoantennas of different widths. The Ge$_{3}$Sb$_{2}$Te$_{6}$ layer is sandwitched between an array of Al nanoantennas and an Al back reflector (inset: conceptual illustration of phase transition in Ge$_{3}$Sb$_{2}$Te$_{6}$ molecules).
	Middle: Experimental measurements and corresponding simulated reflectance spectra from the two representative MSs.
	Right: Representative reflectance spectra for two distinct perfect absorber pixels with design wavelengths of 3.4 and 3.9 $\mu$m, respectively (top). Color-coded reflectance from a supercell of 20 $\mu$m at the design wavelengths. The arrangement of pixel is distinguishable from the colormap (bottom). 
	(C) Switchable chirality \cite{yin2015active}.
	Top: Perspective view of a dynamic chiral plasmonic dimer stack embedding a layer of Ge$_{3}$Sb$_{2}$Te$_{6}$ (red box) and SEM image of the fabricated right-handed MS (scale bar = 1 $\mu$m).
	Bottom: Overall experimental response of the bias layer overlaid with the active chiral dimer. Upon amorphous to crystalline transition of the PCM, the CD signal sign becomes opposite at the wavelength of 4200 nm for both fabricated enantiomers. 
	(D) Plasmonic transition in GST. \cite{gholipour2018phase}.
	Left: Cut-away section view of the hybrid ZnS-SiO$_{2}$/GST/ZnS-SiO$_{2}$ MS and top SEM image of the fabricated MS (removed areas by focused ion-beam milling are depicted in dark). 
	Middle: Relative permittivity of Ge$_{2}$Sb$_{2}$Te$_{5}$ in its amorphous and crystalline states achieved from ellipsometric measurements (top). FOM comparison for polycrystalline GST and common plasmonic materials (bottom).
	Right: Experimentally measured spectral dispersions of the hybrid MS with different grating periods (as labeled) and GST states under excitation of a TM-polarized illumination.
}
	\label{figS3}
\end{figure*}

The introduction of new PCMs with higher optical contrasts and lower loss in the mid-IR regime (such as Ge$_3$Sb$_2$Te$_6$) made another avenue for the realization of dynamic MSs with a stronger tunability range. Authors in Ref.~\cite{michel2013using} studied the behavior of PCM-plasmonic MSs with three different configurations: Al nanoantennas arrays underneath, inside, or on top of a sputtered 50 nm-thick PCM layer (see Figure~\ref{figS3}A). The proposed structure relies on the Wood's anomaly to achieve a narrowband resonance for switching applications. Indeed, in a lattice of resonant nanoantennas, the interference between the nanoantenna resonance and the Bragg resonance of the lattice generates narrow resonances. Experimental results show that, upon the structural transition of the PCM, a large tuning range to the full width at half maximum (FWHM) ratio (about 1.03) can be achieved. Michel \textit{et al.} later demonstrated the reversible tuning of these resonances by using ultrafast laser pulses as the stimulation source for the phase conversion in Ge$_3$Sb$_2$Te$_6$ \cite{michel2014reversible}. They used 800 nm femtosecond pulses with a 1.65 mJ energy per pulse, a 50 fs pulse width, and repetition rates up to 960 Hz focused at 230-270 $\mu$m area on the sample. Moreover, to improve the optical functionalities and simplify the fabrication process, they avoided the deposition of capping and buffer layers.

In several applications, such as perfect absorption and thermal emission, Ohmic losses within the constitutive materials are favorable to enhance the overall performance of the optical device and its figure of merit (FOM). In this regard, Tittl \textit{et al.} presented the first experimental demonstration of a band- and consequently temperature-selective switchable mid-IR perfect absorber \cite{tittl2015switchable}. As depicted in Figure~\ref{figS3}B, their proposed structure is a MIM architecture composed of a spacer layer of Ge$_{3}$Sb$_{2}$Te$_{6}$ sandwiched between an Al back reflector and an array of Al nanosquares. Thanks to the strong gap surface plasmon (GSP) mode, nearly perfect absorption ($>$90\%) independent of the incident angle and polarization occurs. A resonance shift over 500 nm in the reflectance response upon the structural transition of GST from the amorphous to the crystalline state was experimentally realized. By simultaneously increasing the number of pixels and reducing the FWHM of the resonances, such an interesting platform offers diffraction-limited multispectral thermal imaging capability with high resolution (see the right panel of Figure~\ref{figS3}B).

The first experimental demonstration of tunable chiral MSs using PCMs was reported by Yin \textit{et al.} \cite{yin2015active}. The authors leveraged the transparency window of Ge$_{3}$Sb$_{2}$Te$_{6}$ (between 2.8 and 5.5 $\mu$m) for large spectral tunability (from 4.15 to 4.90 $\mu$m) of the circular dichroism (CD) response in the mid-IR regime. They utilized this effect in combination with a static chiral bias-type layer to flip the sign of the CD signal. The underlying system consists of a 50-nm-thick layer of PCM sandwiched between a Born-Kuhn type chiral plasmonic dimer comprising of corner-stacked nanorods (Figure~\ref{figS3}C). The ease of fabrication with no additional lithography step and optimum near-field coupling of the generated localized surface plasmon with the functional materials are the two-fold advantages of this architecture.

So far, most MS optic applications have been incorporated in PCMs as a tunable dielectric medium. However, the inherent plasmonic properties, i.e., negative relative permittivity, of some PCMs upon transition from the amorphous to the crystalline state has gone somewhat overlooked. In 2018, Gholipour \textit{et al.} demonstrated that upon transition of a 70-nm-thick structured layer of Ge$_{2}$Sb$_{2}$Te$_{5}$ from amorphous to polycrystalline phase, the material functionality is switched from dielectric to the metallic (i.e., plasmonic) in the ultraviolet (UV) to near-vis spectral range \cite{gholipour2018phase}. As shown in Figure~\ref{figS3}D (middle panel), while the real part of the relative permittivity of a-GST is positive at wavelengths below 660 nm, it exhibits negative values necessary for the generation of a surface plasmon polariton (SPP) mode. Figure~\ref{figS3}D (middle panel) shows the plasmonic FOM, defined as the ratio of the real part of the propagating surface plasmon polariton wavevector to its imaginary part which is a representation of the SPP decay length, for the polycrystalline GST and a series of widely used noble metals. It is clear that in the UV to near-vis range, polycrystalline GST has comparable FOM values to their counterparts' albeit surpassed by Al. The authors fabricated a hybrid MS using focused ion beam milling of a multistack ZnS-SiO$_{2}$/GST/ZnS-SiO$_{2}$ layer to create a 1D array of nanogratings on top of the glass substrate. In the case of TM-polarization (incident electric field perpendicular to the grooves), the MS introduces resonances for both extreme states of GST, which manifest themselves in the transmitted and reflected visible colors (inset in Figure~\ref{figS3}D). In the as-deposited amorphous phase, there exist displacement current resonances owing to the contrast between the GST as a high index dielectric and the low-index ambient. However, for the crystalline phase, the fundamental mode is a plasmonic resonance regarding the opposing signs of relative permittivity at the interfaces between the GST and the surrounding dielectric medium. Thanks to the anisotropic nature of nanogratings, the resonant peaks disappear when the MS is illuminated with a TE-polarized (incident electric field parallel to the grooves) light. Accordingly, the patterned structure functions as an effective medium with a non-dispersive response.

\begin{figure*}[t]
	\centering
	\includegraphics[trim={0cm 1cm 0cm 0cm},width=\textwidth, clip]{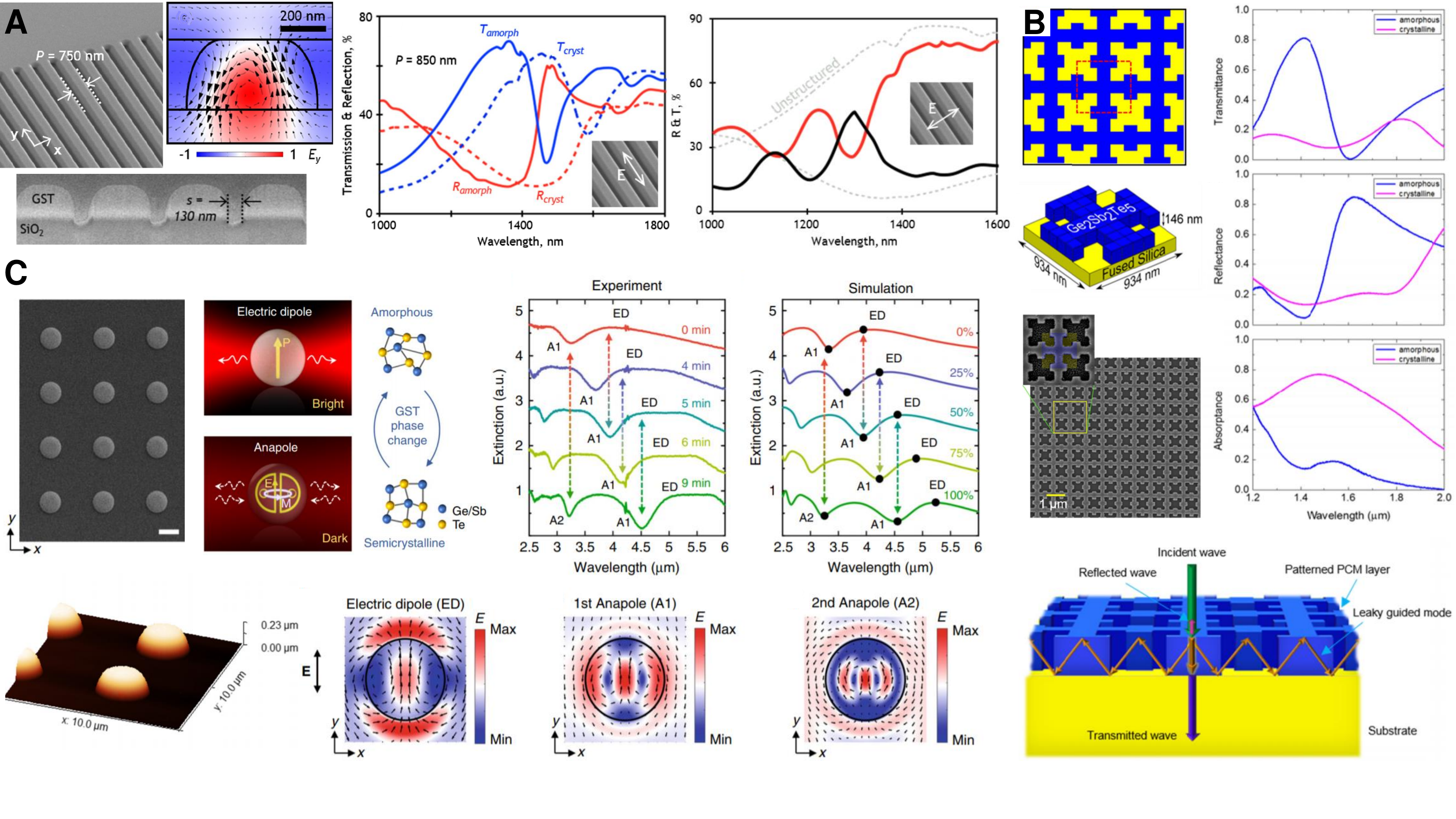}
	\caption{All-PCM MSs for active tailoring of the scattering light.
	(A) All-dielectric PCMs \cite{karvounis2016all}.
	Left: Oblique incidence and cross-sectional SEM images of the fabricated MS using focused ion beam milling. The electric field profile overlaid with arrows representing the direction and magnitude of the magnetic field under TE-polarized illumination.
	Middle: Microspectrophotometrically measured reflection and transmission spectra for the amorphous and polycrystalline MS with p = 750 nm and s = 130 nm under TE-polarized illumination.
	Right: Micro-spectrophotometrically measured reflection (red) and transmission (black) spectra for the amorphous (solid lines) and unstructured (dashed lines) MS with p = 850 nm and s = 130 nm under TM-polarized illumination.
	(B) Tunable guided mode resonance \cite{pogrebnyakov2018reconfigurable}.
	Left: Schematic of the optimized design and SEM images of the fabricated bistable transmittive/absorptive MS.
	Right: Experimental transmittance, reflectance, and absorbance spectra of the MS with amorphous and crystalline states.
	Bottom: Formation of the guided mode resonance in the so-called fishnet MS using ray tracing illustration.
	(C) Active control of anapole modes \cite{tian2019active}.
	Left: Top SEM image (scale bar = 1 $\mu$m) and 3D AFM image of the fabricated GST nanodisks. Middle: Conceptual illustration of the excited electric dipole resonance and the anapole mode in a GST nanosphere in the amorphous and polycrystalline state, respectively.
	Right: Evolution of extinction spectra from experimental measurement and numerical simulations of GST nanodisks with radius of 1 $\mu$m and height of 220 nm. The good agreement between the experimental and simulation results justifies that appropriate baking durations were adopted for the phase transformation of GST to the intermediate states.
	Bottom: Near electric field distribution of the fundamental modes in the cross-section xy-plane for amorphous, 50\% crystalline, and fully crystalline nanodisk, respectively.  
}
	\label{figS4}
\end{figure*}

\subsection{Hybrid dielectric/PCM metasurfaces for global amplitude control}

Plasmonic MSs governed by localized and lattice surface plasmon modes exhibit high dissipation losses, limited scattering cross-sections, and low efficiency. To mitigate these drawbacks, CMOS-compatible high-index and semiconductor MSs with a wealth of distinct optical resonances have been introduced as promising alternatives. In this regard, dynamic dielectric MSs that rely on high-contrast PCM nanostructures enabling adaptive functionalities have gained increased interest recently.

Karvounis \textit{et al.} experimentally demonstrated bistable, rather high-quality transmission and reflection resonances using a 300-nm-thick array of non-diffractive sub-wavelength Ge$_{2}$Sb$_{2}$Te$_{5}$ nanogratings (see Figure~\ref{figS4}A) \cite{karvounis2016all}. Upon TE-polarized excitation, anti-phase displacement currents along the GST core and a circulating magnetic field in the cross-section of the core occur (see Figure~\ref{figS4}A). To tailor the electromagnetic response of the MS, GST nanogratings are converted from the amorphous to the crystalline state by raster scanning of a 532 nm continuous wave laser with a spot diameter of 0.5 $\mu$m and intensity of 3 mW$\mu$m$^{-2}$. Such a structural transitions shift the initial near-IR resonance by as much as 10\% providing switching contrast ratios of up to 5:1 (7 dB) in reflection and 1:3 (-5 dB) in transmission under a TE-polarized illumination (see Figure~\ref{figS4}A, middle panel).

Pogrebnyakov \textit{et al.} presented an optimized design and the experimental realization of a tunable polarization-insensitive MS using 150-nm-thick Ge$_{2}$Sb$_{2}$Te$_{5}$ nanostructures to switch the optical functionality from a high transmittive filter to a highly absorptive device in the near-IR wavelength range \cite{pogrebnyakov2018reconfigurable}. As depicted in Figure~\ref{figS4}B, the transmission resonance of the structure in the amorphous state is governed by a leaky guided-mode resonance induced by the periodic pattern of air voids within the GST layer. Due to the increased intrinsic loss of crystalline PCM, the field intensities in the structure is greatly reduced which results in transmission drop with a 7:1 contrast ratio.

Tian \textit{et al.} performed a comprehensive mode analysis of structured phase-change alloys using the rigorous multipole decomposition technique \cite{tian2019active}. The study shows that the high refractive index of Ge$_{2}$Sb$_{2}$Te$_{5}$, coming with low loss in the mid-IR regime, empowers its nanostructure (which is a nanodisk) to support a diverse set of multipolar Mie resonances including ED, MD, and anapole state (see Figure~\ref{figS4}C). Moreover, the dramatic optical contrast of GST enables dynamic controllability of these resonances leading to progressive spectral shifting of fundamental resonance modes. As a proof-of-concept, they showed that the ED-to-anapole shifting can be achieved by inducing 50\% phase-change at any given wavelengths in the mid-IR spectral range (see Figure~\ref{figS4}C). Multimodal shifting between the scattering bright and dark modes over a broadband region facilitates multispectral optical switching with high ERs. Notably, the intermediate phase transformation of the 220-nm-thick GST nanodisks was carried out by baking the sample on a hotplate with fixed temperature of 145 $^{\circ}$C and different durations up to 9 minutes. Using a similar configuration, the same group demonstrated transmittance contrast up to 30 dB between two extreme phases of a 400-nm-thick GST nanodisk near the multipolar resonances in the mid-IR spectral range \cite{tian2018reconfigurable}.

Leveraging the optical contrast of PCMs, Petronijevic \textit{et al.} demonstrated optical tuning of the electromagnetically induced transparency (EIT) effect in a Ge$_{2}$Sb$_{2}$Te$_{5}$/Si MS \cite{petronijevic2016all}. The constitutive unit-cell is composed of a coupled dipole nanoantenna, which supports an electric dipole-like low-Q bright mode, and a square shape nanoring, which supports a magnetic dipole-like high-Q dark mode, hybridized by a thin (17 nm) layer of GST. When GST is in its amorphous state, an EIT-like transmission peak due to destructive interference of coupled modes in the transmission dip occurs. Crystallization of GST redshifts the resonances and destroys the EIT-like effect. Simulation results show a contrast ratio up to 10 in the telecommunication range. In a follow-up work, the same group utilized a 100-nm-thick layer of GeTe with structured Si nanobars to experimentally demonstrate switching of the collective dipole-like resonance of a hybrid MS with moderate contrast ratio ($\sim$5) at the wavelength of 1.55 $\mu$m \cite{petronijevic2019near}. The crystallization process was carried out through slow heating of GeTe above its crystallization temperature ($\sim$ 200$^{\circ}$C). More recently, several innovative hybrid MSs for active manipulation of fundamental resonance modes of constituent meta-atoms through the phase transition of embedded PCMs have been proposed \cite{jafari2016zero,lan2019highly,bai2019tunable,bai2019near}.  

Tian \textit{et al.} demonstrated a polarization-insensitive broadband perfect absorber in the vis-near-IR region using a PCM-based MS. The MS is composed of an array of Ge$_{2}$Sb$_{2}$Te$_{5}$ nanosquares separated from the Au mirror by a stack of low-index dielectric and high-index PCM layers. Thanks to the excited dipole/quadrupole resonances from the patterned GST structure and the cavity resonance mode from the GST planar cavity, near-perfect absorption peaks are achieved in the wavelength range from 350 to 1500 nm. A study on temporal variation of temperature through a heat transfer model revealed that the temperature of GST can be raised from room temperature to $\sim$480 K in just 0.56 ns with a light fluence of 1.11$\times$10$^{8}$ Wm$^{-2}$. This not only lowers the power requirements for the phase transformation of PCMs but also decreases the crystallization process time to less than 1 ns.

\begin{figure*}[t]
	\centering
	\includegraphics[trim={0cm 6cm 0cm 0cm},width=\textwidth, clip]{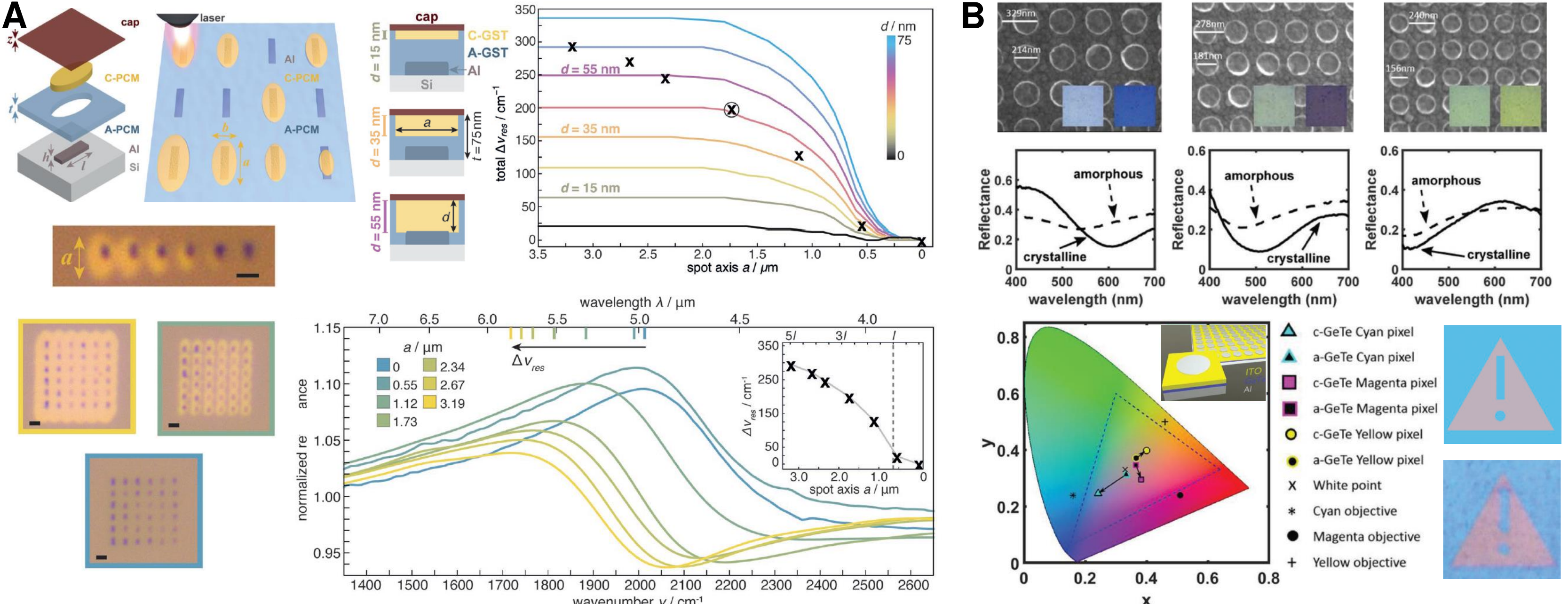}
	\caption{Dynamic PCM/plasmonic MSs for local amplitude control.
	(A) Resonance tuning by local programming \cite{michel2019advanced}.
	Left: Schematic of the stacked layers of the hybrid MS. A-PCM and C-PCM stand for amorphous and crystalline PCM, respectively. Selected nanorod antennas are locally addressed (or crystallized) using a semiconductor laser generating an elliptical beam profile. Bright field image of locally addressed nanoantennas with increased spot sizes (scale bar = 1 $\mu$m).
	Bottom: Normalized reflectance spectra of 6$\times$6 unit-cells addressed with different crystalline spot sizes. Optical microscope images of the super cells written with three different spot sizes of a = 0, 1.73, 3.19 $\mu$m (scale bars = 1 $\mu$m).
	Right: Cross section view of a unit-cell simulated for the same spot size but different crystallization depths (i.e., d). Colored spectra of resonance shift versus the writing spot size and the crystallization depth calculated from numerical simulations. Black crosses are achieved from experimental measurements.
	(B) Tunable color display \cite{carrillo2019nonvolatile}.
	TOP: SEM images of the fabricated color display MS for generating different colors (insets: optical microscope images from the MS under white light illumination).
	Middle: Experimental reflectance spectra from cyan, magenta, and yellow color pixels, respectively.
	Bottom: Representation of the reflectance spectra in the middle panel in the CIE 1931 chromaticity space (inset: sketch of the simulated MIM meta-atom and the PCM-based MS). A binary image generated by a combination of pseudo-white amorphous and cyan crystalline pixels and the corresponding optical microscope image of the fabricated MS.
}
	\label{figS5}
\end{figure*}

Reconfigurable nanoscale optical cavities are promising candidates for applications such as displays and artificial retina devices. In a distinct, wide-angle display framework, Schlich \textit{et al.} reported full color-switching by utilizing a thick multilayered reflective display incorporating a 18-nm-thick layer of Ge$_{2}$Sb$_{2}$Te$_{5}$ on top \cite{schlich2015color}. In their approach, a train of femtosecond laser pulses is employed to detune the resonance condition of the Fabry-Perot (F-P) cavity by transforming the phase of the GST layer. In another work, Yoo \textit{et al.} leveraged a stack of PCM layers in an optical F-P cavity to generate a distinct artificial color spectrum \cite{yoo2016multicolor}. The highly absorbing nature of PCM layers combined with the strong optical interference effect governed by the cavity enables multiple color appearances by the selective phase transition of PCM layers. This concept was pursued by integrating a subwavelength optical cavity, which incorporated two separated layers of GeTe, with a joule heater element to produce a tunable micro-display rendering 4 different colors \cite{jafari2019reconfigurable}. 

\subsection{Hybrid plasmonic/PCM metasurfaces for local amplitude control}

In the previous subsection, the adaptive functionalities are governed by global control on the amplitude of the incident light through a uniform structural transition of PCMs all over the metadevice. To harness the high potential of PCMs, local tuning of the individual meta-atoms in a reversible fashion is indispensable. In this subsection, recent developments enabling pixel-by-pixel programming of the MS using finely focused optical beams and localized electrical currents are discussed. We distinctly discuss both plasmonic and all-dielectric MS platforms with resonant and non-resonant meta-atoms exploiting PCMs to enable dynamic meta-optics. 

A bistable transmittive MS with a periodic arrangement of coupled Al nanopatches and 70-nm-thick Ge$_{2}$Sb$_{2}$Te$_{5}$, as the dielectric spacer, was proposed to tune far-field radiation patterns in the mid-IR frequencies range \cite{alaee2016phase}. Numerical results show that GST with as-deposited state satisfies Kerker's condition, in which constructive (destructive) interference occurs between the radiation of the electric and magnetic dipoles in forward (backward) direction, while the crystalline state only supports electric dipole resonance mode. As a result, the directive radiation pattern in the amorphous state is transformed into the omnidirectional scattering upon switching the state of the PCM layer to the crystalline.

Recently, Michel \textit{et al.} demonstrated that phase switching of the PCM with localized laser spots can arbitrarily tune the resonance of an individual hybrid meta-atom up to one FWHM \cite{michel2019advanced}. The investigated MS is composed of a 2D arrangement of Al nanorods deposited on a Si substrate and covered by a 75-nm-thick Ge$_{3}$Sb$_{2}$Te$_{6}$ layer which is protected by a thin capping film of ZnS-SiO$_{2}$ (see Figure~\ref{figS5}A). A semiconductor laser diode (660 nm central wavelength, sub-microsecond pulse duration) generating an elliptical beam spot, that is matched to the elongated shape of the nanorod, was employed for the writing process. While the aspect ratio of the long to the short axis (i.e., a/b) was kept constant during the experiments, the size of the elliptical beam spot was modified by changing the laser pulse power. As shown in Figure~\ref{figS5}A (right bottom), by increasing the laser spot size, the resonance frequency redshifts from 4.98 $\mu$m (vertical light blue line) in the amorphous state to 5.84 $\mu$m (vertical light yellow line) in the fully crystalline case. Comprehensive simulations and experiments were performed to reveal the complicated relationship between the crystallization depth and the temperature gradient in the PCM layer. Figure~\ref{figS5}A (right-top) depicts the resonance shifts for each crystallization depth (i.e., d) as a function of spot size (i.e., a) achieved from self-consistent multiphysics simulations. These findings show that through simultaneous control of the lateral size, crystallization depth, and position of the optically induced crystalline spot within each unit-cell, precise tuning of the reflection amplitude shift (less than 1 $\mu$m here) and phase shift (less than $\pi /6$ here) can be achieved.

Utilizing a MIM structure confining a thin layer of GeTe, a tunable optoelectronic color generation system was proposed more recently \cite{carrillo2019nonvolatile}. In the crystalline state, the gap plasmon mode characterized by mirrored currents in the top and bottom Al layers excites a transversal magnetic dipole mode that absorbs the incident light (see Figure~\ref{figS5}B). Given the optimized geometrical parameters, such a resonant absorber can selectively absorb the red, green, and blue spectral bands of the visible spectrum and generate vivid cyan, magenta, and yellow pixels. By switching the state of the PCM to the amorphous phase, the resonant peak is suppressed and a reflective pseudo-white color is realized. Figure~\ref{figS5}B illustrates a binary image composed of amorphous (with pseudo-white color) and crystalline (with cyan color) pixels generated using the experimentally measured spectra (right-top) and experimentally achieved from the fabricated MS written using a scanning laser (right bottom).

\begin{figure*}[t]
	\centering
	\includegraphics[trim={0cm 9.3cm 1.5cm 0cm},width=\textwidth, clip]{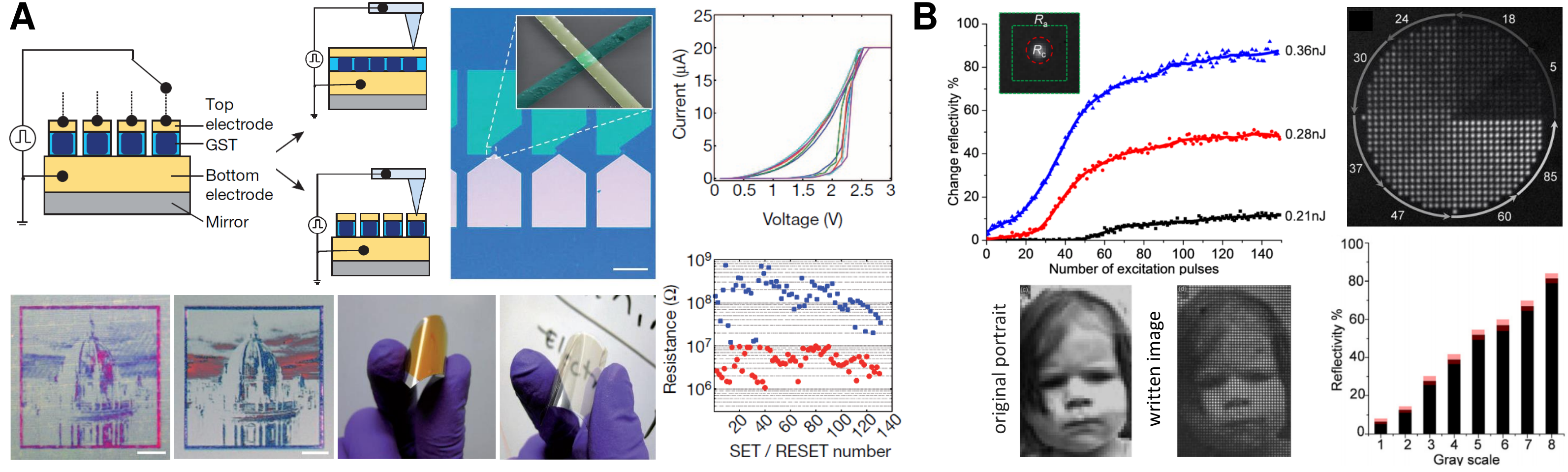}
	\caption{Pixel-by-pixel crystallization for local control of the amplitude response.
	(A) Local addressing approach by an electrical current \cite{hosseini2014optoelectronic}.
	Left: Cross-section view of two different GST-based MS platforms used as electronic displays as well as the schematic of a conductive AFM tip used for localized phase transformation of GST.
	Bottom: Constructed image on a continuous ITO/GST/ITO stack and lithographically defined 300 nm$\times$300 nm pixels using electrical current pulses (white pixels are amorphous regions. Scale bar = 10 $\mu$m). Reflective and semi-transparent type electronic displays sputtered on a flexible boPET substrate.
	Right: Optical microscope image of ITO/GST/ITO crossbar-like devices (inset: false-colour SEM image of one single pixel). I-V curve demonstrating the crystalline switching behavior. Resistance evolution of a single pixel after several SET (low resistance) and RESET (high resistance) cycles.
	(B) Multilevel crystallization using ultrashort laser pulses \cite{wang20141}.
	Left: Measured change in the optical reflectivity [(R$_{\textrm{c}}$-R$_{\textrm{a}}$)/R$_{\textrm{a}}$]of the GST sample versus the number of induced pulses (for 3 different pulse energies). R$_{\textrm{c}}$ and R$_{\textrm{a}}$ are the peak reflectivity in the center of the crystalline spot (enclosed in the red circle) and the average reflectivity of the amorphous region (enclosed in the green square). The original grayscale portrait of a kid and the written 8-level grayscale image into the GST film. 
	Right: 8 levels of crystallization shown in the spiral plate pattern with 34 $\mu$m diameter. Pulse numbers are increased counter-clockwise (from 5 to 85) as shown by arrows. Average reflectivity change of each level with the corresponding standard deviation bar.
}
	\label{figS6}
\end{figure*}

\subsection{Hybrid dielectric/PCM metasurfaces for local amplitude control}

To highlight the potential of reprogrammable pixelated MSs, Hosseini \textit{et al.} \cite{hosseini2014optoelectronic} demonstrated the first representation of a high resolution, high speed, and low power consumption microdisplay (see Figure~\ref{figS6}A). They employed a simple optoelectronic reflective framework in which a 7-nm-thick layer of Ge$_{2}$Sb$_{2}$Te$_{5}$ is sandwiched between two layers of ITO as a transparent conductive oxide material. By applying electric current pulses through a conductive tip of an atomic force microscope (AFM), they switched the phase of GST from the amorphous to the crystalline in the nanometer scale. Due to the high optical contrast between the two solid states of GST, a vivid pattern of a color image can be imprinted using this technique. The authors improved the resolution of such color displays by replacing continuous films of ITO/GST/ITO with a crossbar type pixelated array of the same stack where each pixel can be randomly accessed and manipulated through the conductive AFM probe. To verify this, an array of subwavelength pixels (300 nm $\times$ 300 nm with 200 nm pitch) was fabricated and electrically switched using a nanoscale conductive tip. Figure~\ref{figS6}A shows the resulting optical images in which the optical contrast between the amorphous and crystalline pixels is striking. Reflective and semi-transparent type color displays with wide-viewing-angle were demonstrated on both rigid and flexible substrates. Also, to assess the electrical switching characteristics of a single pixel within a more easy-to-measure architecture, vertically crossbar-like ITO/GST/ITO devices with an active area of 300 nm$\times$300 nm were fabricated (see Figure~\ref{figS6}A). The experimental current-voltage measurements show 350 times increase in the resistance between amorphous and crystalline phases at the threshold voltage of 2.2 V. Notably, a SET direct current can transform the device to its low resistance (or crystalline phase) while the amorphization process can be performed by applying a 100 ns RESET pulse and amplitude of 5 V to set back the device to its high resistance state. Later, the same group reported improvements in the off-line color depth modulation and resolution leveraging a similar approach and substituting Ge$_{2}$Sb$_{2}$Te$_{5}$ with a growth dominant phase-change alloy Ag$_{3}$In$_{4}$Sb$_{76}$Te$_{17}$ (AIST) \cite{rios2016color}.

Wang \textit{et al.} took benefit of multilevel crystallization state of PCMs in a stacked platform of dielectric/GST/dielectric to realize high-density optical data storage and grayscale holography \cite{wang20141}. Using a home-built optical system including an ultrafast pump-probe laser connected to a beam scanning apparatus, a grayscale image was written to the sandwiched 50-nm-thick Ge$_{2}$Sb$_{2}$Te$_{5}$ layer. By reading out the reflectivity of a single pixel in the center of the written spot (0.64 $\mu$m FWHM, 50 $\times$ objective lens, NA = 0.8), a mapping between the crystallization level and number of induced laser pulses (with different pulse energies of 0.21, 0.28, and 0.36 nJ) was achieved (see Figure~\ref{figS6}B). Given 730 nm wavelength writing laser and $\sim$1 $\mu$m gird size, they also demonstrated a storage density of 1.7 Gbit/in.$^{2}$ considering 8 distinct crystallization levels (equivalent to 3 bits/mark). Recently, the same group used a similar approach to imprint a grayscale image on a 70-nm-thick Ge$_{2}$Sb$_{2}$Te$_{5}$ film, which was used later as a photomask with multilevel absorption \cite{wang2017reconfigurable}. This mask enables submicron lateral resolution grayscale photolithography to finely control the local exposure dosage necessary for 3D sculpting of photoresists used for fabrication of 3D MSs. Other groups have reported grayscale image recording by taking advantage of the multilevel structural evolution of PCMs induced by focused laser beams with different pulse energies \cite {wei2017grayscale,wen2018multi}. Also, computer-generated hologram patterns with 1 $\mu$m pixel pitch and 16 k$\times$16 k resolution were realized by the local phase transition of a 20-nm-thick Ge$_{2}$Sb$_{2}$Te$_{5}$ film \cite{lee2017holographic}. This was accomplished by excimer laser (308 nm wavelength, 30 ns pulse width, 600 Hz repetition rate) photolithography with a fluence of 60–80 mJcm$^{-2}$ on the surface of the PCM layer. The refractive index change due to the crystallization of GST shifts the resonance governed by thin-film interference condition in each pixel which can be applied to realize a broadband, full-color diffracting hologram panel. Recently, a rewritable full-color computer-generated hologram using PCM-based color-selective diffractive optical components has been demonstrated \cite{hwang2018rewritable}. The recorded spatial binary pattern in the Ge$_{2}$Sb$_{2}$Te$_{5}$ layers combined with thickness-tailored substrates generate color-selective optical diffractions that can be tuned by inducing intermediate states in GST.

\section{Active Phase control with tunable phase-change metasurfaces}

The potential functionalities of reconfigurable PCM-based MSs discussed in the previous section are restricted to the amplitude modulation of the scattered light. To enable real-world, on-demand applications, dynamic optical components tailoring the phasefront of the incident light is essential. More recently, several groups have demonstrated tunable phase-gradient MSs employing PCMs, thereby extending the capability of dynamic wavefront shaping, such as reconfigurable beam-steering, tunable focusing, and switchable photonic spin-orbit interactions. In the following section, a review of the recent progress on the implementation of PCM-based MSs for active tuning of the phasefront of the scattered light is provided.

\subsection{Hybrid plasmonic/PCM metasurfaces for global phase control}

\begin{figure*}[t]
	\centering
	\includegraphics[trim={0cm 3.5cm 0cm 0cm},width=\textwidth, clip]{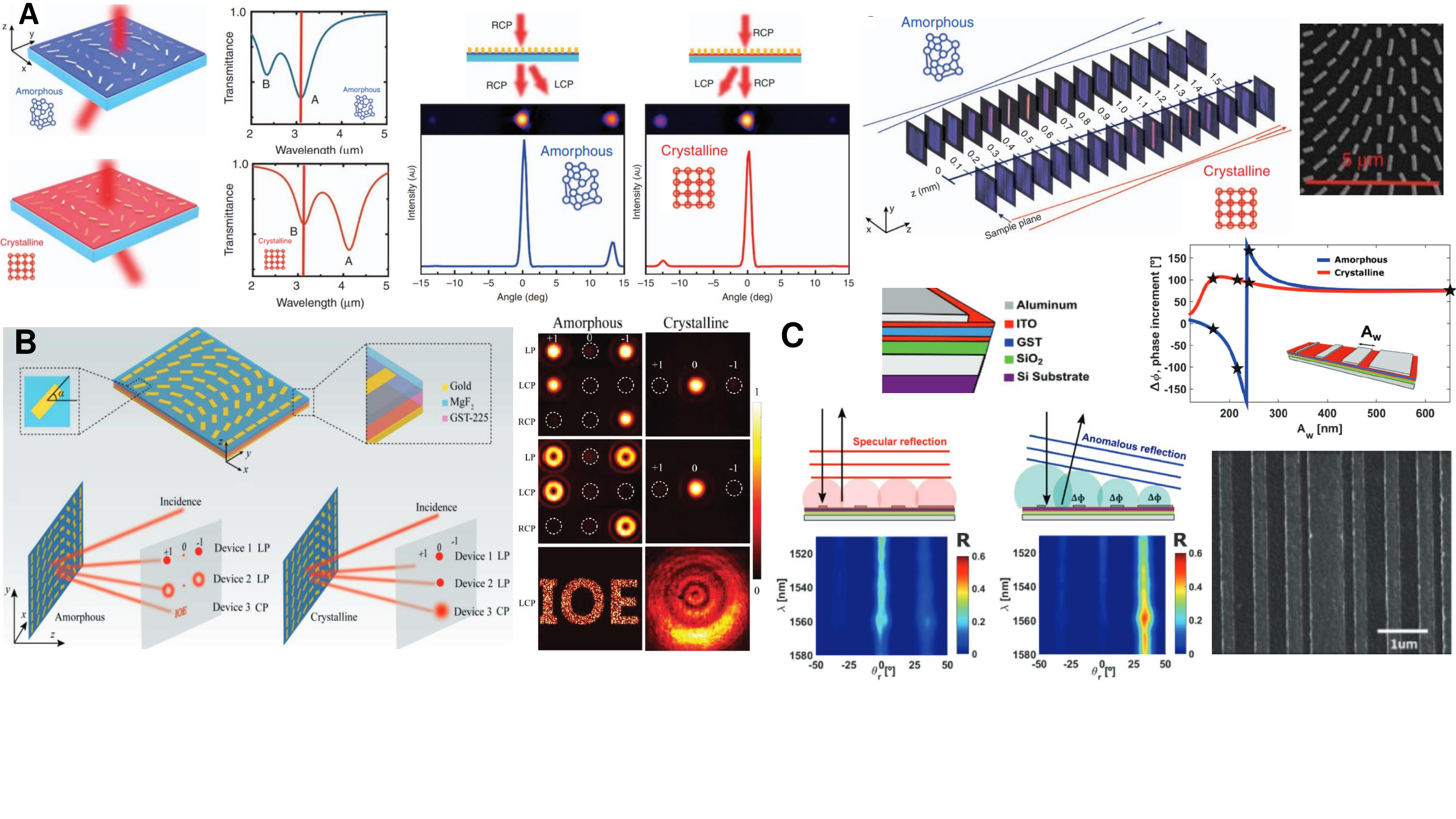}
	\caption{Illustration of adaptive functionalities using hybrid plasmonic/PCM MSs for active control of the global phase profile.
	(A) Dynamic phasefront switching \cite{yin2017beam}.
	Left: 3D illustration of dynamic a P-B phase MS consisting of two types of meta-atoms; large length (A) and short length (B). While in the amorphous state, only type A interacts with the incident light and deviate the beam into the left, in the crystalline phase type B does this job but into the opposite direction.
	Middle: Infrared camera images and intensity plots of the transmitted beams in the amorphous and crystalline state.
	Right: Transmitted beam intensity for cylindrical bifocal metalens imaged at different distances z from the MS. In the amorphous state, a bright line appears at z = 0.5 mm, whereas in the crystalline case the focal line changes to z= 1 mm. SEM image of a portion of the fabricated cylindrical metalens.
	(B) Switchable photonic spin–orbit interactions \cite{zhang2018plasmonic}.
	Left: Schematic representation of the MS topology, the constituent materials, and optical performance of three designed MSs in the amorphous and crystalline states.
	Right: Experimentally measured intensity distributions of the designed MSs for switchable spin Hall effect, vortex beam generation, and holography.
	(C) Reconfigurable beam-steering \cite{de2018nonvolatile}.
	Top: Schematic of the MIM meta-atom and its phase evolution as a function of width (i.e., A$_{\textrm{w}}$) for both amorphous (blue curve) and crystalline (red curve) states. The stars correspond to the widths selected to fabricate the supercell of the MS.
	Bottom: Artistic rendering of Huygens principle for the reconstruction of the reflected light under normal incidence, and measured angular reflectance spectra for both cases of amorphous and crystalline. SEM image of the fabricated sample for beam steering applications.
}
	\label{figS7}
\end{figure*}

So far, the proposed dynamic MSs utilized resonant dispersive meta-atoms, which intrinsically limit the operational bandwidth of corresponding optical functionalities. To facilitate dispersionless phase control while making the scattered amplitude from the phase response decoupled, Pancharatnam-Berry (P-B) or geometric phase MSs have been introduced as an effective paradigm. By spatially varying the orientation of identical meta-atoms around their optical axis, not only wideband gradient MSs can be realized, but also the fabrication tolerance is alleviated. In this regard, Yin \textit{et al.} demonstrated the first dynamic P-B based MS consisting of two differently sized interleaved meta-atoms arranged on top of a low-loss 50-nm-thick layer of Ge$_{3}$Sb$_{2}$Te$_{6}$ \cite{yin2017beam}. As shown in Figure~\ref{figS7}A, when the GST layer is in the amorphous state, the resonance condition is only fulfilled for meta-atoms with longer lengths. However, upon conversion of GST to the crystalline state, only meta-atoms with the shorter length strongly interact with the incident light at the same wavelength. This enables interweaving of two functionalities using two parallel rows of meta-atoms with appropriate lengths interleaved in a supercell architecture. As depicted in Figure~\ref{figS7}A, by rotating the longer and shorter meta-atoms clockwise and counter-clockwise, respectively, an opposite phase gradients can be imparted to the transmitted circularly polarized (CP) light leading to a cross-polarized light with +1 and -1 order diffraction, respectively. Moreover, by encoding the quadratic phase distribution of cylindrical metalens in the spatial orientation of plasmonic meta-atoms, a bifocal focusing can be realized upon switching the state of GST, although the overall efficiencies are less than 10\%. It is notable that resonant meta-atoms in the spatial multiplexed MSs have relative low scattering cross-section thus decreases the overall efficiency compared to the conventional schemes.

Zhang \textit{et al.} have experimentally demonstrated switchable spin-orbit interactions using a hybrid P-B phase MS \cite{zhang2018plasmonic}. A MIM configuration is adopted in which the insulator layer is comprised of a thin layer of MgF$_{2}$ on top of a 600-nm-thick Ge$_{2}$Sb$_{2}$Te$_{5}$ film sandwiched between the bottom Au substrate and the top array of Au nanoantennas (see Figure~\ref{figS7}B). Due to the judiciously engineered meta-atoms, when the GST is in the amorphous state, the reflectance of the cross-polarized light from the MS is as high as 60\%, while the co-polarized reflectance is less than 10\%. Accordingly, the MS exhibits a highly efficient half-wave plate functionality with the polarization conversion ratio (PCR) overcoming 80\% in the mid-IR spectral range. By annealing the a-GST active layer and converting its state to the crystalline, the MS functions as a simple mirror with a relatively low PCR of 10\%. The authors leveraged this effect to dynamically manipulate the coupling between spin and orbit momentums of photons in a broadband wavelength range from 8.5 to 10.5 $\mu$m. As a proof-of-concept illustration, three MSs controlling the spin-orbit interactions were fabricated and characterized, which enabled spin Hall effect, vortex beam generation, and holography in the amorphous case (see Figure~\ref{figS7}B). The state transition of the a-GST to the c-GST simply destroys the geometric phase phenomena and thus deactivates these optical functionalities.

\begin{figure*}[t]
	\centering
	\includegraphics[trim={0cm 0cm 0cm 0cm},width=\textwidth, clip]{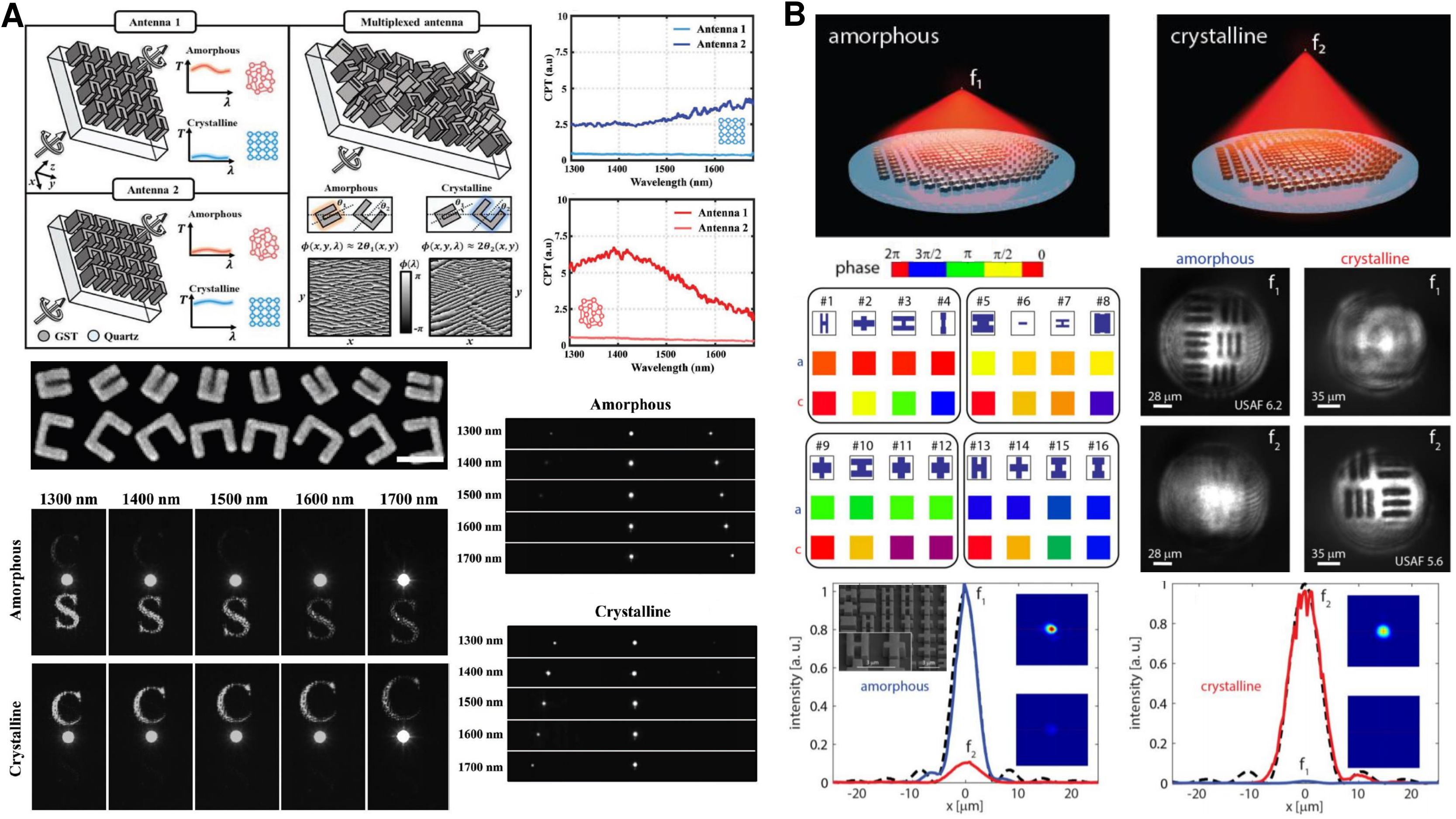}
	\caption{Dynamic high-index MSs for active control of the global phase response.
	(A) Wideband wavefront switching \cite{choi2019metasurface}.
	Top: An arrangement of periodic U-shaped nanoantennas with large/zero CPT for amorphous/crystalline states (type 1) and crystalline/amorphous states (type 2) over wide near-IR wavelength range under illumination of a CP light.	Representation of a multiplexed array of nanoantenna types 1$\&$2 with different orientation angle $\theta_{1}$ and $\theta_{2}$, respectively, into a subwavelength size supercell. The phase of the transmitted cross-polarized light is determined by $2\theta_{1}(x,y)$ and $2\theta_{2}(x,y)$ in the amorphous and crystalline states, respectively, with suppressed chromatic dispersion according to the principle of P-B phase. Experimentally measured CPT spectra for nanoantenna types 1$\&$2 for amorphous and crystalline phases.
	Bottom: SEM image of the fabricated sample (scale bar = 500 nm). Captured Fourier plane images of the reconstructed hologram images and deflected light for both GST phases in five discrete wavelengths.
	(B) Diffraction-limited varifocal focusing \cite{shalaginov2019reconfigurable}.
	Top: Schematic representation of a dynamic bifocal metalens using engineered GSST meta-atoms to focus the light at f$_{1}$ = 1.5 mm and f$_{2}$ = 2.0 mm in the amorphous and crystalline states, respectively.
	Middle: Constructed library of 16 different meta-atoms to provide all 16 possible phase combinations in accordance with 4-level phase discretization (a and c stand for amorphous and crystalline, respectively). Resolved lines of USAF resolution charts in both phases.
	Bottom: Comparison of the aberration-free Airy beam intensity profile (dashed black curve) with the diffraction-limited focusing profile of the bifocal meta-lens in its amorphous (blue solid curve) and crystalline (red solid curve) states, respectively (inset: SEM image of a portion of the fabricated meta-lens and 2D images of the focal spot captured by an infrared camera).
}
	\label{figS8}
\end{figure*}

More recently, de Galarreta \textit{et al.} demonstrated a reconfigurable MS for beam steering applications in the operating spectral range of 1530–1570 nm \cite{de2018nonvolatile}. As shown in Figure~\ref{figS7}C, the MS is comprised of SiO$_{2}$, ITO, GST, and ITO multilayer stack sandwiched between an Al back reflector and a top array of periodically arranged Al nanoribbons with different lateral widths. The choice of Al as the plasmonic material not only reduces the fabrication cost but also makes the whole process CMOS compatible. Furthermore, the spacer layer between the nanoantennas and the PCM layer can be removed due to less diffusion of Al compared to Au in GST. This maximizes the interaction of the PCM layer with the near fields of the plasmonic nanoantennas. However, its relatively low melting temperature (around 660 $^{\circ}$C) makes the amorphization process of the GST layer challenging due to possible deforming or even melting issues. It is notable that the MIM configuration not only supports the fundamentally enhanced GSP mode, characterized by the antiparallel displacement current, but also does the subwavelength feature of the dielectric spacers facilitate the quenching process during the amorphization. Herein, the gradient phase control is achieved by slightly detuning each nanoribbon width from the specified center resonant frequency (see Figure~\ref{figS7}C). Beam steering with simultaneously multiple reflecting angles can also be achieved by pixelating the MS where each pixel can specifically be designed to steer the incident light at a different angle. When the GST is in its amorphous state, the MS functions as an anomalous reflector to a predesigned angle while in the crystalline state the incident beam experiences no phase gradient, and thus the incident light is specularly reflected.

\begin{figure*}[t]
	\centering
	\includegraphics[trim={0cm 9.6cm 0cm 0cm},width=\textwidth, clip]{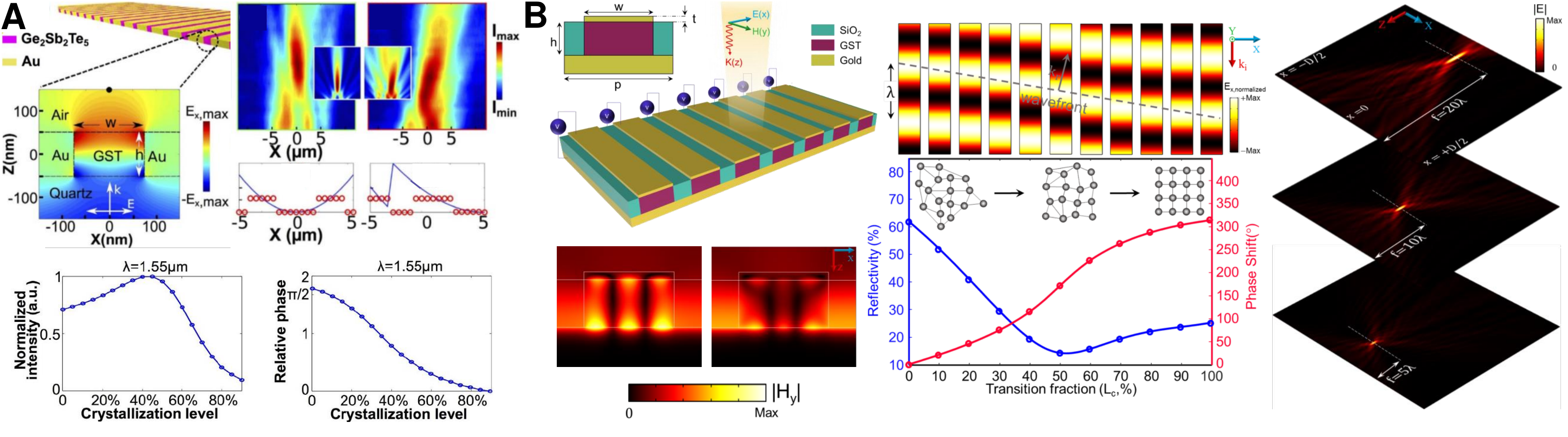}
	\caption{Local phase manipulation using plasmonic MSs in combination of PCMs.
	(A) Phasefront engineering principle \cite{chen2015engineering}.
	Top: Schematic of the planar meatlens and the electric field of the first order F-P mode in the cross-section of the GST slit. Experimentally measured focusing pattern in xz-plane for two different metalenses with on- and off-axis focusing capabilities corresponding to the binarized discrete phase distributions, respectively, (insets: the corresponding simulation results of the metalens using the binarized GST crystallization levels). The calculated phase profiles (blue curves) and the implemented binarized discrete phase distributions (red circles) are depicted for each metalens.
	Bottom: The relative phase and normalized electric field intensity as a functions of the crystallization level at the point monitor (black dot in the top panel).
	(B) Reconfigurable multifunctional MS \cite{abdollahramezani2018reconfigurable}.
	Left: Artistic rendering of the an electrically reconfigurable MS with pixel-by-pixel addressability. Simulated magnetic files profiles for short range and propagative surface plasmon resonance modes.
	Middle: Normalized reflected electric field patterns and evolution of reflection amplitude (left axis) and phase (right axis) as a function of the crystallization level.
	Right: Simulated intensity profiles of the reflected light from the tunable metalens in the transverse cross section for nominal focal lengths of $20\lambda$, $10\lambda$, and $5\lambda$, respectively.
}
	\label{figS9}
\end{figure*}

\subsection{Hybrid dielectric/PCM metasurfaces for global phase control}

Choi \textit{et al.} recently demonstrated wavefront switching in the near-IR spectral range utilizing a high-index MS comprising of an arrangement of rotating 260-nm-thick, U-shaped Ge$_{2}$Sb$_{2}$Te$_{5}$ nanoantennas \cite{choi2019metasurface}. As shown in Figure~\ref{figS8}A, two types of nanoantennas are used, in which type 1 exhibits large cross-polarized light transmittance (CPT) in the amorphous state under CP light illumination while exhibiting near-zero CPT for the crystalline phase. On the other hand, type 2 is optimized to show the same performance in the crystalline and amorphous states, respectively. Leveraging a multiplexed supercell (with 1.16 $\mu$m size) of both types of nanoantennas arranged based on the P-B phase principle, two distinct applications, i.e., anomalous refraction angle switching and dispersionless active hologram were implemented. As shown in Figure~\ref{figS8}A, the nonresonant scattering characteristic of these engineered meta-atoms grants wide (over 500 nm) operational bandwidth and high signal-to-noise ratio (above 7 dB) adaptive functionalities. Furthermore, a reliable fabrication process for dry etching of a 260-nm-thick layer of GST to form nanostructures with sharp side walls was also provided.

More recently, Shalaginov \textit{et al.} introduced an active metalens platform comprising of an array of judiciously patterned Ge$_{2}$Sb$_{2}$Se$_{4}$Te$_{1}$ Huygens meta-atoms sitting on top of a CaF$_{2}$ substrate (see Figure~\ref{figS8}B) \cite{shalaginov2019reconfigurable}. A generic design methodology enabling phase minimization error while maximizing the optical efficiency on the transformation of GSST from the amorphous to the crystalline state was presented. The continuous [0, 2$\pi$] phase profiles required for the bifocal focusing in the amorphous (with NA = 0.45) and crystalline (with NA = 0.35) cases are discretized to 4 equidistant phase levels. Accordingly, 16 distinct meta-atoms are necessary to cover all possible combinations of phase jumps upon the structural transition of GSST. Through full-wave simulations, a library of Huygens meta-atoms with distinct geometries, including ``I'', ``H'', and ``+'', was generated by sweeping the structural parameters to achieve the 16 optimal meta-atoms corresponding to 16 combinations of phases (see Figure~\ref{figS8}B). Consequently, a bistable varifocal metalens with diffraction-limited performance (above 20\% at both states) and low cross talk (switching contrast ratio of 29.5 dB) was experimentally demonstrated at the mid-IR wavelength range around $\lambda = 5.2~\mu$m. For further validation, standard USAF 1951 resolution charts images were captured by a meta-lens in its two states. Figure~\ref{figS8}B depicts that USAF 6.2 (half-period 8.8 $\mu$m) and USAF 5.6 (half period 7.0 $\mu$m) are greatly resolved in the amorphous and crystalline states, respectively, which are well-matched with the theoretical resolution limits of 9 $\mu$m and 7 $\mu$m, respectively.

\subsection{Hybrid plasmonic/PCM metasurfaces for local phase control}

\begin{figure*}[t]
	\centering
	\includegraphics[trim={0cm 6.5cm 0cm 0cm},width=\textwidth, clip]{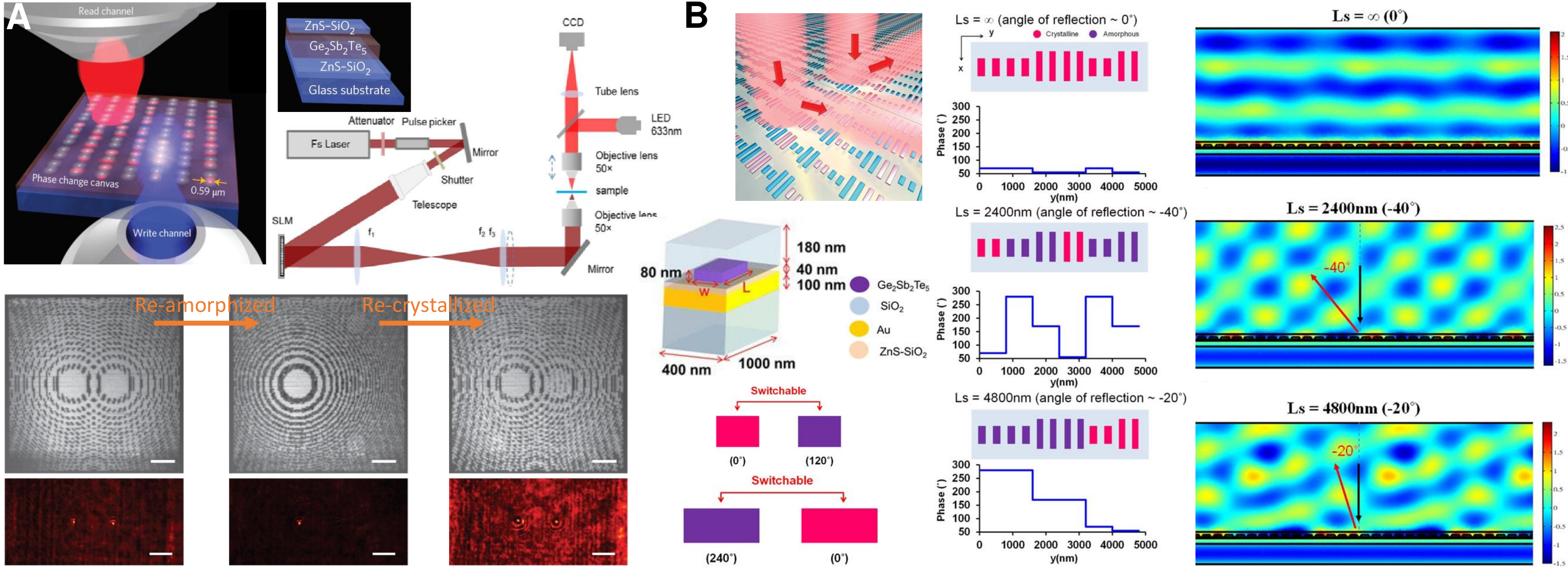}
	\caption{Dynamic manipulation of the phasefront by all-PCM-based MSs.
	(A) Optically reconfigurable MS.
	Top: Writing and reading process of a 2D pattern corresponding to various optical components including lenses and diffractive elements. The write channel carries ultrafast laser pulses for imposing continuous change in the refractive index of the GST film which is under observation through the read channel. Schematic illustration of the multilayered stack and the experimental setup.
	Bottom: Realization of Fresnel zone patterns focusing a plane wave into two different focal points. First, superimposed Fresnel zone patterns are written on the sample, then one of them is erased, and finally, both patterns are restored. Transmission focal spots generated by each of the implemented patterns (scale bar = 10 $\mu$m).
	(B) Dynamic beam deflection. 
	Left: Schematic diagram of the all-PCM gradient MS and its unit-cell consisting of a GST nanorod on top of a ZnS-SiO$_2$ layer and an Au back reflector. Selection rule for the GST meta-atoms; by changing the material state of the two optimized GST nanobars, a binarized phase modulation can be achieved.
	Middle: Design of a supercell of the dynamic gradient MS with different superlattice periodicities (i.e., L$_{s}$) to achieve deflection angles of 0, -20$^{\circ}$, and -40$^{\circ}$, as well as the corresponding spatial phase profiles.
	Right: Numerically simulated scattered electric field intensity plots for different deflection angles corresponding to L$_{s}$ = $\infty$, 2400 nm, and 4800nm, respectively, under normal illumination at $\lambda$ = 1550 nm.
}
	\label{figS10}
\end{figure*}

In 2015, Chen \textit{et al.} demonstrated a hybrid metalens consisting of a 1D array of Au nanoslits filled with 100-nm-thick Ge$_{2}$Sb$_{2}$Te$_{5}$ (see Figure~\ref{figS9}A) \cite{chen2015engineering}. Theoretical results show that the F-P resonance mode supported inside each slit can be spectrally adjusted by changing the crystallization fraction of the GST. This way, the phase of the transmitted light at the operating wavelength of 1.55 $\mu$m can be controlled over a 0.56$\pi$ range. However, in experiments, only two states of amorphous and fully crystalline can be achieved that limits the number of feasible optical phase patterns to two. The crystallization process is carried out through scanning of a laser beam (532 nm, 4 mW, focused by a 100 $\times$ objective lens) along the slits at a speed of 0.2 $\mu$/s. As a proof-of-concept demonstration, two distinct phase profiles were encoded into two different samples by selectively crystallizing the GST slits. Figure~\ref{figS9}A illustrates the far-field patterns of on- and off-axis metalenses in comparison to an amorphous reference sample.

In Ref.~\cite{abdollahramezani2018reconfigurable}, the authors proposed a hybrid PCM-plasmonic MS to locally tailor the amplitude, phase, and polarization responses of the reflected light using a unique addressable MIM structure (see Figure~\ref{figS9}B). They leveraged the two fundamental modes of the structure, i.e., short-range surface plasmon polariton coupled to the intermediate 180-nm-thick Ge$_{2}$Sb$_{2}$Te$_{5}$ nanostripe, and the propagative surface plasmon polariton mode tunneling to the GST nanospacer (see Figure~\ref{figS9}B). By taking advantage of the multistate phase transition of the GST, they have shown a drastic modification in the mutual interaction of such enhanced modes leading to an inherently broadband response. Electro-thermal simulations results show considerable control on the key properties of the reflected light in the near-IR spectral range; broadband, high-efficiency phase shift up to 315$^{\circ}$, and reasonable amplitude modulation up to 60\%. A reconfigurable, high numerical aperture, and diffraction-limited varifocal metalens was implemented by addressing crystallization fraction of each meta-atom. The focal point can change from $\sim 5\lambda$ to $\sim 20\lambda$ while preserving the high-resolution feature comparable to the Airy disk (see Figure~\ref{figS9}B). 

\subsection{Hybrid dielectric/PCM metasurfaces for local phase control}

Employing high-speed optical pulses is an alternative approach for local conversion of PCM-based MSs as nanosecond and microsecond laser pulses were conventionally used in the optical data storage technology for robust reversible switching between the extreme state of PCMs. In 2015, Wang \textit{et al.} successfully demonstrated multilevel switching of Ge$_{2}$Sb$_{2}$Te$_{5}$ enabled by careful controlling of the energy and the number of stimulating optical pulses from a femtosecond laser \cite{wang2016optically}. They implemented this technology by developing a set-up consisting of a spatial light modulator for writing the optical pattern (with a diffraction-limited resolution of 0.59 $\mu$m using an objective lens with NA = 0.8) and an imaging system for reading the exposed zones to optical pulses (see Figure~\ref{figS10}A). The system used a pulse picker connected to a high-repetition-rate femtosecond laser (85 fs pulses with 85 MHz repetition rate at the wavelength of 730 nm) to control the intensity and duration of the femtosecond pulse trains. The experimental results show light-induced phase transition within an extremely small volume, down to 0.02 $\mu$m$^{3}$, of a 70-nm-thick GST layer sputtered on a glass substrate covered with a ZnS–SiO$_{2}$ film,  which is promising for realization of dynamic three-dimensional MSs. Several on-demand optical functionalities including bichromatic and multi-focus Fresnel zone plates, super-oscillatory lens, wavelength-multiplexed focusing lens, and hologram were imprinted to the same area without changing the structure of the optical system. Such a reprogrammable MS is enabled via writing, erasing, and rewriting of two-dimensional binary or greyscale patterns induced by sub-microscale phase transition of GST (see Figure~\ref{figS10}A).

In 2016, Chu \textit{et al.} proposed an all-dielectric MS consisting of an array of different-length Ge$_{2}$Sb$_{2}$Te$_{5}$ nanobars that can be selectively reconfigured to locally control the phase profile of the incident light (see Figure~\ref{figS10}b) \cite{chu2016active}. In this sense, dynamic gradient MSs can be realized by varying the period of constituent super cells comprising of meta-atoms with amorphous, partial, and/or full crystalline states. Simulation results indicate high performance specular and anomalous reflection angle controlling by switching the material state of 180-nm-thick GST nanobars. Each GST nanobar here works as a dipole antenna, whose resonance condition is imposed by the refractive index of the induced crystallinity, capable of abruptly modulating the phase of the reflected wave (see Figure~\ref{figS10}b).

Recently, a hybrid MS consisting of dielectric/PCM meta-atoms in which GST nanoposts are enclosed by Si nanorings was proposed to locally tailor the phase front of the transmitted light \cite{abdollahramezani2018dynamic}. Such a novel scheme provides an enhanced electromagnetic field inside the core via the strong interference of optically induced electric and magnetic dipoles. The resulting rather high-Q resonance mode together with the large accessible refractive index changes due to the state transition of GST offer a remarkable phase shift of 325$^{\circ}$ and large transmittance (more than 0.6) at the operational wavelength of 1340 nm. Several optical functionalities using such a unique MS including bifocal focusing and anomalous transmission have been demonstrated.

Forouzmand \textit{et al.} theoretically investigated a high-index MS consisting of a 1D array of Ge$_2$Sb$_2$Se$_4$Te$_1$ nanoribbons separated from the Au substrate with a thin film of low-index dielectric \cite{forouzmand2018dynamic}. By locally transforming the material state of each meta-atom using a focused optical beam, the strength and spectral position of ED and MD resonance modes governed the overall response of the meta-device can be modified at will. Relatively large phase agility ($\sim 270^{\circ}$) and high performance (more than 45\%) are granted through operation in the the so-called off-resonance regime (around 1.55 $\mu$m), where the inherent dissipative loss due to the pronounced light-matter interaction within the lossy PCM is suppressed. Dynamic control over the adaptive functionalities such as beam deflection can be achieved with such a platform.

\section{phase-change photonic Integrated circuits} 

The recent development of large-scale programmable Si photonic devices with small footprints, low power consumption, and high bandwidth (in contrast to the bandwidth-limited electronic circuits) is attributed to the massive progress in PICs \cite{sun2015single, graydon2016birth}. Considering the features listed above, PCM-based PICs have the potential to revolutionize the architecture of next-generation computers and data storage systems. In the following subsections, we will first review recent progress on the implementation of integrated phase-change switches and modulators. Then, a comprehensive overview on the usage of these key elements in realization of on-demand applications including all-photonic memories, arithmetic processors, non-von Neumann computing platforms, and brain-inspired synaptic neural networks will be presented.

\subsection{Integrated phase-change photonic switches and modulators} 

Switches and modulators play key roles in the PICs. Generally, these components are implemented using Mach-Zehnder interferometers (MZIs) (with wide operational bandwidth) \cite{zhuang2015programmable, liu2016fully, perez2017multipurpose} and microring resonators (with high modulation strength and narrow operational bandwidth but high sensitivity to the fabrication imperfections) \cite{reed2010silicon} through the plasma dispersion effect (with weak modulation of the refractive index) \cite{lu201616}, the thermo-optic effect (suffering from large footprint, low speed, and high power consumption) \cite{chen2016low}, or the electro-optic effect (with higher speed but suffering from less switching efficiency and the thermal drift). To overcome the aforementioned constraints for realization of high-performance, large-scale reconfigurable integrated photonic networks, an ideal switch needs to have three important features: low static and dynamic power consumption, high switching contrast, and ultrafast switching rate. In this regard, PCMs, due to their desirable characteristics such as non-volatility (which retains the material phase without keeping the external stimulus active), high refractive index contrast between the amorphous and crystalline states (resulting in high on-off ratio), ultrafast transition down to sub-nanoseconds \cite{loke2012breaking}, data retention for years, and high cyclability up to $10^{12}$ switching cycles \cite{burr2010phase} have been successfully demonstrated for realization of near-ideal integrated photonic switches \cite{tanaka2012ultra, rude2013optical, rios2015integrated, zheng2018gst, wu2018low, xu2019low}.

The crucial issue in most integrated phase-change photonic switches is the phase transition of the incorporated PCM between its different phases. This phase transition can be performed by heating the PCM element using external heaters (thermal effect), optical pulses (photothermal effect), and electrical pulses (electrothermal effect). Each of these phase transition mechanisms determines the switching performance of the device enabling specific applications. Therefore, in this subsection, with a focus on the mechanism used for the phase transition of the PCM, we present a comprehensive review on the development of integrated phase-change photonic switches and modulators.


\begin{figure*}
	\centering
	\includegraphics[trim={0cm 0cm 0cm 0cm},width=\textwidth, clip]{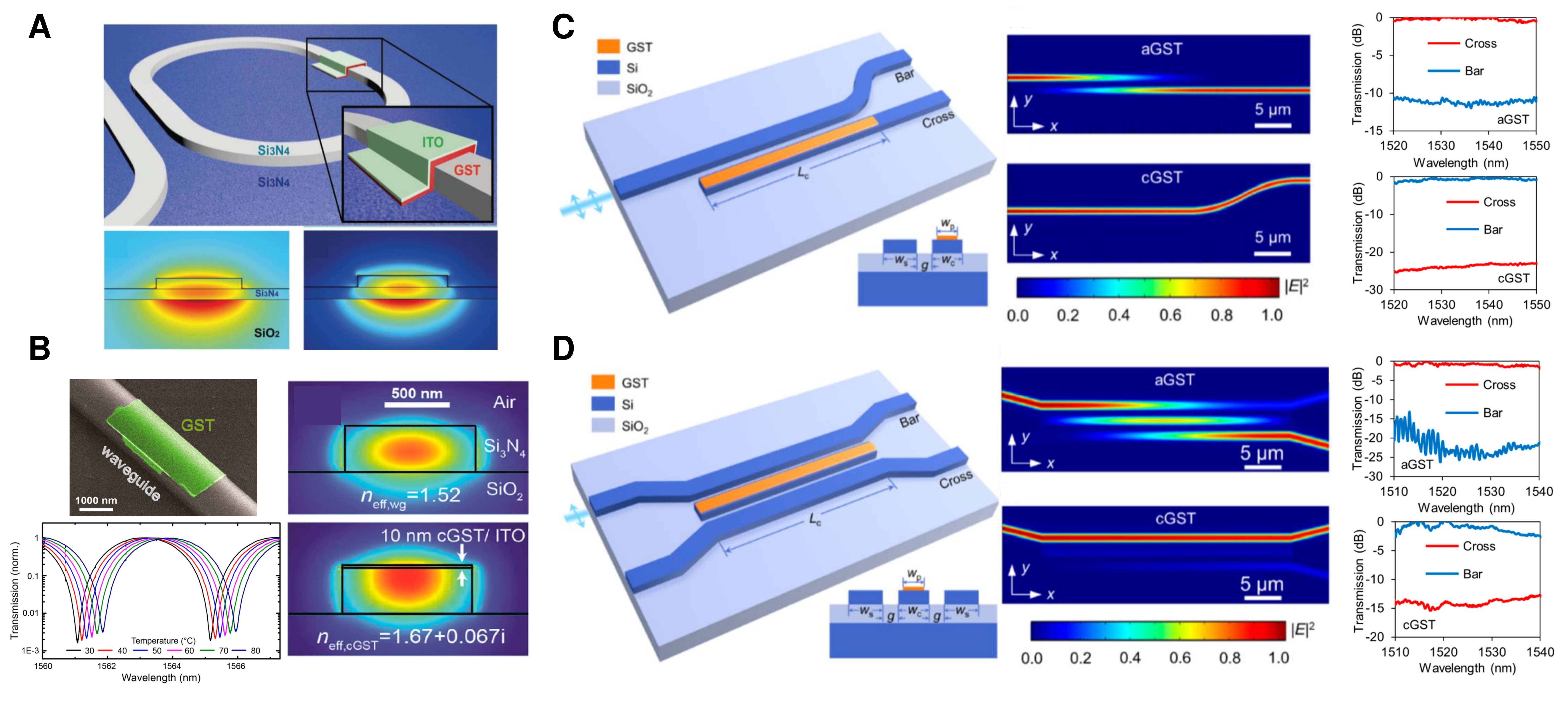}
	\caption{Integrated photonic phase-change switching based on thermo-optic effects.
	(A) Top: Schematic of the integrated photonic SiN-on-insulator platform for wideband switching operation. 
	Bottom: The simulated field profile of the TM mode showing the interaction of the propagating mode inside the waveguide with the GST thin film on top in the amorphous (left) and crystalline (right) states \cite{rios2014chip}.
	(B) Left top: The SEM image of a Si$_3$N$_4$ waveguide partially covered by a 10-nm-thick GST layer.
	Right: The waveguide mode without (upper panel) and with (lower panel) the c-GST layer.
	Left bottom: Optical transmission spectra of a MZI incorporated with a GST element which is initially in its amorphous state. Upon thermal annealing of the sample, a spectral shift and a decrease in the ER are observed \cite{stegmaier2016thermo}.
	(C, D) Operation principle of the integrated photonic 1$\times$2 (C) and 2$\times$2 (D) directional coupler switches. 
	Left: The schematic of the integrated photonic switches. 
	Middle: Simulated normalized optical field intensity of the switch for amorphous (top) and crystalline (bottom) state of the GST. 
	Right: The measured transmission spectra at the cross and bar ports with a-GST (top) and c-GST (bottom) \cite{xu2019low}.}
	\label{omid_fig1}
\end{figure*}


We start with integrated phase-change photonic switches based on the thermal annealing approach in which the incorporated PCM element is initially in its amorphous state, and switching occurs by slowly (using a hotplate) or quickly (using a rapid thermal processing (RTP) system) heating the sample so that the phase of the PCM is set to the crystalline state. This was experimentally demonstrated using an integrated photonic silicon nitride-on-insulator platform consisting of a nanophotonic Si$_3$N$_4$ bus waveguide which was coupled to a racetrack resonator partially covered by a Ge$_{2}$Sb$_{2}$Te$_{5}$ thin film (see Figure~\ref{omid_fig1}A) \cite{rios2014chip}. The input light that is set at the resonance wavelength of the racetrack resonator (on-resonance) can be coupled to or isolated from the racetrack resonator based on the phase state of GST. In fact, the dramatic difference in the complex refractive index of the GST upon its phase change from amorphous to the crystalline state (performed by heating the sample at 200 $^{\circ}$C for 3 min on a hotplate) strongly affects the attenuation inside the racetrack resonator, and in turn, changes the resonance condition by detuning the resonance wavelength of the racetrack resonator. 
In the low-loss amorphous case, the resonance condition of the racetrack resonator remains unchanged and the light propagates through the racetrack resonator as the a-GST film does not present (bottom left panel in Figure~\ref{omid_fig1}A). Therefore, the on-resonance input light traveling through the bus waveguide is critically coupled (where the round-trip loss rate is equal the coupling rate) to the racetrack resonator, and consequently, no power is transmitted to the output of the bus waveguide (transmission `0'). However, state transition to the highly absorptive crystalline phase increases the attenuation coefficient of the propagating mode inside the racetrack. This alters the resonance condition of the racetrack resonator, and thus, the off-resonance input light in the bus waveguide can no longer be coupled to the racetrack resonator and is completely transmitted into the output port of the bus waveguide (transmission `1').

It is noteworthy that the determining factor in the operation of photonic phase-change switches is the heat-induced refractive index change of the PCM \cite{rios2014chip,rios2015integrated,stegmaier2017nonvolatile}. A discussion on the effect of annealing processes on the switching dynamics to optimize the switching performance is presented in Ref.~\cite{stegmaier2016thermo}. The authors studied the thermo-optical effect of a thin film of Ge$_{2}$Sb$_{2}$Te$_{5}$ deposited on top of a Si$_3$N$_4$ waveguide as shown in Figure~\ref{omid_fig1}B. The waveguide is designed to ensure a single-mode operation when there is no GST on top (see the right top field profile). Even after the deposition of a thin layer of GST, the propagation condition does not tangibly change due to the low loss feature of the amorphous state. Upon transition of a-GST to its highly-absorptive crystalline phase, the effective complex refractive index changes which influences the evanescent coupling between the propagating light in the waveguide and the c-GST film and thus alters the waveguide mode (see the right bottom field profile). To study this thermo-optical effect, the authors used a transmission spectrum analysis of MZIs and a transient optical pump/probe measurement scheme for extracting the change in the imaginary and real parts of the refractive index of GST, respectively. The left bottom panel in Figure~\ref{omid_fig1}B shows the transmission spectrum of a MZI integrated with a GST element. A redshift in the spectrum and a decrease in the ER are observed upon heat-induced phase transition of GST.

While the resonator-based configuration presented in Figure~\ref{omid_fig1}A enables narrowband switching applications, other configurations have been investigated for broadband switching performance. In 2008, Ikuma \textit{et al.} proposed a 2$\times$2 integrated photonic switch using a short PCM waveguide placed between two Si waveguides in a directional coupler configuration in which the transmitted power through the bar and cross ports are controlled by the state of the Ge$_{2}$Sb$_{2}$Te$_{5}$ \cite{ikuma2008proposal}. Inspired by this work, some groups have demonstrated broadband directional coupler switches relying on the dramatic contrast in the extinction coefficient of the incorporated PCMs \cite{zhang2018broadband, xu2019low, olmo2019novel}. In 2019, Xu \textit{et al.} experimentally implemented phase-change Si photonic 1$\times$2 (see Figure~\ref{omid_fig1}C) and 2$\times$2 (see Figure~\ref{omid_fig1}D) one-way switches by using an asymmetric directional coupler configuration incorporated with Ge$_{2}$Sb$_{2}$Te$_{5}$ \cite{xu2019low}. High bandwidth (more than 30 nm with -10 dB cross talk) and relatively low loss ($\sim$1 dB) switching performance was demonstrated in the telecommunication band. The accurate phase transition of the PCM from its initial amorphous state to the crystalline state was performed by performing rapid thermal annealing of the device at 200 $^{\circ}$C for 10 min under the flow of nitrogen to prevent oxidation. The switching operation for the case of the directional coupler 1$\times$2 switch relies on the phase-match condition between the GST-on-Si hybrid waveguide (cross output) and the Si waveguide (bar output). For the low-loss a-GST, the effective refractive index of the TE mode in the hybrid waveguide is close to that of the Si waveguide (which is designed for the single-mode operation) resulting in an evanescent coupling of the input light from the Si waveguide (bar input port) to the hybrid waveguide (cross port) as shown in the middle top and right panels in Figure~\ref{omid_fig1}C. For the highly-absorptive c-GST, however, a strong phase-mismatch between the two waveguides happens due to the huge difference between their effective refractive indices. As a result, the input light in the Si waveguide is isolated from the hybrid waveguide and fully transferred to the bar output (see the middle bottom and right panels in Figure~\ref{omid_fig1}C). The operation principle for the directional coupler 2$\times$2 switch can be explained similarly where coupling between the supermodes traveling through the three-waveguide system (two Si waveguides as cross and bar waveguides with one GST-on-Si hybrid waveguide as shown in the left panel of Figure~\ref{omid_fig1}D) can be modified by controlling the GST phase. When GST is in its amorphous state, the symmetric modes of the Si waveguides can be efficiently coupled to the antisymmetric mode of the hybrid waveguide due to the phase-matching condition. As a result, the input light from the first Si waveguide (bar waveguide) is completely coupled to the GST-on-Si hybrid waveguide, and then is fully coupled to the second Si waveguide (cross waveguide), and finally, fully delivered to the cross port (see the top middle and the right panels of Figure~\ref{omid_fig1}D). On the other hand, in the crystalline phase, the phase-matching condition is not met anymore which leads to the isolation of the input Si waveguide from the GST-on-Si hybrid waveguide, and thus transfers the light into the bar output (as shown in the bottom middle and the right panels of Figure~\ref{omid_fig1}D).

\begin{figure*}
	\centering
	\includegraphics[trim={0cm 0cm 0cm 0cm},width=\textwidth, clip]{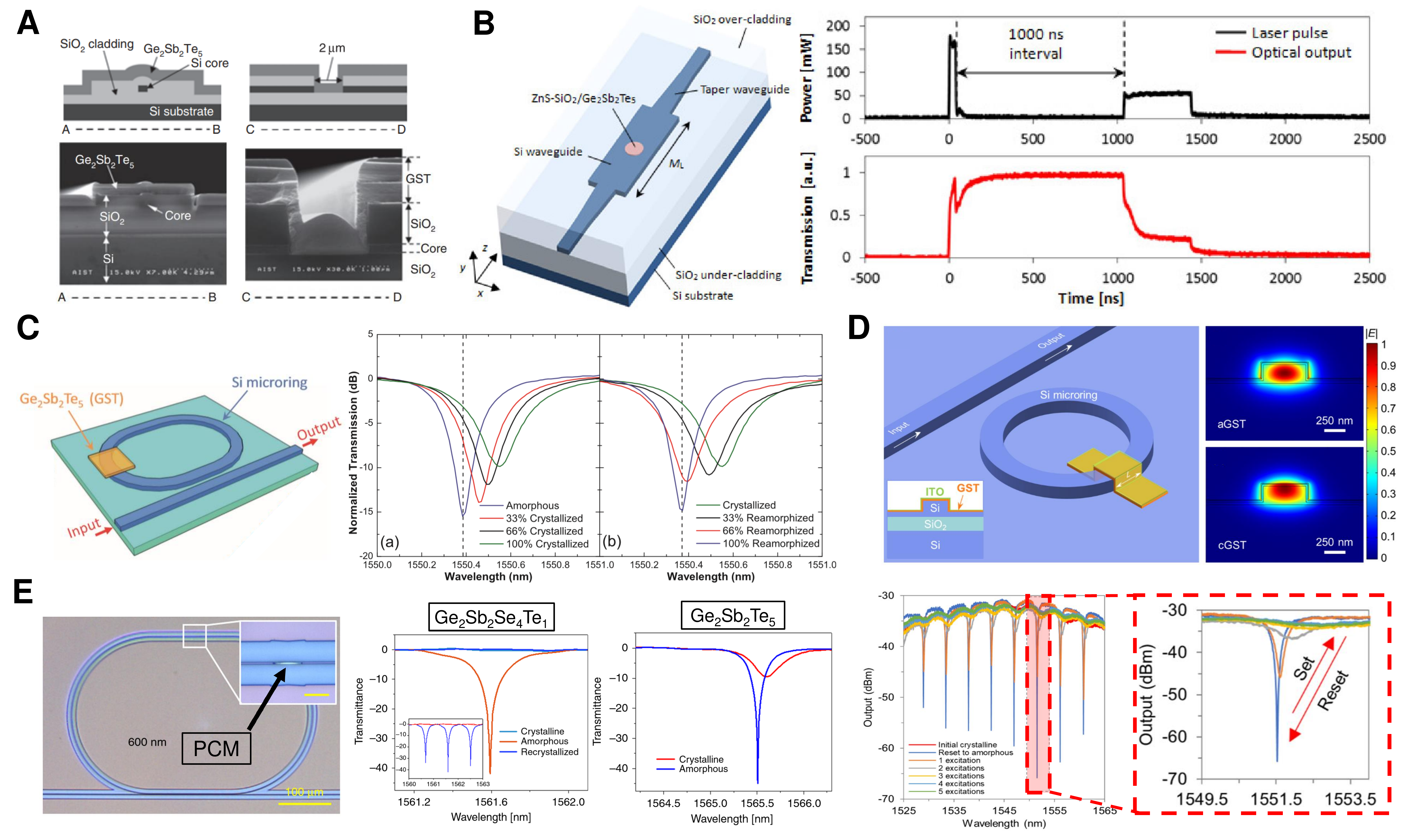}
	\caption{Photonic integrated phase-change switches and modulators triggered with free-space illuminations.
	(A) The schematic (top) and SEM images (bottom) of the pioneering work on one-way (left) \cite{ikuma2010small} and reversible (right) \cite{ikuma2010reversible} optical gate switching.
	(B) Left: Schematic of the optical gate switch consisting of a $1\times1$ MMI waveguide with a circular-shaped GST cell on top exited by external out-of-plane laser pulses. Right: The waveforms of laser pulses (upper diagram) and optical transmission at output (lower diagram) with a sequence of off/on/off operations \cite{tanaka2012ultra}. 
	(C) Left: A reversible optical switch consisting of a Si racetrack resonator (partially deposited by a thin film of Ge$_{2}$Sb$_{2}$Te$_{5}$) coupled to a Si waveguide. 
	Middle and right: The evolution of the transmission spectrum during crystallization (middle) and re-amorphization (right) showing the reversible switching operation \cite{rude2013optical}. 
	(D) Left: Schematic of the GST-on-Si photonic modulator (inset: cross-section of the hybrid waveguide). 
	Right: The mode profiles of the hybrid waveguide at 1550 nm for amorphous (top) and crystalline (bottom) states. 
	Bottom: Transmitted output spectra of the modulator employing a 2 $\mu$m width GST element excited with a train of pulses with different pulse numbers \cite{zheng2018gst}.
	(E) Left: Optical image of the resonant switch in which a low-loss GSST nanostrip deposited on top of a SiN ring resonator as shown in the inset. Middle and Right: The transmission spectra of the switch showing complete on/off modulation of the resonant peaks when integrated with Ge$_{2}$Sb$_{2}$Se$_{4}$Te$_{1}$ (middle) outperforming the traditional resonant switch with the similar configuration but integrated with Ge$_{2}$Sb$_{2}$Te$_{5}$ (right) \cite{zhang2019broadband}.}
	\label{omid_fig2}
\end{figure*}

In a distinct work, a non-conventional chalcogenide PCM, namely GSST, with lower extinction coefficient than that of traditional GST alloys, was employed to demonstrate a similar functionality using a directional coupler configuration \cite{zhang2018broadband}. The experimentally measured results show broadband integrated photonic 1$\times$2 and 2$\times$2 switches with low insertion loss (0.01 to 0.4 dB in the C-band) and cross talk (i.e., the contrast ratio between the on/off- state at the output ports) over 15 dB ($>$ 25 dB at 1550 nm).

Thermal treating of the PCM using a hotplate or RTP system is a slow process that makes such devices impractical for ultrafast switching applications. More importantly, it limits the device functionality to the one-way switching (from initial amorphous to crystalline) since the re-amorphization mechanism is a fast melt-quenching process, which cannot be performed using such simple bulky thermal heaters. Therefore, to realize the desired ultrafast and reversible (two-way) switching application, other means including optical and electrical pulses have been considered. We continue the review by discussing the recent developments using short optical and electrical pulses to reversibly change the state of PCMs, and then point out some recent researches investigated mixed-electro-optic effects for switching and modulation applications.


The optical excitation of the integrated PCM element can be performed using an out-plane focused laser beam or an in-plane guided laser light. While the former enables ultrafast switching performance, the latter adds many interesting fully-integrated on-chip applications such as on-chip all-photonic memories and mathematical operators. In the following, we first review those works used the free-space optical excitation approach, and then present some recent works utilizing an in-plane stimulus.

Ikuma \textit{et al.} demonstrated a pioneering work on one-way optical gate switching by incorporating a thin layer of Ge$_{2}$Sb$_{2}$Te$_{5}$ with a Si core waveguide (see the left panel in Figure~\ref{omid_fig2}A) working in the 1525-1600 nm wavelength window with MD of 12.5 dB \cite{ikuma2010small}. They used 660 nm free-space laser pulses with peak power of 89 mW and FWHM of 500 ns to switch the GST from its amorphous (on) state to crystalline (off) state. Later, they successfully demonstrated a reversible optical gate switch consisting of a shallow-etched Si core waveguide formed on a silicon-on-insulator (SOI) substrate and covered by a thin layer of Ge$_{2}$Sb$_{2}$Te$_{5}$ (see the right panel in Figure~\ref{omid_fig2}A), working in the same operational wavelength range \cite{ikuma2010reversible}. They used an out-plane illumination with a peak power of 19 mW/94 mW (453 ns/8.4 ns FWHM) for crystallization/re-amorphization processes.

Since in Refs.~\cite{ikuma2010small} and \cite{ikuma2010reversible} the size of GST elements are larger than the laser spot size, a semi-crystalline zone appears around the beam spot that increases the insertion loss. To overcome this drawback, the authors leveraged a multimode interference (MMI) Si waveguide integrated with a Ge$_{2}$Sb$_{2}$Te$_{5}$ nanodisk (with a diameter of 1 $\mu$m) covered by a ZnS-SiO$_2$ capping layer (see Figure~\ref{omid_fig2}B) \cite{tanaka2012ultra}. The operation principle of such a reversible optical gate switch with a wide operating wavelength range of 100 nm around 1575 nm is shown in the right panel of Figure~\ref{omid_fig2}B. The GST is initially in its off-state where c-GST with high extinction coefficient significantly absorbs the propagating light resulting in a near-zero power transmission. On the other hand, the low-loss a-GST allows propagation of light in the waveguide with a high optical transmittance ($\sim 1$). The amorphization (switching from off- to on-state) is performed by irradiating the GST element using a laser pulse with a width of 40 ns and a peak power of 160 mW. To set back the state of the switch to its initial off-state (i.e., the re-crystallization process), a laser pulse with a width of 400 ns and a peak power of 50 mW was used. 

Instead of relying on the difference in the imaginary part of the refractive indices of the two material phases, Rude \textit{et al.} benefited from the remarkable change in the real part of the refractive index of Ge$_{2}$Sb$_{2}$Te$_{5}$ ($\Delta n\approx2.5$ while $\Delta k\approx1$ at $\lambda$ = 1.55 $\mu$m as shown in Figure~\ref{figH2}). They experimentally implemented a reversible optical switch comprising of a Si racetrack resonator partially covered by a thin layer of GSt and coupled to a bus waveguide as shown in Figure~\ref{omid_fig2}C \cite{rude2013optical}. An out-plane focused illumination from a $975$ nm diode laser with laser power of 12 mW (45 mW), pulse duration of 300 ns (20 ns), FWHM of 80 ns (8 ns) for crystallization (re-amorphization) of a 1 $\mu$m diameter GST element was used. This way,  the authors could reversibly modulate the optical path of the microring resonator and consequently, tune the resonance frequency of the racetrack resonator. Experimental measurements show switching operation at telecommunication wavelengths (around $\lambda=1.55$ $\mu$m) with MD of 12 dB (see Figure~\ref{omid_fig2}C). 

Recently, Zheng \textit{et al.} demonstrated broadband, fast, and reversible optical amplitude and phase modulation in a hybrid integrated photonic platform (see Figure~\ref{omid_fig2}D) \cite{zheng2018gst}. The simulated mode profiles of the hybrid waveguide at 1550 nm show a dramatic modification of the light mode attributed to the phase transition of the evanescently coupled Ge$_{2}$Sb$_{2}$Te$_{5}$ nanostrip. The GST is initially in the highly-absorptive crystalline state resulting in a high optical transmission at the output (see the lower panel in Figure~\ref{omid_fig2}D). The state of GST can be controlled by short (440 ps) optical pulses in a two-way process. While a single reset pulse with effective energy (with respect to the effective area of the GST strip under illumination of a $120~\mu$m diameter laser spot) of $\sim$620 pJ ($\sim$9 aJ/nm$^3$) was used for amorphization, a train of same duration pulses with $\sim$200 pJ energy ($\sim$3 aJ/nm$^3$) could re-crystalline GST. By varying the number of pulses, reversible (from amorphous to initial crystalline state) and quasi-continuous (from amorphous to several intermediate states) switching operation with a high ER up to 33 dB was shown (see the lower panel in Figure~\ref{omid_fig2}D). However, due to the high absorptive behavior of c-GST at $\lambda =1550$ nm, this device rendered low quality factor resonances \cite{rios2014chip, zheng2018gst, cao2019fundamentals}.

So far, the majority of works have benefited from the dramatic refractive index and extinction coefficient contrast (i.e., $\Delta n$ and $\Delta k$) due to the unique non-volatile phase-transition in PCMs. However, this limits the switching performance of PCM-based nanophotonic devices due to the low figure of merit (FOM), defined as $\Delta n/\Delta k$, of PCMs. To outperform previous integrated photonic switches, more recently, Zhang \textit{et al.} used Ge$_{2}$Sb$_{2}$Se$_{4}$Te$_{1}$ featuring low optical losses in a wide wavelength range of 1-18.5 $\mu$m with a large refractive index contrast of $\Delta n\sim2$ resulting a large FOM. By using a hybrid SiN ring resonator (see the left panel in Figure~\ref{omid_fig2}E) reversible switching operation with a low insertion loss of $<$0.5 dB and record contrast ratio of 42 dB was experimentally demonstrated. They compared their results with a similar photonics platform employing Ge$_{2}$Sb$_{2}$Te$_{5}$ as demonstrated by the middle and right diagrams in Figure~\ref{omid_fig2}E \cite{zhang2019broadband}. To show the repeatability of the switching operation, the GSST film was reversibly converted from amorphous to crystalline with 30 ns width (and 1$~\mu$s periodicity) pulses and crystalline to amorphous with a single 100-ns-width pulse.

\begin{figure*}
	\centering
	\includegraphics[trim={0cm 0cm 0cm 0cm},width=\textwidth, clip]{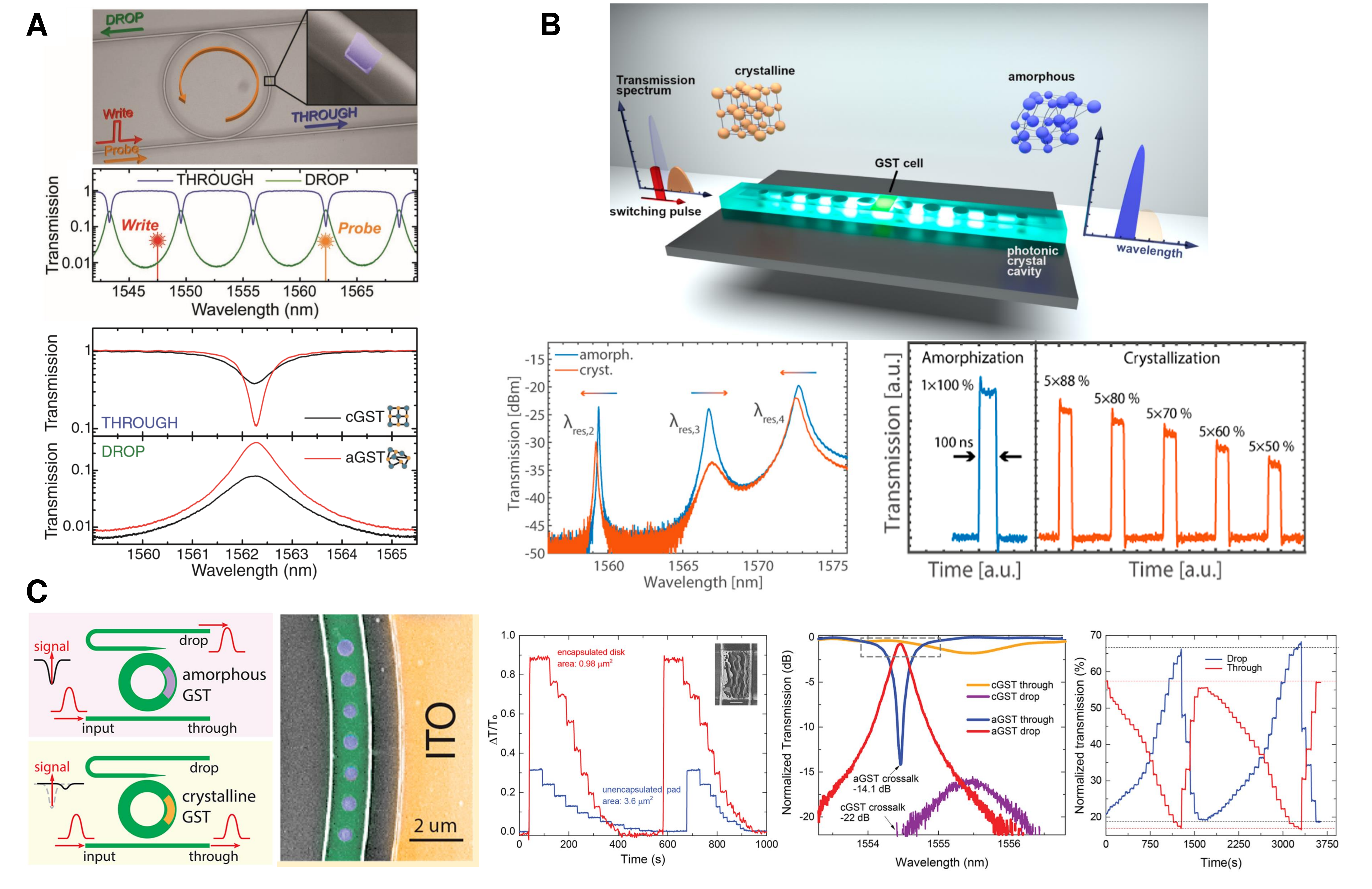}
	\caption{Fully integrated phase-change switches and modulators enabled by on-chip optical pulse triggers. 
	(A) Top: Operation principle of an all-photonic integrated phase-change switch consisting of a GST element (inset) on top of a ring resonator which is evanescently coupled to two parallel waveguides. For the on-resonance condition (when the laser wavelength is equal to the resonance wavelength of the ring resonator), the power is divided between THROUGH and DROP ports, while for the off-resonance condition (the laser wavelength is detuned from the resonance wavelength of the ring resonator), all power is delivered to the THROUGH port. Therefore, to control the state of GST optically, the ``write'' signal wavelength should be adjusted to that of the ring resonator (on-resonance condition). 
	Bottom: Transmission spectra of the ring resonator in the THROUGH (upper panel) and DROP (lower panel) port showing that the on-resonance light is delivered to either the THROUGH port in the crystalline state (black curve) or DROP port in the amorphous state (red trace) \cite{stegmaier2017nonvolatile}. 
	(B) Top: Illustration of a reversible all-photonic switch consisting of a photonic crystal cavity waveguide integrated with a GST cell evanescently coupled to the center of the cavity.
	Bottom left: The change in both transmission and spectral position of the resonance modes of the cavities due to phase transition of GST.
	Bottom right: The switching scheme used for amorphization (using a single 100 ns optical pulse with 100\% peak poser) and re-crystallization (using five groups of pulses with decreasing energy, in which each group contains five pulses with identical peak power) \cite{von2018reconfigurable}.
	(C) Left: Operation principle and the false-colored SEM image of the 1$\times$2 optical switch working based on optically induced phase transition of the seven subwavelength GST nanodisks patterned on a SiN microring resonator. 
	Right (left pannel): The stepwise change in the optical transmission of a simple SiN waveguide covered by 5 GST nanodisks (the structure is not shown here) as the phase of the GST nanodisks changes using optical pulses. Multiple transmission levels are achieved by step-wise re-crystallization of the encapsulated GST nanodisks area (red) and the unencapsulated unpatterned GST patch (blue) (inset: SEM image of the GST patch deformed after one optical pulse). 
	Right (middle pannel): The transmission spectra at the through and drop ports of the 1$\times$2 optical switch near a resonance of the microring resonator when the phase of GST changes from amorphous (high (low) transmission at drop (through) port) to crystalline (redshifted high (low) transmission at through (drop) port). 
	Right (right diagram): Using a mixed electro-optic switching approach in which optical pulses are used for amorphization (set), while electrical pulses are applied to the ITO heater for stepwise re-crystallization (reset). Up to 20 intermediate transmission levels are achieved enabling multilevel switching \cite{wu2018low}. }
	\label{omid_fig3}
\end{figure*}


Hitherto, integrated photonic switches have employed an external bulky source to enable phase transformation of the functional material, which inhibits the realization of a fully integrated on-chip platform. In 2017, Stegmaier \textit{et al.} demonstrated a pioneering fully integrated, reversible all-optical 1$\times$2 switch \cite{stegmaier2017nonvolatile}. The proposed device is shown in Figure~\ref{omid_fig3}A in which a sub-micrometer Ge$_{2}$Sb$_{2}$Te$_{5}$ element is placed on top of a ring resonator which is evanescently coupled to two parallel waveguides. The widths of the waveguides are appropriately designed to ensure the single-mode (TE-like) operation. If the input probe light wavelength is the same as the resonance wavelength of the ring resonator (on-resonance condition), the propagating light is partially coupled to the ring resonator and divided between DROP and THROUGH ports (see the orange marker in Figure~\ref{omid_fig3}A). On the other hand, the probe light with off-resonance wavelength is not coupled to the ring resonator and is fully directed to the THROUGH port. Therefore, by setting the probe signal on-resonance (at 1562.3 nm) and switching the phase of the GST element between low-loss amorphous state and highly-absorptive crystalline state, the ratio of the power coupled to the DROP and TROUGH ports can be controlled (see the lower panel in Figure~\ref{omid_fig3}A). For the initial case, the evanescent interaction between the c-GST element and propagating light in the ring resonator increases significantly. Accordingly, the loss within the ring resonator becomes more than the coupling loss to the waveguide leading to a weakly coupling regime between the ring resonator and the waveguide. Therefore, in the crystalline state, the on-resonant light is directed to the THROUGH port (black curve in the lower panel of Figure~\ref{omid_fig3}A). On the other hand, the on-resonance light (when GST is in the amorphous state) is completely coupled to the ring resonator and then fully transmitted to the DROP port (red traces in the lower panel in Figure~\ref{omid_fig3}A). This stems from the fact that the ring resonator is designed to be critically coupled to the waveguide (i.e., when the loss in the ring resonator is equal to the coupling loss) when the GST is in its low-loss amorphous phase. The probe signal used in this work consists of 500 ps optical pulses with 2.5 pJ energy (which are wide enough in width and low enough in energy to prevent any possible phase transition). The amorphization of the GST was performed by ``write'' signals consisting two 1 ps optical pulses (with 25 ns interval) of 95 pJ energy, while the re-crystallization was achieved using six groups of pulses with decreasing energy (each group contains one hundred 1 ps pulses with identical peak power) as described in Ref.~\cite{stegmaier2017nonvolatile}.

With a simpler concept, later, Zhang \textit{et al.} used a Si asymmetric MZI (AMZI) integrated with a micrometer-scale Ge$_{2}$Sb$_{2}$Te$_{5}$ element on top of the ring resonator to boost the switching ER to more than 20 dB \cite{zhang2018all}. By controlling the number of pulses, the intermediate states of GST that enable tunable switching operation were demonstrated.

To improve the transmission contrast and power consumption of such integrated phase-change photonic switches, recently, Keitz \textit{et al.} proposed to integrate a PCM element with a photonic crystal waveguide instead of a conventional bare photonic waveguide \cite{von2018reconfigurable}. Figure~\ref{omid_fig3}B shows the proposed all-photonic reversible switch in which a sub-micrometer Ge$_{2}$Sb$_{2}$Te$_{5}$ cell is placed on top of a photonic crystal waveguide and evanescently coupled to the center of a high quality factor photonic crystal cavity. The authors properly designed the photonic crystal cavities to guarantee high quality factor resonance modes (i.e., with minimum scattering losses) to enhance the interaction between the guided modes in the photonic crystal waveguide with the GST cell. As a result, when the GST cell is in its highly absorptive crystalline state, it more effectively absorbs the guided light. This enables a higher switching contrast with lower power consumption ($\sim$14\%) in comparison with the conventional integrated phase change photonic switches using a bare waveguide. Moreover, relying on the high refractive index contrast granted by GST, the authors showed the shift of resonance wavelengths that enables a tunable wavelength filter (see the bottom left panel in Figure~\ref{omid_fig3}B). While a single high-power pulse with 100 ns width can perform amorphization, a train of five consecutive groups of 100-ns-width pulses (each group contains five pulses with identical peak power that monotonically decreases in following groups) can re-crystallize the GST element (see the bottom right panel in Figure~\ref{omid_fig3}B). 

The high-power optical pulses used in the previous works can cause deformation in the PCM cells which significantly increases the optical loss. To address this issue, recently, Wu \textit{et al.} presented an innovative configuration by replacing the common rectangular GST cell with an array of subwavelength GST nanodisks (the seven blue circles in the false-colored SEM image shown in the left panel in Figure~\ref{omid_fig3}C) on top of a SiN microring resonator of an add-drop filter \cite{wu2018low}. These Ge$_{2}$Sb$_{2}$Te$_{5}$ nanodisks (with diameters of 500 nm) are encapsulated with a 50-nm-thick Al$_2$O$_3$ layer conformally grown by atomic layer deposition (ALD). The advantages of this configuration are twofold: (i) the pump light can heat the small GST nanodisks more uniformly than the relatively large micro-scale GST cell resulting in more efficient phase transformation, and (ii) the conformally deposited Al$_2$O$_3$ layer can minimize the surface energy when high-power optical pulses are used for phase transition of GST. Therefore, patterning and encapsulating processes can prevent the GST film from deformation. The authors also employed an array of five 10-nm-thick GST nanodisks placed on top of a simple Si$_3$N$_4$ waveguide to demonstrate reversible switching with less than 1 dB insertion loss and more than 20 dB contrast ratio in the telecommunication window. When the GST nanodisks are in their crystalline (amorphous) state, they significantly (negligibly) absorb the propagating probe light in the waveguide resulting in a low (high) optical transmission at the output. A single 50-ns-width optical pulse (220 pJ energy) was used for the amorphization of the five initially crystalline nanodisks. Also, trains of laser pulses for the stepwise re-crystallization process were employed in which each intermediate state was achieved by a train of 50 gradually decreased energy laser pulses with 50 ns duration each. This enables multilevel switching as shown in the left diagram of Figure~\ref{omid_fig3}C. For a systematic comparison, a traditional device in which a rectangular GST cell that fully covers the waveguide width was fabricated. The SEM observation (inset in the left diagram of Figure~\ref{omid_fig3}C) reveals that the GST film is deformed after the first high-power amorphization pulse which significantly reduces the transmission contrast (more than 50\%). Figure~\ref{omid_fig3}C shows the operation principle of a $1\times2$ switch designed near the critical coupling regime in which light is mostly delivered to the drop port when GST is in its low-loss amorphous state. The crystallization process largely redshifts the resonance of the microring resonator leading to high transmission in the through port (see the middle diagram in Figure~\ref{omid_fig3}C). In this work, the phase of the GST nanodisks can be stepwise controlled both optically and electrically. In the all-optical switching scheme, the amorphization is achieved by using a single ``set'' pulse with a width of 50 ns and total energy of 250 pJ (see the middle diagram in Figure~\ref{omid_fig3}C). On the other hand, in a stepwise re-crystallization process, the intermediate steps are realized using a train of 50 ``reset'' pulses (each with 50 ns duration) gradually decaying energy (see the middle diagram in Figure~\ref{omid_fig3}C). Due to the low absorption nature of a-GST, the re-crystallization process relies on high-energy pulses. To address this issue, a mixed electro-optic switching approach was proposed in which optical pulses with a width of 50 ns and total energy of 500 pJ are used for amorphization, while electrical pulses of 5 s duration and 1 V amplitude are applied to the underlying ITO heater (to provide a prolonged heating) for stepwise re-crystallization. This way, up to 20 intermediate transmission states enabling multilevel switching operation are achieved (see the right diagram in Figure~\ref{omid_fig3}C).

\begin{figure*}
	\centering
	\includegraphics[trim={0cm 0cm 0cm 0cm},width=\textwidth, clip]{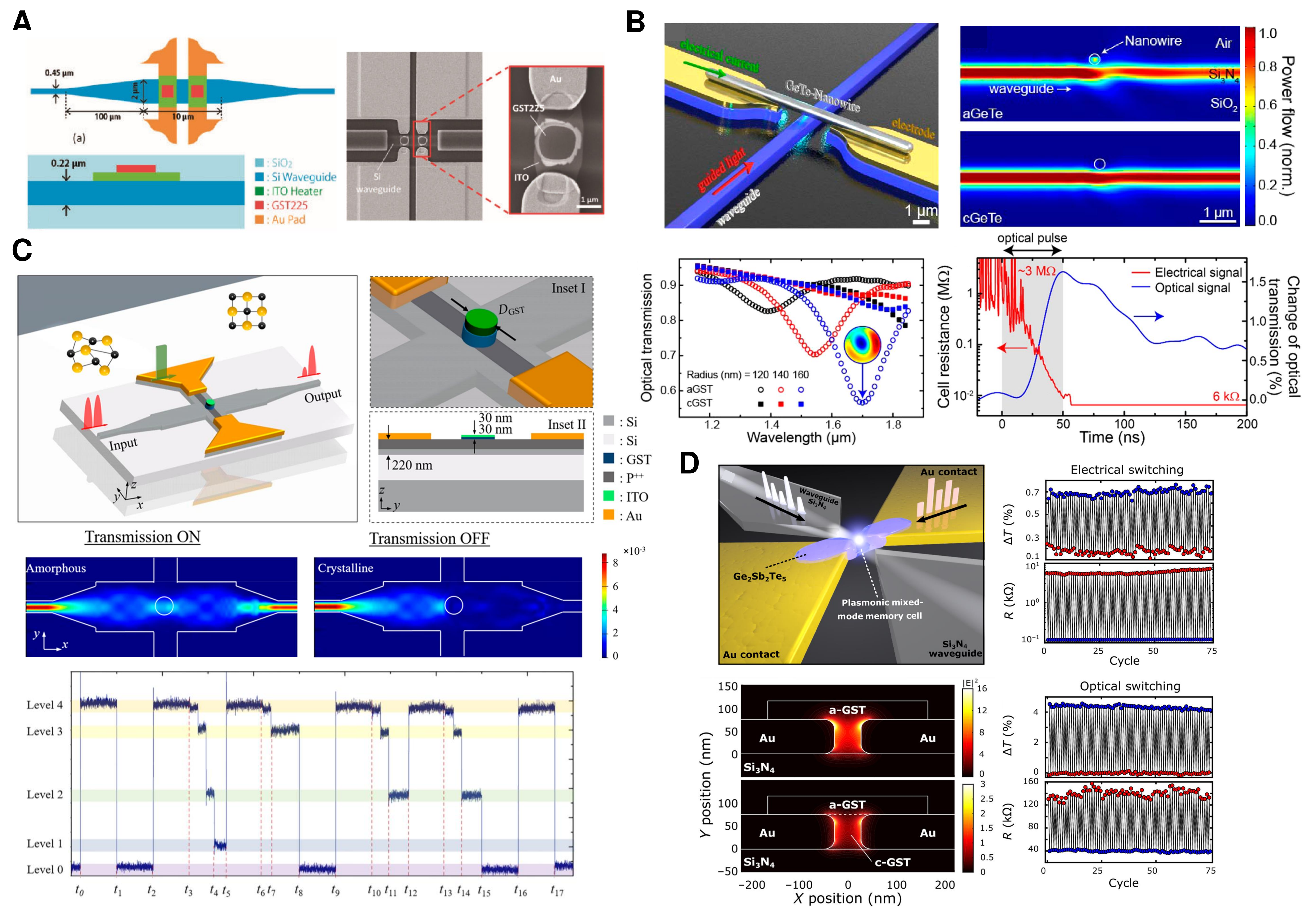}
	\caption{Electro-optic and mixed-mode elecro-optical integrated photonic phase-change switching. 
	(A) The schematic (left) and SEM image (right) of the a current-driven phase-change optical switch in which a Gs$_2$Sb$_2$Te$_5$ thin film is placed on top of an ITO microheater incorporated with a Si waveguide \cite{kato2017current}.
	(B) Operation principle of the mixed-mode measurement. 
	Top left: The structure of the on-chip mixed-mode switch in which the GeTe nanowire (with diameter of 300 nm) is evanescently coupled to a photonic Si$_3$N$_4$ waveguide and at the same time is electrically contacted to two Au electrodes enabling optical and electrical measurements, respectively.
	Top right: Power distributions in Si$_3$N$_4$ waveguide and GeTe nanowire. For low amorphous GeTe (upper panel), the Mie scattering in the GeTe nanowire enhances the evanescent interaction and thus, the propagating light is highly absorbed resulting in a low optical transmission. Upon transition to the highly-absorptive crystalline state, the propagating light is no longer absorbed by the GeTe nanowire leading to an increase in the optical transmission.
	Bottom left: Transmission spectra of the mixed-mode electro-optical switch in both amorphous and crystalline states.
	Bottom right: Evolution of the GeTe nanowire resistance and corresponding optical transmission during crystallization process \cite{lu2016mixed}.
	(C) Top: The structure and operation principle of an optical memristive switch consisting of a GST nanodisk covered by a thin layer of ITO both on top of a MMI structure.
	Middle: Simulated field distributions of the propagating light in the MMI waveguide for ON-state (a-GST) and OFF-state (c-GST) switching, respectively.
	Bottom: Multilevel switching operation by partially crystallizing the GST nanodisk \cite{zhang2019miniature}.
	(D) Top left: The structure and operation principle of a mixed-mode plasmonic/phase-change integrated photonic device which consists of two tapered Si$_3$N$_4$ waveguides (for delivering the light) and two Au electrodes.
	Bottom-left: The pronounced field enhancement inside the plasmonic gap for low-loss a-GST (upper panel) while an attenuation in the field enhancement is observed for highly-absorptive c-GST (lower panel).
	Right: Electrical (upper panel) and optical (lower panel) programming of the device which can be readout with both electrical means (the change in the device's resistance between two Au electrodes) and optical means (the change in the optical transmission) \cite{farmakidis2019plasmonic}.}
	\label{omid_fig4}
\end{figure*}

So far, realization of ultrafast, reversible phase-change photonic switches/modulators using on-chip and free-space laser pumps has been successfully demonstrated. However, these approaches cannot be practical for large-scale integration. For the former case, the alignment and focusing of the out-plane optical beam on the PCM inclusion is a slow process and constrained by diffraction-limitation of light. For the latter case, the routing process of the pump light and switching of a large-area PCM are cumbersome \cite{feldmann2017calculating}. To address these limitations, recently, electro-optic \cite{kato2017current,zhang2019nonvolatile,zhang2019miniature,zheng2019nonvolatile} and mixed-mode electrical/optical \cite{lu2016mixed, farmakidis2019plasmonic} approaches have been employed to enable reversible phase transformation of large PCM inclusions. 


In 2017, Kato \textit{et al.} experimentally demonstrated a current-driven reversible optical switch by integrating a $2.25\times3$ $\mu$m$^2$ ITO microheater with a 30-nm-thick Ge$_{2}$Sb$_{2}$Te$_{5}$ film on top of a MMI Si waveguide (see Figure~\ref{omid_fig4}A) \cite{kato2017current}.
The thickness of ITO was optimized to be 30 nm to reduce the switching energy while keeping the ER above 30 dB. By amorphization of both GST patches using a 100 ns current pulse of 20 mA, the propagating light experiences low loss resulting in high optical transmission (on-state). On the other hand, by re-crystallization of both GST patches using a 100 ms pulse of 12 mA, the probe light is highly absorbed resulting in a low optical transmission (off-state). An intermediate state was also obtained by amorphization of one GST patch, while maintaining the other one in the crystalline state. An average ER of 1.2 dB was experimentally demonstrated over the wavelength range of 1525-1625 nm.

In another work, Lu \textit{et al.} experimentally studied a mixed-mode electro-optical switching operation by embedding a GeTe nanowire into a nanophotonic circuit \cite{lu2016mixed} (see Figure~\ref{omid_fig4}B). The GeTe nanowire is evanescently coupled to a Si$_3$N$_4$ photonic waveguide and electrically contacted to two Au electrodes. While the former guides the optical signals sent by an off-chip fiber-coupled pump-probe setup for triggering/probing the state of GeTe nanowire, the latter enables electrical probing of GeTe phase through a RF setup. The reversible phase transition between amorphous and crystalline states of GeTe was carried out using nanosecond optical pump pulses. The state of the GeTe nanowire can simultaneously be measured through monitoring the intensity of the transmitted probe light and the slope of the I-V cure. As shown in the top right panel of Figure~\ref{omid_fig4}B, in the amorphous state, the resonant Mie scattering observed in the GeTe nanowire significantly enhances the light-matter interaction between the light propagating through the waveguide and evanescently-coupled GeTe nanowire (on-resonant). As a result, the transmitted power at the resonance wavelength is significantly reduced as depicted by transmission dips in the bottom left panel of Figure~\ref{omid_fig4}B. On the other hand, due to the dramatic refractive index change upon the state transition, the light is not coupled anymore to the crystalline GeTe nanowire (which is off-resonant) resulting in a high optical transmission. The bottom right panel in Figure~\ref{omid_fig4}B shows a decrease in the measured resistance (from $\sim$3 M$\Omega$ to 6 k$\Omega$) and an increase in the optical transmission upon the full cross-sectional amorphization of the initially crystalline GeTe nanowire (with diameter of 300 nm) by using optical pump pulses with 50 ns duration and $\sim$10.7 nJ energy. To reset the switch, a 50 ns optical pump pulse with 6.2 nJ energy was used to re-crystallize the GeTe nanowire.

More recently, Zhang \textit{et al.} experimentally demonstrated an optical memristive switch consisting of a MMI Si waveguide accommodating a Ge$_{2}$Sb$_{2}$Te$_{5}$ nanodisk on top (see Figure~\ref{omid_fig4}C), which can be set/reset employing electrical pulses \cite{zhang2019nonvolatile,zhang2019miniature}. The top panels in Figure~\ref{omid_fig4}C show the structure of the switch in which a Si strip orthogonally crosses the center of the MMI, and is heavily P$^{++}$-doped at the center, where GST presents. The 30-nm-thick GST nanodisk with a diameter of 1 $\mu$m is covered by a 30-nm-thick layer of ITO film to prevent oxidation. By applying an electrical pulse, the doping stripe behaves as a resistive heater and transforms the state of GST. According to the self-imaging principle, the propagating light is focused into the center of the MMI, and so significantly interacts with the GST nanodisk leading to a high-performance switching. The optical power distributions in the MMI in the middle panels of Figure~\ref{omid_fig4}C show that the initially low-loss a-GST does not absorb the light resulting in a high optical transmission (on-state) while the highly-absorptive c-GST significantly absorbs the light leading to a low optical transmission (off-state). The authors showed that multilevel optical transmission levels in the off-state can be achieved through a stepwise crystallization process by either varying the pulse duration from 40 to 200 ns and fixing the amplitude to 3.5 V, or by varying the pulse amplitude from 3 to 6 V while fixing the pulse duration at 100 ns. This enables multistep switching performance with five distinct intermediate steps as shown in the bottom panel of Figure~\ref{omid_fig4}C. On the other hand, multiple optical transmission levels in the on-state can be achieved through a stepwise amoy7rphization process by using high power short electrical pulses with fixed duration of 20 ns and varying amplitude from 11 to 11.8 V, or a fixed amplitude but varying duration from 10 to 20 ns.

Although the architecture presented in Ref.~\cite{lu2016mixed} enables mixed-mode readout, it is based on using either optical or electrical set/reset pulses, and not both, to realize switching. In Ref.~\cite{farmakidis2019plasmonic}, the authors demonstrated not only both optical (transmission) and electrical (resistance) readout, but set/reset operations in a single device suitable for hybrid optoelectronic computing platforms. In this work, Farmakidis \textit{et al.} leveraged a small-footprint platform of an electrically connected nanoplasmonic structure intersecting with an integrated photonic waveguide. This not only benefits from the strong light-matter interaction and low-loss switching in a subwavelength regime but also enables both electrical and optical reversible switching and readout operations in a hybrid integrated phase-change binary/multilevel memory. The top left panel in Figure~\ref{omid_fig4}D shows the schematic of the proposed device in which a Si$_3$N$_4$ rib waveguide is coupled to a plasmonic nanogap (with a tapered geometry) filled with a 75-nm-thick Ge$_{2}$Sb$_{2}$Te$_{5}$ film protected from environmental oxidation by a 5-nm-thick SiO$_2$ capping layer. The state of GST can be reversibly switched between its amorphous (high resistance and low transmission) and crystalline (low resistance and high transmission) by sending either optical pulses through the waveguide or electrical pulses through the Au electrodes. The Si$_3$N$_4$ waveguide is coupled to a plasmonic MIM waveguide (formed by two Au electrodes and the portion of GST in the nanogap region as shown in the cross-section view in the bottom left panel of Figure~\ref{omid_fig4}D). The field enhancement inside the nanogap depends on the state of the GST. For the low-loss a-GST, the light propagating in the Si$_3$N$_4$ waveguide is strongly coupled to the nanogap region resulting in a strong field enhancement in that region (as shown in the bottom left panel in Figure~\ref{omid_fig4}D) and low optical transmission at the output. For highly absorptive c-GST, however, this coupling is highly reduced, resulting in a lower field enhancement (as the lower electric field intensity in Figure~\ref{omid_fig4}D demonstrates) and in turn, higher optical transmission. The right panels in Figure~\ref{omid_fig4}D show both the electrical (upper panel) and optical (lower panel) switching of  the device which can be readout with both electrical means (i.e., the change in the device’s resistance between two Au electrodes) and optical means (i.e., the change in the optical transmission of the waveguide). For optical switching, partial crystallization of GST is performed by sending piecewise optical set pulses (7.5 mW peak power for 8 ns followed by 3 mW for 400 ns) through the waveguide while a single rectangular reset pulse with 7.5 mW peak power and 8 ns duration is used for re-amorphization. Moreover, the authors successfully demonstrated the electrical switching of the device by sending an electrical reset pulse with 10 ns duration (5 ns rise/fall time) and 350 mV peak amplitude through the Au contacts for re-amorphization. A triangular pulse with 5 ns rise and 500 ns fall time and 350 mV peak amplitude was used for the crystallization process. In both cases, a CW optical probe signal and a constant voltage source ($V_\textrm{bias} = 50$ mV) were used to simultaneously monitor the change in transmission and resistance, respectively.

\subsection{Integrated phase-change photonic binary and multilevel memories} 

We live in an era with the explosive growth of information. The volume of data generated from smart devices, internet, fundamental research, and advanced technologies is exponentially increasing and reaching to a volume beyond the capacity of available computing/storage paradigms. Classical computers use the von Neumann framework in which data are transferred between the central processing unit (CPU) and memory units for processing and storage, respectively. Thus, the constant exchange of data between CPU and memory units limits the overall computing efficiency and leads to a traffic jam widely referred to as the von Neumann bottleneck \cite{zhang2019designing, zhai2018toward}. This bottleneck can be efficiently addressed through the introduction of a new paradigm that enables simultaneous storage and processing (i.e., arithmetic operations) of data in a single unit. PCMs seem a perfect fit for such a paradigm. First, the non-volatile nature of the phase change can supply the storage need. Second, the stimulation pulses, used for the phase conversion in PCMs, can be cleverly employed as the command signals for the implementation of logic operations. In other words, the width, height, and even shape of a pulse can be used as various knobs for the manipulation of a PCM phase, in which the data is stored. Third, the use of optical pulses (as the stimulation) brings the superiority of the optical domain over the electrical domain, namely the ultrawide bandwidth, multiplexing capability, low residual crosstalk, and high-speed operation. In addition, with the integration of PCMs with the rich portfolio of optical components (e.g., waveguides, resonators, and switches) the data processing capabilities can be extended from a unit cell (i.e., a PCM segment) to the system level. Thus, reading, writing, manipulation, and routing of the information can be all performed in an all-photonic scheme, a trend that will set the stage for the next-generation processors including emerging quantum computers.

\begin{figure*}
	\centering
	\includegraphics[trim={0cm 0cm 0cm 0cm},width=\textwidth, clip]{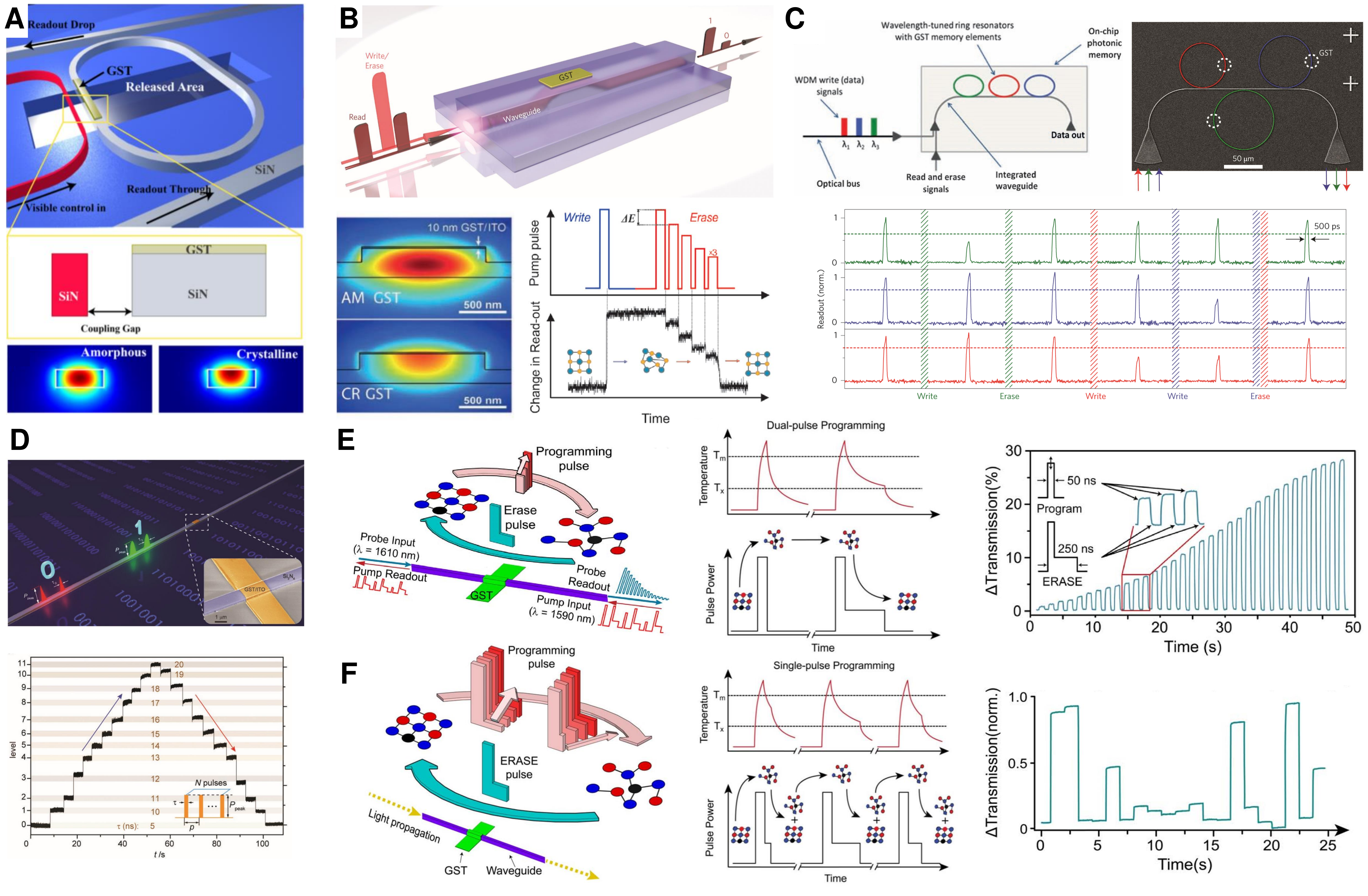}
	\caption{All-optical fully integrated on-chip phase-change binary and multilevel memories. (A) Top: The schematic of an all-optical multilevel memory consisting of a SiN waveguide coupled to a SiN ring resonator partially covered by a thin film of GST. Bottom: The simulated modal profile of the SiN waveguide covered by GST for the amorphous and crystalline states \cite{pernice2012photonic}. (B) The operation principle of an all-optical fully integrated on-chip multilevel memory. Top: The schematic of the memory including a straight waveguide partially covered by a GST element which can be programmed by ultrafast optical pulses (passed through the waveguide) and read by monitoring the amplitude of the optical signal at the exit port. Bottom-left: Simulated field distribution showing the low (high) absorption of propagating light in the waveguide by amorphous (crystalline) GST resulting in large (small) transmission at the exit port upon writing (erasing) the GST cell. Bottom-right: Demonstration of the write/erase process (upper panel) and resulting multilevel transmission in the readout port (lower panel) \cite{rios2015integrated}. (C) A multi-bit and multi-wavelength architecture. Top-left: The schematic of a WDM-enabled memory device with write signal at three different wavelengths ($\lambda_1$, $\lambda_2$, $\lambda_3$) associated with three different resonance wavelengths of the three ring resonators with three different radii. Top-right: The SEM image of the fabricated device shown in top-left \cite{rios2015integrated, wright2019integrated}. Bottom: The wavelength-selective readout of three PCM cells shown in the top panel by using 500 ps pulses. (D) Top: Schematic of an all-optical memory in which optical pulses with PWM are used to alter the pulse widths ($\tau_0$ and $\tau_1$ for example) which enables a control over the transmission through the waveguide. Bottom: The experimental demonstration of 12 different levels obtained by using a train of pulses with different widths $\tau$ (inset) from 5 to 20 ns \cite{cheng2018device}. (E) Left and middle: Concept of dual-pulse programming in which a single rectangular pulse with different amplitudes is used for amorphization (writing), while a double-step pulse is used for re-crystallization (erasing). Right: 34 memory levels achieve by using fixed-width programming dual-pulses (inset) following the approach shown in the middle panel \cite{li2019fast}. (F) Left and middle: Concept of single double-step programming in which a single dual-power pulse is used for arbitrary amorphization and re-crystallization of the GST by varying the amplitude and/or width of the second portion of the double-step pulse. Bottom: Arbitrary switching between transmission levels independent of the previous state \cite{li2019fast}.}
	\label{omid_fig5}
\end{figure*}


Optical property contrast, switching speed and minimum size of different optical states are the crucial parameters of PCMs for memory operation. Due to the dramatic optical contrast of GST over the visible and near-infrared spectral range, and its sub-nanosecond switching rate with high reproducibility and durability obtained over $10^{12}$ switching cycles \cite{lencer2008map}, this alloy has been used as a suitable candidate for an integrated PCM photonics non-volatile memory. As the first demonstration, in 2012, Pernice \textit{et al.} proposed an all-optical multilevel memory (in which the state of the GST can be changed from amorphous to different intermediate states) consisting of a microring resonator partially covered by a thin film of GST and coupled to a nanophotonic waveguide (see Figure~\ref{omid_fig5}A) all in a SiN platform to enable broadband optical applications \cite{pernice2012photonic}. A control port (the red SiN waveguide in Figure~\ref{omid_fig5}A) is used to deliver high-intensity optical pulses (with 600 fs width and less than 5.4 pJ for writing) for the phase switching (i.e., writing/erasing), and also to provide a low-intensity readout pulse for reading the states of the operating memory. When GST is in its low-loss amorphous state, the light traveling inside the microring resonator does not experience significant loss, so the propagating mode remains confined inside the SiN portion of the waveguide (see the bottom panel in Figure~\ref{omid_fig5}A) and the resonance condition of the microring resonator does not change. As a result, the input light is critically coupled to the ring resonator, and then, coupled to the Readout Drop port, resulting in a ~0$\%$ transmission in the Readout Trough port (level ‘0’ of the memory or ‘erasing’). On the other hand, upon the switching of the GST to its highly-absorptive crystalline state, the traveling light is highly absorbed by the GST (see the mode profile in the lower panel in Figure~\ref{omid_fig5}A), which changes the resonance condition of the microring resonator (the resonance wavelength is different from that of the input light). Therefor, the input light can no longer be coupled to the microring resonator resulting in a 90$\%$ transmission in the Readout Through port (level ‘1’ of the memory or ‘writing’).

Recently, multilevel data storage was also achieved by Rios \textit{et al.} via controlling the crystalline fraction in a PCM film. The authors used a simpler structure and experimentally demonstrated an eight-level memory with switching speeds of $\sim$1 GHz and switching energies of 13.4 pJ with the ability to arbitrarily switch between different intermediate states \cite{rios2015integrated}. Figure~\ref{omid_fig5}B shows the schematic of this functional layer. Since the GST cell is evanescently coupled to the waveguide, the phase state of the GST cell can determine the effective refractive index (see the bottom-left panel in Figure~\ref{omid_fig5}B), and accordingly, the optical attenuation through the waveguide. For switching the GST cell between its different states, high-power optical pulses are used. To amorphize (write) the GST cell, a single rectangular high-power pulse (with short width of 10 ns and switching energy of 13.4 pJ) is used (see the bottom-right panel in Figure~\ref{omid_fig5}B). Upon transition to the amorphous state, the low-loss GST cell does not absorb the passing light in the waveguide, leading to a high-transmission state (level ‘1’). On the other hand, a train of consecutive pulses with decreasing energy is used to perform a step-wise partial re-crystallization (erase) of the GST cell. When GST cell is in its fully crystalline state, the evanescent coupling between the waveguide and the highly-absorptive c-GST lead to a significant attenuation in the traveling light, resulting in a low-transmission state (level ‘0’). Therefore, the content of the memory cell is encoded in the state of the GST layer (i.e. ‘0’ in fully crystalline state and ‘1’ in amorphous state), or equivalently, stored in the intensity of the transmitted light at the readout port (‘0’ for low-transmission and ‘1’ for high-transmission). It is worth mentioning that partial re-crystallization of the GST using a single rectangular pulse is practically challenging because if the phase transition happens before the end of the pulse, the remaining optical energy can heat up the GST further to the melting temperature, and cause immediate re-amorphization. Therefore, by applying the erase scheme described above, they prevented the re-amorphization of already crystallized regions of the GST cell, and enabled access to intermediate re-crystallized states (or storage levels). For the readout of these states, we need to measure the degree of attenuation of the transmitted light associated with the level of re-crystallization of the GST cell. To do so, a single low-power light pulse (with 500 ps width and 0.48$\pm$0.03 pJ energy) is sent out along the waveguide, and the amount of the transmitted power identifies the state of the memory.

Authors in Ref.~\cite{rios2015integrated} further used the wavelength-division multiplexing (WDM) to demonstrate multi-wavelength operational schemes for accessing individual memory elements (see Figure~\ref{omid_fig5}C) \cite{wright2019integrated}. To do so, they employed the wavelength-filtering property of three ring resonators with three different resonance frequencies (~1 nm separation around 1550 nm) coupled to a bus waveguide. The design contains GST cells with a footprint of $1\times1$ $\mu$m$^2$ deposited on top of the ring resonators as shown in Figure~\ref{omid_fig5}C. Since only on-resonance lights can be coupled to the ring resonators (i.e., no off-resonant coupling), the three memory cells can be addressed selectively for writing, erasing, and reading cycles. Using such WDM technique as well as the multilevel addressing, very recently, Feldmann \textit{et al.} experimentally demonstrated the addressing of 256 memory elements to realize an all-photonic non-volatile memory with a storage capability of up to 512 bits of data in a $16\times16$ array of memory elements.  

In contrast to the approach in Ref.~\cite{rios2015integrated} (i.e., single pulse writing and multipulse erasing), Cheng \textit{et al.} demonstrated that using a pulse-width modulation (PWM) approach, the efficiency of the re-crystallization process in terms of speed, energy and process control can be improved \cite{cheng2018device}. Figure~\ref{omid_fig5}D demonstrates the concept of PWM switching in which optical pulses with different widths ($\tau$), but with a fixed peak power (amplitude), are traveling through a waveguide to switch a GST cell evanescently coupled to the waveguide. By changing $\tau$, the re-crystallization and amorphization of the GST can be controlled, leading to different levels of the memory (i.e, $\tau_0$ for level ‘0’, and $\tau_1$ for level ‘1’). More importantly, in their approach, each specific PWM sequence of pulses results in a particular re-crystallization state, independent of the starting state, meaning that each particular memory level can be directly accessed from any other stored level. As an example, Figure~\ref{omid_fig5}D shows a reversible 12-level memory implemented by a train of multiple ($N=10$) pulses with identical peak powers of $P_{\textrm{peak}}=1.4$ mW and the periodicity of $p=30$ ns (inset in Figure~\ref{omid_fig5}D) but different widths from 5 to 20 ns. Each particular width is used for reaching to a different level. More interestingly, this approach was used for the realization of ‘OR’ and ‘NAND’ logic using integrated photonic devices, which is discussed in detail in the Ref.~\cite{cheng2018device}.

All the proposed all-photonic data storage/addressing schemes listed above are promising as they can reduce the latency related to the electronic memories, and the high power consumption for back-and-force electrical-to-optical data conversion. However, the use of multiple high power pulses for re-crystallization (erase command) not only slows down the programming operation, but more importantly, lowers the power efficiency and the signal-to-noise ratio, limiting the number of intermediate levels in the memory \cite{rios2014chip, feldmann2017calculating, rios2019memory}. To overcome these shortcomings, Li \textit{et al.} proposed the same integrated photonic configuration to \cite{rios2015integrated}, but with a single double-step (dual-power) pulse for the erasing command (see Figure~\ref{omid_fig5}E). Using this modification, they successfully implemented a high-efficient multilevel memory with the ability to store up to 34 nonvolatile levels associated with over 5 bits \cite{li2019fast}. This multilevel operation was achieved by using a dual-pulse programming approach (see the left panel in Figure~\ref{omid_fig5}E) in which a fixed double-step erase pulse (with 250 ns width) was used for the step-wise crystallization, while a single rectangular writing pulse was used for amorphization (see the middle panel in Figure~\ref{omid_fig5}E). By monotonically increasing the amplitude of the programming pulse, they achieved 34 resolvable memory levels as shown in the right panel in Figure~\ref{omid_fig5}E. More interestingly, to enable arbitrary switching between the memory levels, authors used only one double-step pulse for both amorphization and re-crystallization (see the lower-left panel in Figure~\ref{omid_fig5}F). In this approach, the first part of the double-step pulse is used to remove the previous state of the memory by bringing the PCM above its melting temperature before re-crystallization (lower middle panel in Figure~\ref{omid_fig5}F). Then, by varying both the amplitude and width of the second part of the double-step pulse, arbitrary switching between different crystallization levels can be achieved (middle and right panel in Figure~\ref{omid_fig5}F).

We note that in studies listed above, GST was not included in the light path, but it was evanescently brought to its vicinity. The reason is that by placing the GST cell directly in the light path in the waveguide, the light passing through the GST cell would be significantly attenuated by highly absorptive c-GST, which significantly increases the power consumption of the device. Therefore, the GST cell is placed on top of the waveguide so that the light evanescently interacts with the GST cell.

\subsection{Integrated phase-change photonic arithmetic processors}

\begin{figure*}
	\centering
	\includegraphics[trim={0cm 0cm 0cm 0cm},width=\textwidth, clip]{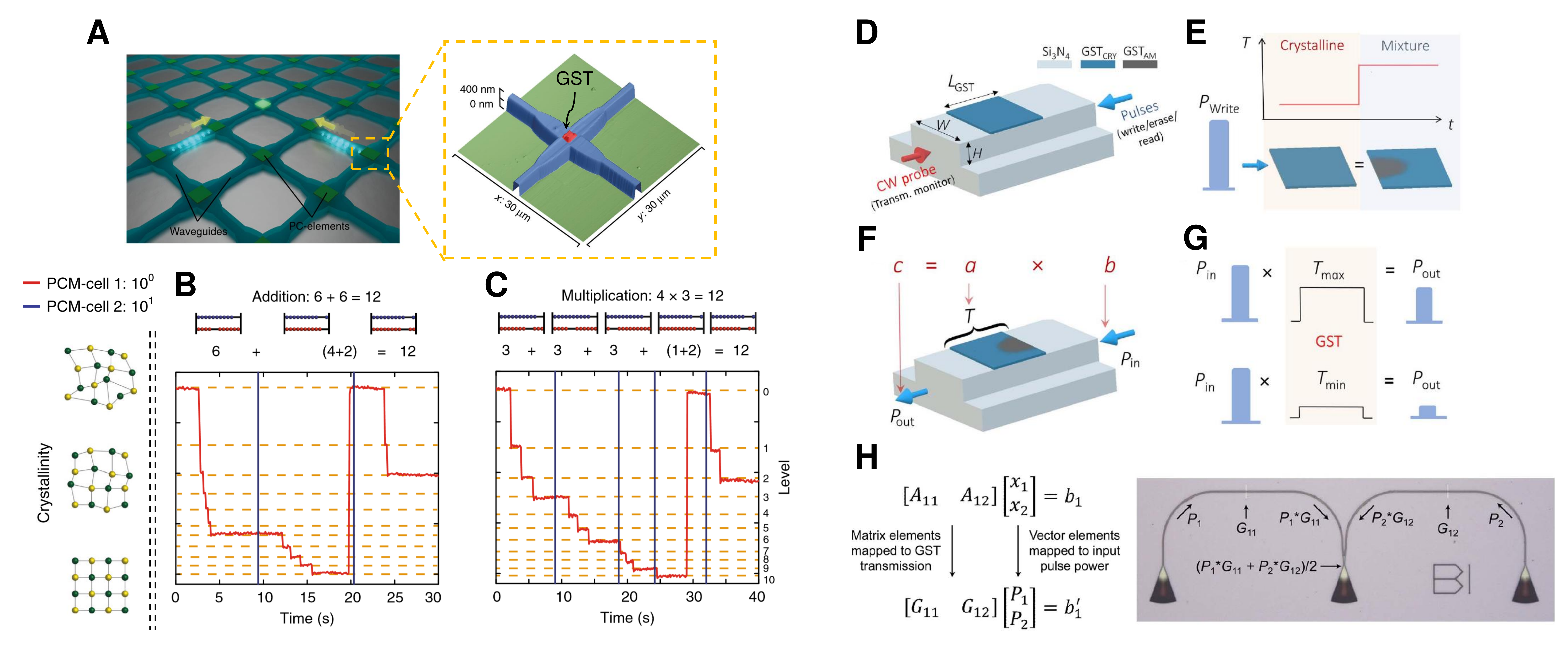}
	\caption{All-photonic integrated phase-change arithmetic processing. (A) An all-photonic abacus. Left: The schematic of the waveguide crossing array with a PCM cell placed at each waveguide crossing point, which can be addressed by two optical pulses. Right: AFM image of a single waveguide crossing \cite{feldmann2017calculating}. (B) and (C) Elementary arithmetic operations in base 10. (B) The operation principle of ‘$6+6=12$’. Resembling the operation of an abacus, when the tenth level (upper row in abacus, in the inset) is reached, a carryover is performed and the cell is reset to its initial state. (C) The operation principle of addition of multiplication ‘$4\times3=12$’, which is calculated by performing successive additions \cite{feldmann2017calculating}. (D) The schematic of the integrated GST-based photonic memory which is programmed and read by optical pulses and is monitored by a CW probe \cite{rios2019memory}. (E) Determination of the transmittance state, $T$, of the device by applying a write pulse $P_{\textrm{write}}$ for changing the phase-state of the GST cell from crystalline (base-line) to any intermediate state \cite{rios2019memory}. (F) The scheme for multiplication of two scalars $a$ (coded in the transmittance level associated to the phase state of the GST cell) and $b$ (coded into the input power $P_{\textrm{in}}$) \cite{rios2019memory}. (G) The readout power $P_{\textrm{out}}$ represents the results of the multiplication of low-power read pulse $P_{\textrm{in}}$ (which does not change the phase state of the GST cell) by the modulated transmittance, $T$, calculating the  $c=a\times b$ \cite{rios2019memory}. (H) Left: Extension of the operation principle shown in (D)-(G) to MV multiplication in which the matrix elements $A_{\textrm{ij}}$ (first element) are coded into the transmittance of the GST cells, i.e. $G_{\textrm{ij}}$, while the vector elements $x_\textrm{n}$ are coded into the input power of readout pulses $P_{\textrm{n}}$. Right: The optical image of the fabricated device used for the implementation of the MV multiplication of $(1\times 2) \times (2\times 1)$ \cite{rios2019memory, wright2019integrated}.}
	\label{omid_fig6}
\end{figure*}

The multi-state scheme offered by the integration of non-volatile PCMs with nanophotonics can set the stage for an all-photonic computer in which the processing and storage operations can simultaneously occur in the same location. This is in contrast to the von Neumann approach in conventional computers in which the data is being continuously transferred between a CPU (for processing) and an external memory (for storage). In a representative demonstration, Feldmann \textit{et al.} incorporated a GST cell at the crossing points of a rectangular waveguide array to enable independent manipulation of individual cells as shown in Figure~\ref{omid_fig6}A \cite{feldmann2017calculating}. This abacus works based on the progressive crystallization of embedded PCMs and can perform arithmetic operations such as addition, multiplication, subtraction, and division, a set of operations required for a non-von Neumann arithmetic calculator. As an example, the operation principle of the addition of ‘$6+6=12$’ in the base ten is represented in Figure~\ref{omid_fig6}B. Two PCM cells are used to represent the quantity of the first (red beads in the lower row of Figure~\ref{omid_fig6}B) and second (blue beads in the upper row of Figure~\ref{omid_fig6}B) digits of a two-digit system. These GST cells are initially in the amorphous states (high transmission) that represents the number zero. By performing a stepwise crystallization of the GST cell by using groups of identical picosecond pulses (each sets a predetermined crystallization fraction) different numbers are generated. In this case, each group consists of five consecutive pulses each with pulse energies of 12 pJ, and each group causes one crystallization step which is equal to one unit number. Due to operation in the base-10 system, the width and energies of the pulses are chosen such that 10 groups of pulses can fully crystallize the cell from its initial amorphous state, mimicking numbers 0 to 9 . Therefore, to perform ‘$6+6$’, first, the level of the first PCM cell is set to six by sending six groups of pulses down to the waveguide. Then, to add the second summand of ‘$6$’, more six groups of pulses are sent through the waveguide. Once the cell reaches to the 10th level, the device is reset (re-amorphized) to level 0 by using ten 19 pJ pulses. During the resetting of the first PCM-cell (to zero), the carryover of 10 is stored in the second PCM-cell by sending the corresponding pulse sequence and setting this cell to level 1. The resetting of the first cell and setting the second cell are carried out before applying the rest of the input sequence. Then, after applying remaining input sequences, the final states of the first and second cells are 2 and 1, respectively, representing the expected answer of 12 at the end of the calculation. This operation can be used for implementing other elementary operations  by sequential addition (multiplication), addition of the nine’s complement  of the second number to the first number (subtraction), and by sequential subtraction (division) \cite{wright2019integrated}. An example of multiplication of ‘$4\times3=12$’ is shown in Figure~\ref{omid_fig6}C.

More recently, the implementation of scalar, and more interestingly, matrix-vector (MV) multiplication using integrated phase-change photonics are demonstrated \cite{rios2019memory}. The MV multiplication is an important operation for modern data science such as image processing and machine learning. Figure~\ref{omid_fig6}D shows a photonic memory device consisting of a Si$_3$N$_4$/SiO$_2$ waveguide with a thin layer of GST (10 nm) and an ITO capping layer (10 nm) atop. The transmittance state $T$ is determined by a write pulse $P_{\textrm{write}}$ used for amorphization and re-crystallization of the GST cell (Figure~\ref{omid_fig6}E). For the MV multiplication operation, several scalar multiplications each in the form of ‘$c=a\times b$’ should be performed. To implement this multiplication, instead of the successive addition in two PCM cells (as explained above), authors used a single PCM cell to directly carry out this multiplication. In their approach, the multiplier, $a$, is mapped into the transmission state $T$ of the single cell (same as the previous approach), but the multiplicand, $b$, is mapped into the power of an input pulse $P_{\textrm{in}}$ (see Figure~\ref{omid_fig6}F). Therefore, the multiplication is performed when the propagating light passes the evanescently coupled PCM cell, and the calculated result can be readout from the output power $P_{\textrm{out}}$ (see Figure~\ref{omid_fig6}G). 

By designing an appropriate integrated photonic circuitry consisting of multiple PCM cells, the authors experimentally implemented the MV multiplication of a ($1\times2$) matrix by a ($2\times1$) vector which is shown in Figure~\ref{omid_fig6}H. Moreover, other form of arithmetic operations, such as logic operations can be performed, as described in details in Refs.~\cite{cheng2018device, youngblood2019tunable}.

\begin{figure*}
	\centering
	\includegraphics[trim={0cm 0cm 0cm 0cm},width=\textwidth, clip]{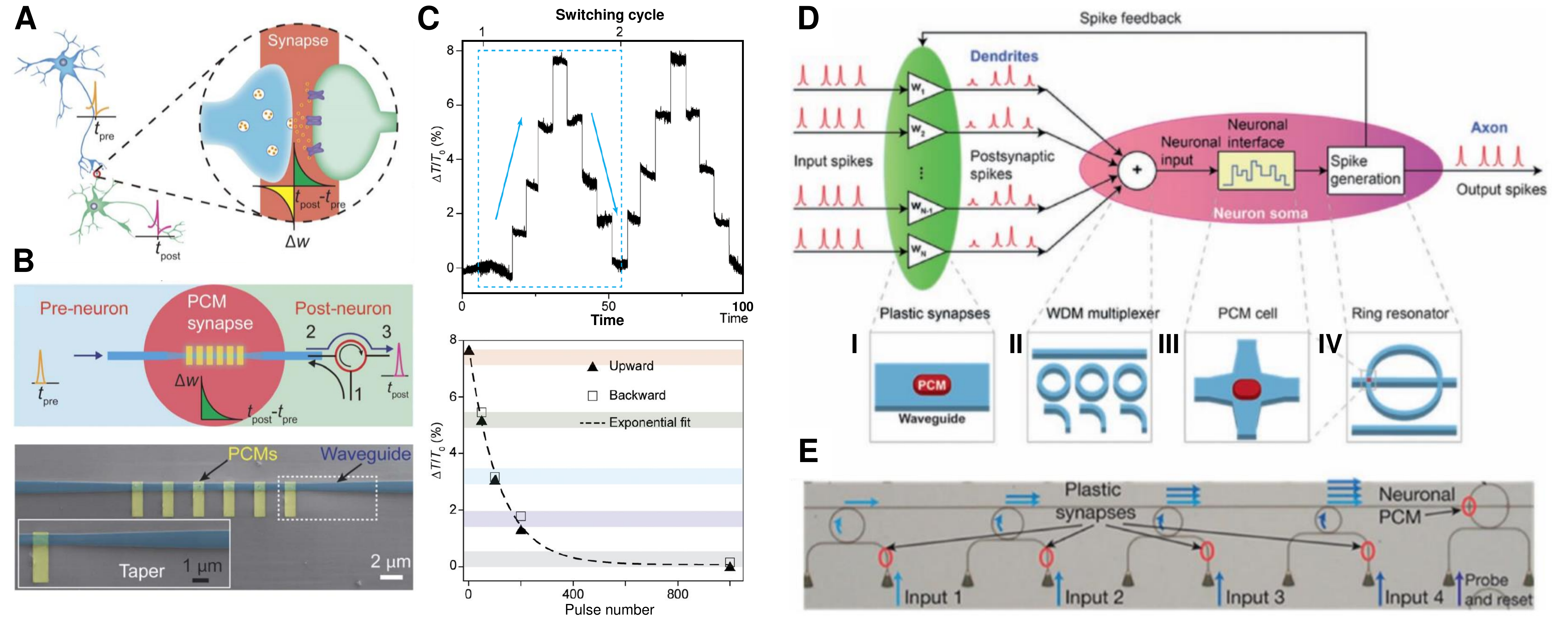}
	\caption{All-photonic fully integrated phase-change synapse and neuromorphic processor.
	(A) Conceptual representation of neurons and synapses. The inset shows a synapse junction and synaptic plasticity ruled by STDP, i.e., $\Delta w = Ae^{-\Delta t/\tau}$, in which $\Delta t = t_{\textrm{post}}-t_{\textrm{pre}}$, where $A$ and $\tau$ are constants \cite{cheng2017chip}.
	(B) Schematic of an on-chip integrated photonic synapse consisting of a tapered waveguide (dark blue) covered by discrete GST islands (yellow). GST islands are initially crystalline (i.e., $T_0$ = ``0''). Change in the transmission level (i.e., $\Delta T/T_0$), by modifying the crystallization level of GST, is attributed to the synaptic plasticity of the device ($\Delta w$). $\Delta T/T_0$ can be controlled by the number of triggering pulses similar to those in the STDP in the neural synapse shown in A \cite{cheng2017chip}. (C) Top: Five optical transmission levels of the devices related to the five weights of the photonic synapse which are obtained by optical pulses with 50 ns duration and 404.5 pJ energy. Bottom: The dependency of the synaptic weight ($\Delta T/T_0$) to the number of pulses which can be fitted by an exponential fitting function of $\Delta T=Ae^{r\times N}$ to resemble the STDP in a neural synapse shown in A \cite{cheng2017chip}. (D) Top: Schematic of the all-photonic neurosynaptic system implemented in \cite{feldmann2019all} comprising of N pre-synaptic neurons (generating pre-synaptic spikes) and one output neuron (the red region) which interconnected by N synapses (green region). Bottom: The input spikes are weighted by plastic synapses (I), and combined and directed into an output waveguide using the WDM technique (II), and fed into a PCM cell (III) on a large ring resonator (IV) mimicking the biological neural functionality of an actual neuron. (E) Optical microscope image of a fabricated neuron in which four inputs are weighted by four synapses and coupled to four small ring resonators to be delivered to the output waveguide \cite{feldmann2019all, wright2019integrated}.}
	\label{omid_fig7}
\end{figure*}

\subsection{Integrated phase-change photonic synapses and neuromorphic processors}

Another attractive capability of the integrated phase-change photonic platform is its ability to implement all-photonic synapses \cite{cheng2017chip, kuzum2013synaptic} and neurons \cite{feldmann2019all} as the central components for the hardware realization of brain-inspired neuromorphic computing which can enhance the efficiency of many important computational tasks, such as image processing, speech recognition, artificial intelligence, and deep learning, in terms of operation speed and energy consumption. The physical implementation of an integrated phase-change photonic synapse was first presented by Cheng \textit{et al.} \cite{cheng2017chip}. Figure~\ref{omid_fig7}A presents the conceptual illustration of neurons and synapses. A neuron (pre-neuron) generates action potentials (or spikes with fire time $t_{\textrm{pre}}$) that propagates along the axon and transmits through a junction to the next neuron (post-neuron) that generates the post-synaptic action potentials (fire time $t_{\textrm{post}}$). The junction is called a synapse (inset in Figure~\ref{omid_fig7}A) with the synaptic weight $w$ that determines the connection strength between the two neurons. The change in synaptic weight, $\Delta w$, is called synaptic plasticity which is determined by neural co-activities and spike timing based on the well-known Hebbian learning or spike timing-dependent plasticity (STDP) rule. According to an asymmetric form of STDP, this synaptic plasticity has the form of $\Delta w = Ae^{-\Delta t/\tau}$, in which $\Delta t = t_{\textrm{post}}-t_{\textrm{pre}}$, and $A$ and $\tau$ are constants, meaning that a smaller (larger) spike timing difference results in a larger (smaller) increase in synaptic weight (inset in Figure~\ref{omid_fig7}A). The structure of the photonic synapse proposed by Cheng \textit{et al.} is shown in Figure~\ref{omid_fig7}B which consists of a tapered SiN waveguide with evanescently coupled discrete Ge$_{2}$Sb$_{2}$Te$_{5}$ islands on top enabling control over the optical transmission level which represents the synaptic weight. A low-power probe light (pre-synaptic spike) from pre-neuron passes through the photonic synapse and is directed to the post-neuron (from port 2 to port 3) by using an optical circulator (the red circle in Figure~\ref{omid_fig7}B). This circulator also guides the high-power optical pump pulses (working in a different wavelength from probe pulses) entered from port 1 to port 2 to change the synaptic weight. Before measuring the optical transmission, the PCM islands are at fully crystalline state (by annealing the device on a hotplate at $\sim$250 $^{\circ}$C for 10 min) resulting in the optical transmission of $T_0$ which is defined as the baseline for readout and associated to a synaptic weight ‘$0$’. The reason behind using multiple PCM islands (instead of a single patch) is the need for precise control of synaptic weighting, and more importantly, the ability to move between different crystallization levels using fixed-duration, fixed-power pulses (see the upper panel in Figure~\ref{omid_fig7}C). The authors in Ref.~\cite{cheng2017chip} showed that by varying the number of optical pulses sent down to the waveguide, fine control over the optical transmission change ($\Delta T= T-T_0$) and thus synaptic plasticity ($\Delta w$) can be achieved. More importantly, they showed that the change in the synaptic weight (or equivalently $\Delta T$) is exponentially and monotonically dependent on the number of optical pulses ($N$) applied, as shown in the lower panel in Figure~\ref{omid_fig7}C, which resembles the STDP rule. Therefore, by using an all-photonic structure based on an interferometer and fitting the exponential function of $\Delta T=Ae^{r\times N}$, where $A$ and $r$ are fitting constants, to the synaptic plasticity governed by the STDP rule (i.e., $\Delta w = Ae^{-\Delta t/\tau}$), a correlation between $\Delta t$ an $N$ to mimic the STDP behavior can be developed.

In a follow-up work, Feldmann \textit{et al.} implemented an all-photonic spiking neural network consisting of four neurons, sixty synapses, and 140 optical elements in total, a platform capable of supervised and unsupervised learning \cite{feldmann2019all}. Figure~\ref{omid_fig7}D shows the schematic of the presented all-photonic network consisting of N pre-synaptic input neurons (which are not shown here) to generate the pre-synaptic input spikes, N interconnecting synapses (green region) to weight the pre-synaptic spikes, and one post-synaptic neuron (the red region) to receive the post-synaptic spikes and generate the output spikes. In this work, each synapse is implemented by a PCM cell placed on top of an optical waveguide in which the synaptic weighting (plasticity) is attributed to the phase transition of the synaptic PCM cell as described previously (panel I in Figure~\ref{omid_fig7}D). Then, the N input waveguides are coupled to N small ring resonators, each with a specific resonance wavelength (diameter), to combine and feed the input pulses (i.e., post-synaptic spikes) into a single output waveguide based on a WDM scheme shown in panel II in Figure~\ref{omid_fig7}D. This upper output waveguide guides the light into a neuronal (large) ring resonator with a PCM cell placed on top and at the waveguide crossing as shown in panels III and IV in Figure~\ref{omid_fig7}D. Upon phase switching of this neuronal PCM cell between amorphous and crystalline states, the resonance condition of the neuronal ring resonator shown in panel IV in Figure~\ref{omid_fig7}D is changed. This way, the coupling between this neuronal ring resonator and the lower probe waveguide can be controlled. The probe light in the lower waveguide is properly adjusted to be on-resonance when neuronal PCM cell is in the crystalline state. Therefore, when the incoming combined pulses are weak enough so that the neuronal PCM cell stays in crystalline state, the probe light is strongly coupled to the neuronal ring resonator. Consequently, no output spike is transmitted into the output waveguide. On the other hand, when the incoming pulses are strong enough to switch the PCM cell to its amorphous state, the probe light becomes off-resonance. As a result, it will be completely directed to the output waveguide without coupling to the ring resonator, resulting in output spikes. Therefor, this neuron resembles basic integrate-and-fire functionality of a biological neuron in which an output spike is generated only when the power of weighted sum of input spikes becomes more than a certain threshold. 

Figure~\ref{omid_fig7}D shows the optical image of one of these all-photonic neurons fabricated on a SiN platform. Using the neurosynaptic platform described above, authors in Ref.~\cite{feldmann2019all} employed both supervised (when the spiking feedback in Figure~\ref{omid_fig7}D is open) and unsupervised (when the spiking feedback in Figure~\ref{omid_fig7}D is closed) learning to perform the pattern recognition task.

\section{emergence of deep learning in analysis, design, and optimization of phase-change nanophotonics}

Despite striking advancements in meta-optics, considering it as a shifting paradigm is highly dependent on complex architectures realizing adaptive functionalities with unprecedented performances. Accordingly, employing new optimization algorithms to facilitate design of non-conventional MSs with large number of structural parameters is indispensable. Herein, we outline the fundamental challenges of traditional design approaches and discuss deep learning algorithms as a new paradigm in intelligent design and analysis of light-matter interactions of reconfigurable metadevices based on PCMs.

Increasing the number of design parameters helps the designer to search over a larger (i.e., higher dimensional) optimization landscape and find the desired structure more accurately. However, this accuracy is provided for a cost of increasing the design complexity. This has necessitated the development of new approaches for the design of MSs with a large number of design parameters. Additionally, new techniques are urgently needed for knowledge discovery, i.e., for obtaining valuable insight about the physics of light-matter interaction, in these nanostructures. Despite their importance in revolutionizing the realization of new MSs (and other optical devices), fundamental study of the dynamics of wave propagation and the role of sub-wavelength device features in these nanostructures on their spatial, spectral, and temporal responses to an incident electromagnetic wave is still at the early stages. As a result, the majority of demonstrated devices are formed by very time-consuming trial-and-error or blind optimization techniques with cumbersome numerical calculations. In addition, existing optimization approaches for MSs and other electromagnetic nanostructures rely on a significant amount of iterations to get to a final design from an initial guess. Even by allocation of remarkable computational resources, such time-consuming techniques are not guaranteed to converge to the global optimum. Moreover, they do not provide much insight into the dynamics of wave propagation inside these nanostructures unless a large set of simulations is performed. Thus, the development of new efficient approaches for rapid, accurate, and detailed analysis, design, and optimization of nanostructures is urgently needed and long overdue.

The widely used current approaches for design and optimization of MSs can be divided into two categories: 1) algorithms that rely on an initial guess and iterative search for an optimal result (i.e., sub optimal) in the predefined design space. This category includes the brute-force and evolutionary techniques (e.g., genetic algorithms) \cite{bossard2014near,khorasaninejad2017achromatic}. Such algorithms are customized for specific designs, i.e., for any arbitrary design problem, all the steps must be repeated to find the optimal structure. Although, they have been widely used in nanophotonic design and optimization problems, they suffer from huge computation complexity and often result in local optimum (rather than global optimum) solutions; 2) algorithms employing artificial neural networks (ANN) to optimize topologies of nanophotonic structures. While being more reliable in providing the global optimum designs, such data-driven algorithms require a significant amount of training instances to be practical for the real-world applications \cite{yao2019intelligent,tahersima2019deep,ma2018deep,sajedian2019optimisation,baxter2019plasmonic,long2019inverse,an2019generative,kudyshev2019rapid,qiu2019deep,hegde2019photonics, hegde2019accelerating,gao2019bidirectional,nadell2019deep, melati2019mapping, sajedian2020design, an2020freeform}.

\begin{figure}[t]
	\centering
	\includegraphics[trim=0cm 0cm 11cm 0cm,width=0.45\textwidth,clip]{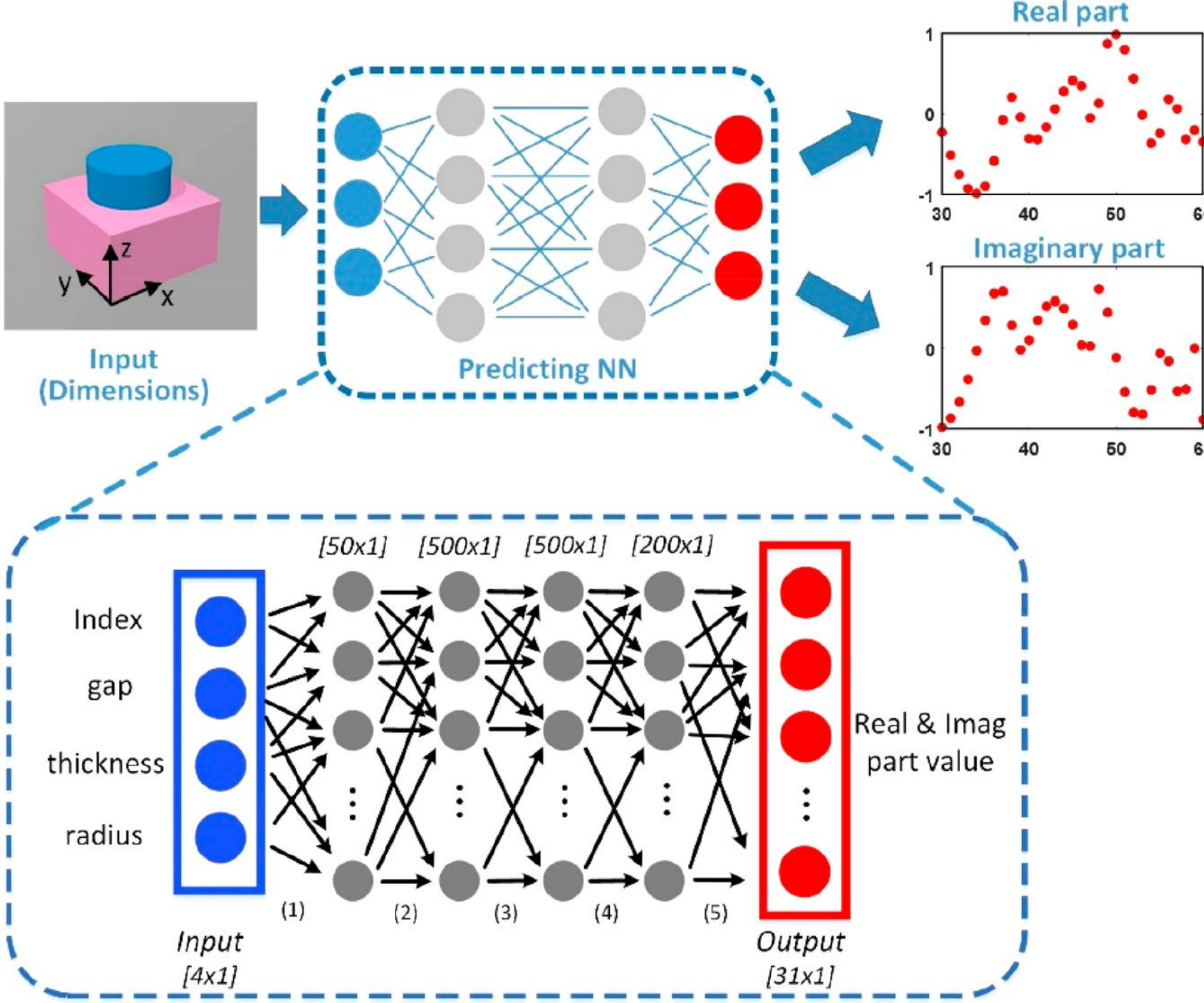}
	\caption{Training a fully connected DNN to relate geometrical and material features to the corresponding real and imaginary parts of the complex transmission coefficient \cite{an2019generative}.}
	\label{figY1}
\end{figure}

\begin{figure*}[t]
	\centering
	\includegraphics[trim=0cm 4.5cm 0cm 0cm,width=\textwidth,clip]{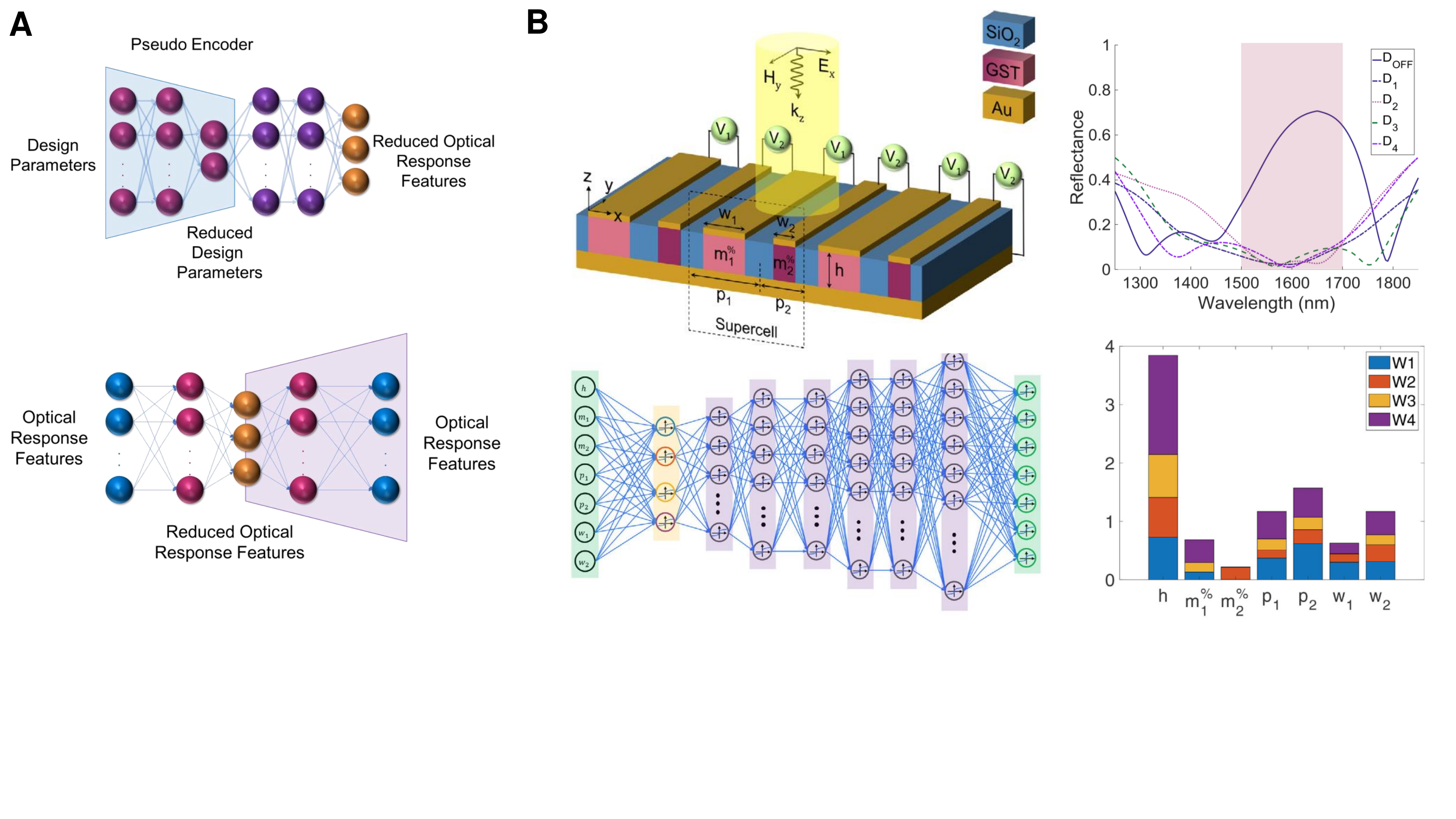}
	\caption{(A) Top: An autoencoder to reduce the dimensionality of the optical response features. Bottom: The pseudo-encoder architecture maps the design space to the reduced response space while extracting the reduced design parameters \cite{kiarashinejad2019deepdim}.
	(B) Left-top: Illustration of the hybrid plasmonic/phase-change material MS made of GST,  Au, and SiO$_2$.
	Right-top: The response of the optimal structure for a reflective band-stop filter for operation in the 1500-1700 nm window using a 5$\times$10 dimensionality-reduced algorithm.
	Left-bottom: Architecture of the single-layer pseudo-encoder for knowledge discovery.
	Right-bottom: The weights of the first layer of the pseudo-encoder \cite{kiarashinejad2019deep}.}
	\label{figY2}
\end{figure*}

Recently, An \textit{et al.} presented a deep learning-based (DL-based) technique to form a forward and inverse model for designing  phase-change reconfigurable MSs \cite{an2019deep}. As shown in Figure~\ref{figY1}, this method utilizes a deep neural network (DNN) to relate design and response spaces. To avoid the adverse effects of the potential abrupt changes in the phase or amplitude response of the MS in the training process, the authors wisely decompose the optical response in real and imaginary parts, which grant the smoothness of the data and thus, a more realistic trained model. As a result, the reported approach in Ref.~\cite{an2019deep} meaningfully accelerates the design of MSs compared to the conventional optimization approaches. Nevertheless, the efficiency of these techniques deteriorates as the number of design parameters (dimensionality of input space) and the response features (dimensionality of output space) increase. Generally, optimization algorithms suffer from high dimensionality of the input data. Therefore, it is of great importance to reduce the dimensionality of the data in the design of functional MSs.


The field of machine learning (ML) has seen extensive progress in the development of dimensionality-reduction (DR) techniques that aim to facilitate classification, data visualization, or to reduce the computational cost. Principal component analysis (PCA) is one of the most used and simple DR methods \cite{scholkopf1997kernel}. By projecting data points on the direction of the maximum variance, this linear method finds the best approximation of the data in the reduced-dimension space. On the other hand, nonlinear DR methods like kernel PCA  \cite{mika1999kernel}, autoencoder \cite{Hinton2006}, etc. can potentially provide a better approximation of the data \cite{perraul2013non, hinton2006reducing}.

Recently, Kiarashinejad \textit{et al.} \cite{kiarashinejad2019deepdim} developed a resolution to the existing challenges in both knowledge discovery and the design/optimization of the phase-change reconfigurable MSs by: 1) extensively reducing the dimensionality of the problem using ML-based and DL-based approaches, and 2) using the unique features of NNs for the analysis and design/optimization of the problem in the reduced dimensionality space. The resulting design tool can take a-priori information to enhance any desired feature (e.g., insensitivity to fabrication imperfections) throughout the process. It also provides a trade-off between the accepted error and the complexity (and time) of the simulations. Thus, it can be used for obtaining quick information about the role of design parameters on the overall device performance, getting detailed information about a specific feature of the device, or designing an optimal device for any given functionality. In addition, the technique in Ref.~\cite{kiarashinejad2019deepdim} is not problem-specific, i.e., once it is trained for a given class of MSs, it can be used to design structures with a wide range of functionalities. The schematic of this method is shown in Figure~\ref{figY2}A. First, an autoencoder (Figure~\ref{figY2}A (top panel)) is trained to reduce the dimensionality of the response space. After that, a feed forward NN (Figure~\ref{figY2}A (bottom panel)) is trained to model the relation between the design space and the reduced response space while reducing the dimensionality of the design space. This network, which is known as the pseudo-encoder, can predict the responses in the reduced response space given an unseen set of design parameters. The bottleneck (i.e., the NN layer with the minimum number of neurons) of the pseudo-encoder represents the reduced design space. Thanks to removing the redundancy in the data using the DR in both the design space and the response space, the relation between the reduced design space and the reduced response space can be considered as a one-to-one relation, and one can analytically solve the inverse problem. By solving the inverse problem, the reduced design parameters for a given desired response are found. Since the relation between the reduced design space and the original design space is not one-to-one, the last step of the optimization (i.e., finding the actual design parameters of the original design space) can be performed using other appropriate techniques. Kiarashinejad \textit{et al.} showed that a feed forward NN modeling the relation between the original design space and the reduced design space can be used to facilitate the exhaustive search of the original design space to find the set of optimal design. This approach is a major advancement in the field as it solves the major problem in NN-based approaches for solving the non-one-to-one MS design problems.

In a following work, Kiarashinejad \textit{et al.} showed that their approach can also provide valuable insights about the underlying physics of light-matter interaction in MSs \cite{kiarashinejad2019deep}. The authors used their technique to assess the roles of different design parameters on the response of phase-change reconfigurable MSs. This knowledge was then used to provide better designs for the same MS functionality. 

Figure~\ref{figY2}B (left-top panel) shows the schematic of a reconfigurable MS formed by an array of nanostripes of GST, SiO$_2$, and Au nanostripes on top of a blanket Au back reflector. The unit cell of the structure has 10 design parameters (7 geometrical design parameters and 3 GST crystalline states or equivalently indices of refraction). The response of the structure is its far-field reflection spectrum that is sampled uniformly at 200 points in the 400-800 nm wavelength range. This constitutes a 10$\times$200 dimensional problem. Using the new techniques in Ref.~\cite{kiarashinejad2019deep}, this problem can be reduced to a 5$\times$10 dimensional space with less than 0.01 $\%$ error using only 5000 simulations of the random structures to train the autoencoder and pseudo-encoder in Figure~\ref{figY2}A \cite{kiarashinejad2019deep}. Figure~\ref{figY2}B (right-top panel) shows the response of a MS designed as a band-stop reflective filter. Figure~\ref{figY2}B (left-bottom panel) shows a simple single-layer pseudo-encoder trained for the same structure (Figure~\ref{figY2}B (left-top panel)) for knowledge discovery using the weights of the first (leftmost) layer of the NN. Figure~\ref{figY2}B (right-bottom panel) shows the weights in the first layer. It clearly shows that the height of the structure is the most important design parameter while the indices of refraction of the GST nanostripes have lower importance. Indeed, these parameters have only non-negligible weights in relating to one node of the bottleneck layer in Figure~\ref{figY2}B (left-bottom panel). In other words, the three design parameters effectively act as one design parameter. Thus, the MS can be considerably simplified by using a blanket GST layer for reconfiguration.

\section {Perspective and Outlook}

Despite the impressive achievements of tunable nanophotonics with phase-change chalcogenide materials, which we outlined in this review, we anticipate further growth of interest in this topic. In this subsection, we describe our vision of further research and perspectives in this actively growing area.

First of all, PCMs will continue playing a critical role in actively tunable functional MSs. As we discussed above, MSs have found many applications from microwave to optical frequencies including ultra-thin lenses, imaging, optical information processing, nonlinear optics, analog computation, and invisibility cloak \cite{holloway2012overview, zhao2014recent, krasnok2018nonlinear}. Changing the direction in which an input electromagnetic wave is traveling is one of the most interesting features that MSs offer. In the past few years, gradient MSs have been extensively used to steer the direction of an incident wave in reflection and/or transmission. Although this class of MSs offers unprecedented possibilities to manipulate the direction of the wave flow, they still face challenges that limit their functionalities. Gradient metal MSs are mostly designed based on the generalized law of reflection and refraction \cite{yu2011light} suffer from low efficiencies \cite{ra2017metagratings, asadchy2016perfect}. The reason is twofold: high material losses of traditional (metallic) meta-atoms and low deflection efficiency arising from the fact that this approach does not take into account the impedance mismatch between the input and desired output waves. Moreover, most of the proposed fabrication techniques result in gradient MSs with static functionalities. In other words, once a MS is fabricated based on these methods its functionality is fixed and cannot be changed.

Recently, the implementation of gradient optical MSs based on high-index dielectrics causes a high interest because of their low optical loss and novel regimes of light-matter interaction \cite{baryshnikova2018optical, staude2017metamaterial, kuznetsov2016optically, krasnok2018spectroscopy}. PCMs are dielectrics and semiconductors with relatively high refractive index, and therefore, the optical response of PCM nanoparticles is governed by Mie resonances \cite{tian2019active}, exhibiting lower dissipation losses especially in near-IR and mid-IR. As we have discussed in previous sections, the ability of PCMs in active tuning along with their relatively weak optical loss makes them very attractive for various gradient MSs-based tunable photonic devices including reflectors \cite{chu2016active}, filters \cite{dong2018tunable}, polarizers \cite{zhu2018controlling}, and holograms \cite{wan2017metasurface}. We envision further growing of research works on this topic in the near future. 

It has been recently shown that the issue of low deflection efficiency of gradient MSs can be overcome with the new approach to the MSs design based on the concept of metagratings \cite{ra2017metagratings}. Metagratings are periodic arrays of polarizable particles where the period is comparable to the wavelength. When such a structure is illuminated with an incident wave, in addition to possible scattering into the specular directions, other Floquet modes can, in principle, carry energy away from the surface. Depending on the angle in which the incident wave is illuminating the surface and also the periodicity of the metagrating, the different number of propagating Floquet modes can be present. The direction of these Floquet modes can be controlled by the periodicity of the structure making such surfaces a rich platform for efficient manipulation of electromagnetic waves. MSs designed based on this approach can reroute the incident wave to the desired direction with 100\% efficiency. Also, MSs designed based on this method do not require subwavelength structures lifting fabrication challenges faced by conventional gradient MSs. The use of reversible PCMs would make it possible to erase and rewrite the pattern on the PCM surfaces to achieve the metagrating operation that can be used for controlling electromagnetic waves in a wide range of wavelengths and frequencies. An integration of this platform with a 2D pulse shaper can be used to imprint the designed pattern on the chosen substrate materials, as it has been done for zone-plate devices in Ref.~\cite{wang2016optically} or for polariton nanophotonics in Ref.~\cite{chaudhary2019polariton}.

Recent studies in light scattering engineering made it feasible to tailor structures with unique scattering properties. One of such unusual scattering phenomena is the bound states in the continuum (BICs) with unboundedly large Q-factors and vanishing scattering lines in open systems, which have been recently discovered in photonics \cite{marinica2008bound, hsu2016bound, monticone2014embedded, monticone2018trapping, krasnok2019anomalies, koshelev2018asymmetric}. The existence of these states allows designing optical MSs with diverging and potentially infinite lifetimes. It has been demonstrated that the use of BICs in lasers, nonlinear optics, and sensors allows significant boosting of the performance of these vital applications. Nanoparticles made of high-index dielectrics governed by Mie modes have been demonstrated to be a powerful platform for tailoring BICs-supporting nanostructures and MSs \cite{koshelev2018asymmetric, liu2019high}. However, the reported devices supporting BICs are rather static, with just a few exceptions \cite{fan2019dynamic}. We belive that BIC-supporting dielectric MSs can be made of PCMs that will make them tunable and reconfigurable significantly expanding the possibilities of their practical application.

Another example of unusual scattering phenomena that can benefit from the use of PCMs is the so-called exceptional points (EPs) arising in non-Hermitian structures \cite{krasnok2019anomalies, miri2019exceptional}. Exceptional points correspond to the case when pairs of eigenstates and corresponding eigenvalues of a non-Hermitian operator (e.g., Hamiltonian or S-matrix) coalesce. In structures comprising a balanced amount of gain and loss known as parity-time symmetric \cite{krasnok2019anomalies}, EPs can arise at a real frequency and hence can be detected via CW monochromatic excitation. The existence of EPs results in a nonlinear dependence of the response of a system to variation of the system’s parameters (e.g., frequency detuning, amount of loss or gain). This peculiarity makes them interesting for ultra-sensitive sensor devices, single-mode lasers and quantum optics \cite{krasnok2019anomalies}. The use of PCMs provides a fruitful approach to precise tailoring material loss in space and time and hence can facilitate the designing and realization of structures supporting EPs \cite{huang2017switching} especially in the integrated on-chip scenario. 

The advances of nanophotonics can facilitate the reduction of the power consumption of PCMs-based photonics structures. Usually, to take full advantage of a pronounced change of the dielectric function of PCMs, bare PCMs have to be transformed from the amorphous to the fully crystalline phase, which requires relatively strong laser fields or strong external electrical bias \cite{tian2019active, wuttig2007phase, alaee2016phase, huang2018switching, wuttig2017phase}. For example, the full transition from crystalline to amorphous phase for bare GST material requires a peak intensity of $\sim$1 mW$\mu$m$^{-2}$ \cite{nvemec2009ge}. Although such intensities are not weak, they are much weaker than those used in other reversible approaches including electron-hole plasma excitation that requires laser intensities of $\sim$10$^3$ – 105$^5$ mW$\mu$m$^{-2}$ \cite{baranov2016nonlinear, makarov2015tuning, shcherbakov2015ultrafast, yang2015nonlinear, paniagua2019active}, and comparable with the intensity of thermo-optical nonlinearity of Si \cite{zograf2017resonant}. As we discussed above, the power consumption and tuning intensities of PCMs-based devices can be significantly reduced via strong optical heating araising at surface plasmonic resonances of metal nanostructures \cite{dong2018tunable, cao2018tuneable, cao2013rapid}. 

The use of PCM in nanoantennas with several employed coupled resonant modes allows efficient tuning of their power pattern \cite{alaee2016phase, lepeshov2019nonscattering}. In contrast to the spectral properties of an antenna, its scattering pattern demonstrates a stronger tunability because it relies rather on the relative phases of involved modes. For example, in paper \cite{lepeshov2019nonscattering}, a hybrid metal-semiconductor core-shell nanoantenna made of silver Ag core and phase-changing GeTe shell has been designed to possess switching between the superscattering regime and nonradiative cloaking state with remarkably low power consumption. It has been shown that tuning of the PCM crystallinity leads to a tremendous change in the total ($\sim$15 times) and forward ($\sim$100 times) scattering with the characteristic intensity of $\sim$10 $\mu$W$\mu$m$^{-2}$ which is $\sim$2 orders of magnitude weaker than for bare GeTe material or Si particles tuning by thermal heating.

To harness the disruptive concept of meta-optics in the realization of on-demand adaptive functionalities, design of dynamically reconfigurable MSs with unprecedented performance is indispensable. Although increasing the number of design parameters spans the optimization landscape (or equivalently space dimensionality) that helps electromagnetic designers find the best possible solution, this freedom adds serious complexities to the design process. While employing conventional approaches such as topology and multi-objective optimization techniques have resulted in high-performance, complicated metadevices, so far, such methods have been limited to the context of non-tunable meta-optics with predefined fixed functionalities. Newly emergent deep learning approaches as a powerful platform for intelligent analysis, design, and optimization of tunable MSs, leveraging functional materials including PCMS, can empower electromagnetic designers to effectively overcome hyper-dimensional problems with reasonable computational resources.


Integrating chalcogenide PCMs with PICs has enabled reversible switching between arbitrary states, which makes them near-ideal candidates for all-photonic implementation of switches and modulators, memories, arithmetic processors and neurosynaptic systems. However, there is still room for improvement of integrated phase-change photonic devices to outperform their electronic counterparts \cite{wright2019integrated}. On top is the switching energy required for phase transition of the PCM. Although the switching energy falls in the range from 50 pJ to a few nJ \cite{rios2015integrated, stegmaier2017nonvolatile, wu2018low, li2019fast}, a 0.1 pJ switching energy has already been demonstrated in an electronic phase-change memory \cite{xiong2011low}. Therefore, it is very crucial to develop new architectures such as plasmonic nanogaps \cite{gemo2019plasmonically, farmakidis2019plasmonic}, photonic crystal cavities \cite{von2018reconfigurable} and crossing waveguides \cite{zhang2019nonvolatile, zhang2019miniature} offering hotspots in which the PCM elements could be exposed to an enhanced electromagnetic field to reduce the required transformation energy. Another aspect is the switching speed. While several types of triggering optical/electical pulses (single pulses with tens of $\sim$ns widths, or groups of $\sim$ps pulses) have been used for stepwise crystallization of the PCM element, the finite crystallization time of PCMs has limited the switching speed \cite{loke2012breaking}. To improve this, pulse shape engineering is necessary. Recently, a qualitative study on the origin and mechanism of the phase switching in the PCMs has been reported in Ref.~\cite{rios2018controlled} to describe the switching dynamics for precise control during amorphization and crytallization. The third area for improvement is the maximum number of cycles that a PCM cell can be switched between states, or the durability of the switching process. An endurance of more than $10^{12}$ cycles has been reported for electronic phase-change memories \cite{kuzum2011nanoelectronic}. Therefore, for the realization of all-photonic ultrafast processing applications, a higher number of cycles is needed. To do so, one may use some new materials development approaches reported in Refs.~\cite{salinga2018monatomic, zhang2018single}.

\noindent \textbf{Acknowledgment:} 
This work was supported by Office of Naval Research (ONR)(N00014-18-1-2055, Dr. B. Bennett).

\noindent \textbf{Conflict of interest:} 
The authors declare no competing financial interest.

\newpage



\providecommand{\noopsort}[1]{}\providecommand{\singleletter}[1]{#1}%

\end{document}